\documentclass[twocolumn,showpacs,prb,citeautoscript,amsmath,amssymb,floatfix,eqsecnum]{revtex4-1}
\usepackage{amsmath,amssymb,color}
\usepackage{bm}
\usepackage{graphicx}
\usepackage{psfrag}
\newcommand{\bd}{\bm}
\allowdisplaybreaks

\begin{document}

\title{Critical pairing fluctuations 
in the normal state of a superconductor: 
pseudogap and quasi-particle damping}

\author{Philipp Lange, Oleksandr Tsyplyatyev, and Peter Kopietz}
  
\affiliation{Institut f\"{u}r Theoretische Physik, Universit\"{a}t
 Frankfurt, Max-von-Laue Strasse 1, 60438 Frankfurt, Germany}

\date{April 18, 2017}

 \begin{abstract}

We study the effect
of critical pairing fluctuations
on the electronic properties in the normal state of a clean superconductor in three dimensions. 
Using a functional renormalization group approach to take the 
non-Gaussian nature of critical fluctuations into account,
we show microscopically that
in the BCS regime, where the inverse coherence length 
is much smaller than the Fermi wavevector, critical pairing
fluctuations give rise to a 
non-analytic contribution to the quasi-particle damping of order
$ T_c \sqrt{Gi} \ln ( 80 / Gi )$,
where the Ginzburg-Levanyuk number $Gi$
is a dimensionless measure for the width of the critical region.
As a consequence, there is a temperature window above $T_c$ where
the quasiparticle damping due to critical 
pairing fluctuations can be larger than the usual $T^2$-Fermi liquid 
damping due to non-critical scattering processes.
On the other hand, 
in the strong coupling regime where $Gi$ is of order unity,
we find that 
the quasiparticle damping due to critical pairing fluctuations is 
proportional to the temperature.
Moreover, we show that in the vicinity of the critical temperature $T_c$
the electronic density of states exhibits a fluctuation-induced
pseudogap. 
We also use functional renormalization group methods to derive and classify 
various types of processes induced by the pairing interaction in Fermi systems close to the superconducting instability.

\end{abstract}

\maketitle

\section{Introduction}
The BCS mean-field theory has been tremendously successful to
explain the physical properties of superconductors,
but the true critical behavior of the classical phase transition between a
normal metal and a superconductor
 is not mean-field like but
belongs to the universality class of the classical XY-model.  
Fortunately, in conventional   
superconductors the critical region where 
fluctuation effects are important is extremely small, so that for all practical purposes
the mean-field approximation is sufficient \cite{Larkin05}.
The smallness of the critical region in weakly coupled BCS
 superconductors is due to the fact that in these systems
the zero temperature coherence length $\xi_0$, which measures the typical size of the 
Cooper pairs, is many orders of magnitude larger than the lattice spacing.
A dimensionless measure of the temperature range $\delta T$ around the
critical temperature $T_c$ where fluctuations are important is given by the 
Ginzburg-Levanyuk number \cite{Ginzburg60,Levanyuk59}  
$Gi$, which for a clean three-dimensional superconductor can be written 
as \cite{Larkin05}
 \begin{equation}
 Gi = \frac{ \delta T}{T_c} \approx  0.8 \left( \frac{\pi T_c}{E_F} \right)^4 ,
 \label{eq:Gi}
 \end{equation}
where $E_F$ is the Fermi energy.
In the weak coupling BCS regime the value of $Gi$ is typically in the range between 
$10^{-14}$ and $10^{-12}$, so that the critical region cannot be resolved experimentally.
On the other hand, in strongly correlated superconductors the inverse coherence length
$1/\xi_0$ can have the same order of magnitude as the Fermi momentum $k_F$. In this case $Gi$ is of the order of unity and the critical regime is experimentally accessible.  
Another class of experimentally tunable systems
where fluctuations of the superfluid order parameter cannot be neglected are
the ultracold fermions with attractive interaction in the vicinity of
the unitary point where the two-body scattering length diverges \cite{Bloch08}.

Although fluctuation effects in superconductors and superfluids have been studied for many 
decades \cite{Larkin05}, there are still some open questions. 
In particular, the renormalization of the electronic single-particle excitations
in the normal state at or slightly above the critical temperature are not completely understood. 
In a seminal work by Aslamazov and Larkin\cite{Aslamazov69}, the transport time and its effect on conductivity were shown to be divergent at 
the transition temperature within the ladder approximation.  
The effect of superconducting fluctuations on the density of states 
and the tunneling resistance has been studied within a perturbative approach to first order in the 
strength of the superconducting interaction\cite{DiCastro90}. This approximation is expected to break down sufficiently 
close to the critical temperature \cite{Larkin05}, where the non-Gaussian nature of the pairing 
fluctuations and the renormalization of the electronic single-particle excitations must be taken into account. 
The single-particle spectral function was calculated numerically in Refs.~[\onlinecite{Palestini12}] and [\onlinecite{Reichl15}]
using the same ladder approximation, which corresponds to treating fluctuations of the superconducting order 
parameter only on the Gaussian level. To the best of our knowledge, a quantitative analysis of the electronic 
density of states and the quasi-particle damping beyond this approximation does not exist in the literature.
In the superconducting phase, the modification of the electronic density of states due to Gaussian order 
parameter fluctuations has been studied by Lerch {\it{et al.}} \cite{Lerch08}, who found an unexpected logarithmic renormalization of the BCS result. 
In the present work we focus on the temperature regime
above the critical temperature $T_c$ where the system is in the normal state and hence 
the anomalous part of the electronic self-energy vanishes.
This simplifies the calculations and 
enables us to include the effect of non-Gaussian critical order-parameter fluctuations
on the single-particle spectrum using renormalization group methods.

Diagrammatically, retaining Gaussian fluctuations of the superconducting 
order parameter is
equivalent to calculating the effective two-body interaction between 
fermions in ladder approximation, which amounts to 
solving the Bethe-Salpeter equation for the effective interaction in the particle-particle 
channel \cite{Larkin05}. Higher order interaction processes involving fluctuations with arbitrary momentum transfer give rise to non-Gaussian order parameter fluctuations.
The next order effect of the pairing coupling on the critical temperature 
of a weakly interacting superconductor in the BCS regime 
has first been calculated by Gorkov and 
Melik-Barkhudarov (GM) \cite{Gorkov61}, who showed that even for arbitrarily weak bare
interaction the fluctuations lead to a finite decrease of the 
critical temperature.
In recent years the effects of induced interactions due to non-Gaussian pairing fluctuations have been studied for various other setups, such as 
systems involving more than two
fermion flavors \cite{Heiselberg00}, effective models
describing the crossover from a BCS superconductor to a Bose-Einstein condensate 
(BCS-BEC crossover) \cite{Chen05,Floerchinger08,Yu09}, 
and multi-band models describing
the iron-based superconductors \cite{Chubukov16}. 
Moreover, it has been shown \cite{Lederer15} that in the vicinity of a nematic
quantum critical point the induced interactions mediated by soft fluctuations
associated with the nematic order parameter can enhance the critical temperature for
superconductivity.

In this work we use a functional renormalization group (FRG) 
approach \cite{Kopietz10,Metzner12}
to derive and classify the induced interactions responsible for the
corrections to BCS theory. Our approach is based on the vertex expansion
and partial bosonization in the particle-particle channel \cite{Kopietz10,Bartosch09}, and is therefore
complementary to recent work by Tanizaki {\it{et al.}} \cite{Tanizaki14}, who
have used a purely fermionic formulation of the FRG to 
calculate the correction to the BCS result for $T_c$ due to
pairing fluctuations. Our main focus is 
the effect of critical pairing fluctuations
on the spectrum of single-particle excitations in the normal state.

Let us give a brief overview of the rest of this work and summarize our main results.
In Sec.~\ref{sec:induced} we derive an effective field theory
describing normal fermions which are coupled to pairing fluctuations.
We also show how the GM correction\cite{Gorkov61}
to the critical temperature $T_c$ can be obtained within our approach, and 
that the GM result for
$T_c$ is modified if the chemical potential (and not the density of the electrons) 
is held constant.
Our FRG approach for this model is developed in Sec.~\ref{sec:FRG}, where
we also explain the emergence of various types of
induced interaction processes due to pairing fluctuations
from the renormalization group point of view.
 
In Sec.~\ref{sec:selfGauss} we then discuss the effect of pairing
fluctuations of the superconducting order parameter on
the fermionic self-energy and the density of states within the 
ladder approximation. 
We show that in this approximation 
the density of states exhibits a finite pseudogap but
the damping of quasiparticles with momenta
on the Fermi surface still diverges 
logarithmically as $\ln [ T_c /(T - T_c)]$ for $T \rightarrow T_c$.
While the emergence of a pseudogap due to fluctuations above $T_c$ 
has been intensely investigated in the past\cite{Stajic04,Chien10,Hu10,Magierski11,Mueller17}, it is somewhat surprising that 
the logarithmic divergence of the quasiparticle damping
in a clean three-dimensional superconductor
has not been noticed in the previous literature on the subject \cite{Larkin05}.
This singularity can be cured by taking into account the 
finite lifetime of the quasiparticles in intermediate states, or by 
including non-Gaussian
 critical pairing fluctuations 
which generate a finite anomalous dimension $\eta$ of the pairing fluctuations.
In Sec.~\ref{sec:FRG} we take both effects consistently into account using 
a specific implementation of the FRG.
We find that the quasiparticle damping at $T=T_c$ 
due to critical order-parameter fluctuations
has in the BCS regime the non-analytic form,
 \begin{equation}
 \gamma_{\rm crit} \approx C  \frac{T_c^3}{E_F^2} \ln \left( \frac{E_F }{ T_c} \right)
  \approx T_c  \sqrt{Gi} \ln \left( \frac{80}{ Gi} \right),
 \label{eq:gammares}
 \end{equation}
where our estimate for the numerical prefactor is $C \approx 30$.
Due to the rather large value of $C$, in a sizable 
regime of temperatures close to $T_c$
the critical contribution (\ref{eq:gammares})
to the quasiparticle damping dominates the usual $T^2$-Fermi liquid behavior due to
non-critical interaction processes, as illustrated in
Fig.~\ref{fig:surface_damping} below.
Moreover, we also show that in the strongly interacting superconductors, where the inverse
coherence length
can have the same order of magnitude as the Fermi momentum $k_F$, the 
quasiparticle damping due to critical order-parameter fluctuations 
is proportional to the temperature.
Finally, in Sec.~\ref{sec:conclusions} we present our conclusions and discuss 
possible extensions of the methods developed in this work.

Further technical details are given in five appendices. 
In Appendix~A we discuss in detail the approximations which are necessary to derive 
the GM result \cite{Gorkov61} for 
the critical temperature from the interaction corrections to the particle-particle bubble.
In Appendix B we write down exact FRG flow equations for the induced interactions
in our model. The momentum-dependence of the non-interacting particle-particle
bubble is derived in Appendix C, while in Appendix D we justify why 
in the vicinity of $T_c$ it is sufficient to retain only the zeroth  Matsubara frequency (associated with classical fluctuations) in the bosonic correlation function. Finally, in Appendix F we improve the
FRG calculation of the quasiparticle damping of Sec.~\ref{subsec:FRGdamp}
by taking into account higher order vertex corrections.

\section{Induced interactions in fermionic superfluids}
\label{sec:induced}

\subsection{Effective field theory for superfluid fluctuations}

\begin{figure}[tb]
  \centering
\vspace{7mm}
 \includegraphics[width=0.3\textwidth]{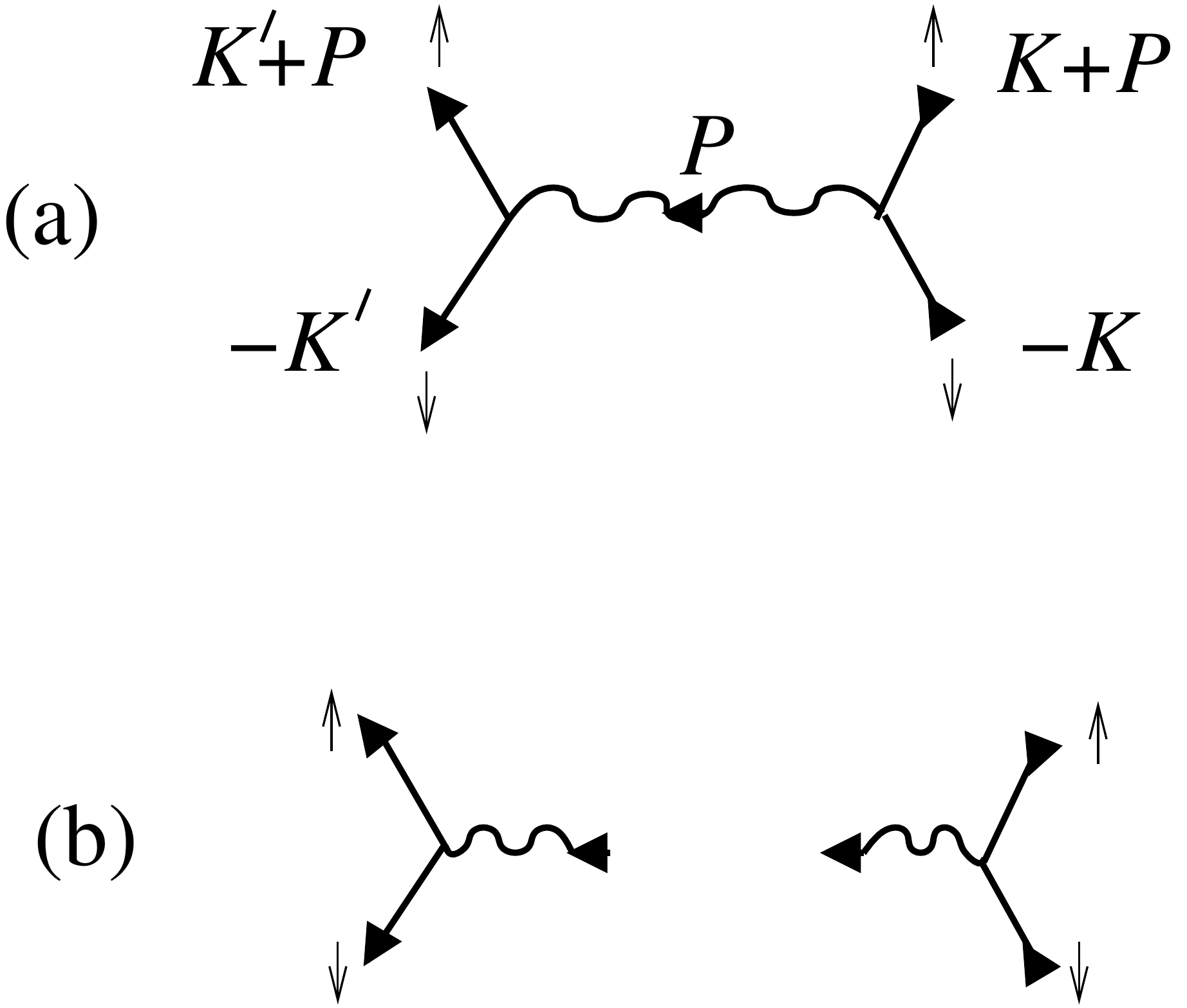}
  \caption{%
(a) Graphical representation of the bare interaction in
Eq.~(\ref{eq:Sbare}). The wavy arrow represents the bare coupling constant
$g_0$, where the arrow indicates the flow of the total energy-momentum
$P$ carried by the interaction.
Incoming external arrows represent ${c}_{\sigma}$ while outgoing
arrows represent $\bar{c}_{\sigma}$. The spin projections $\sigma =
\uparrow, \downarrow$ and the energy-momentum labels
are written next to the legs.
(b) Equivalent three-legged vertices after Hubbard-Stratonovich transformation in the particle-particle channel, see
Eq.~(\ref{eq:Sbare2}). Incoming wavy arrows represent 
the bosonic Hubbard-Stratonovich field ${\psi}$, while incoming wavy arrows represent
the complex conjugate field $\bar{\psi}$.
}
\label{fig:bareint}
\end{figure}

\begin{figure}[tb]
  \centering
 \includegraphics[width=0.45\textwidth]{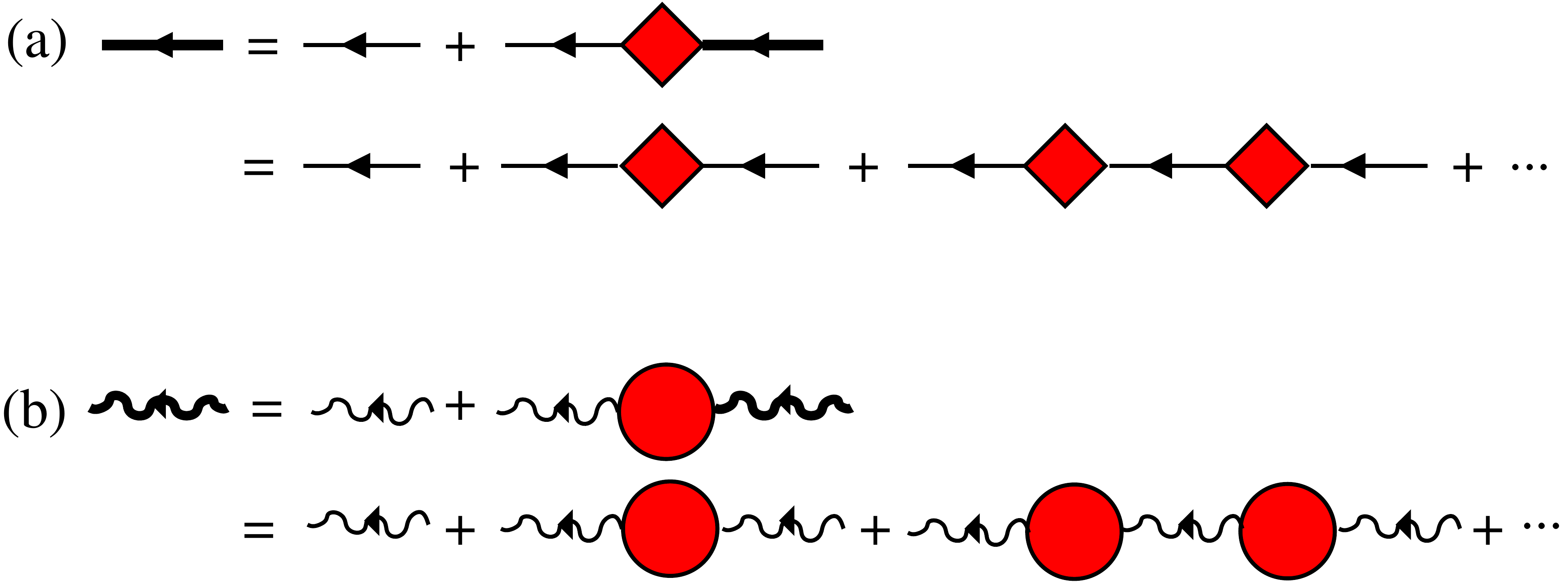}
\vspace{5mm}
  \vspace{-4mm}
  \caption{%
Dyson equations for the fermionic and bosonic propagators.
(a) represents the Dyson equation (\ref{eq:Dysonsigma}) for the
fermionc propagator, denoted by a thick solid arrow. 
The thin solid arrows represent the bare fermionic propagator while the one-particle
irreducible self-energy $\Sigma ( K )$ is represented by a shaded box.
(b) represents the corresponding bosonic Dyson equation (\ref{eq:Dysonphi}).
Here the thick wavy arrow represents the exact bosonic propagator, while the
thin wavy arrow represents the bare interaction. The shaded circle represents the
one-interaction-line irreducible bosonic self-energy $\Phi ( P )$, 
which can be identified with the exact irreducible particle-particle bubble.
}
\label{fig:Dyson}
\end{figure}

\begin{figure}[tb]
  \centering
 \includegraphics[width=0.4\textwidth]{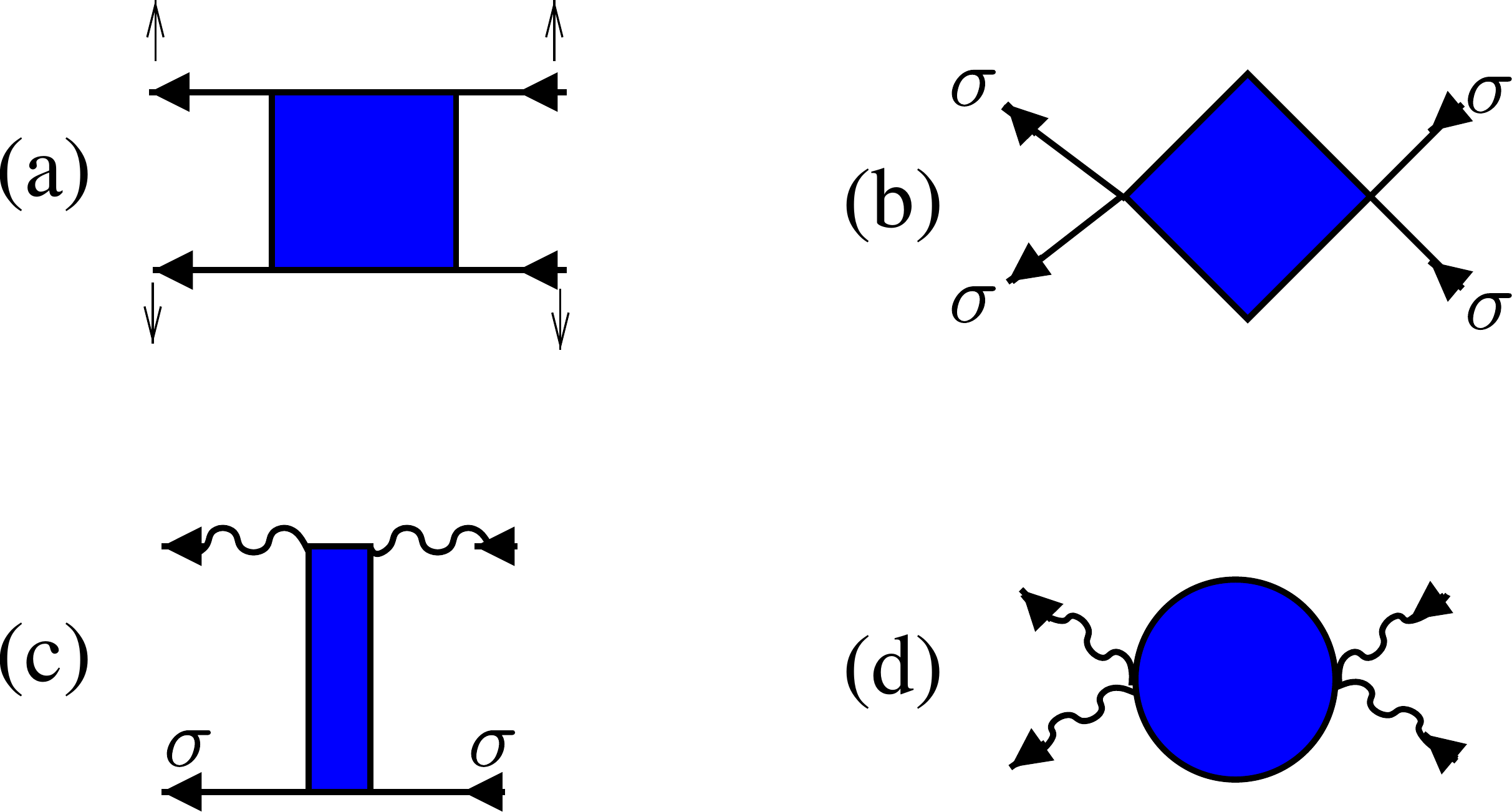}
\vspace{5mm}
  \vspace{-4mm}
  \caption{%
Induced two-body interactions in our model system with bare action
given in Eq.~(\ref{eq:Sbare2}).
(a) Induced interaction between fermions with opposite spin;
(b) Induced interaction between fermions with parallel spin $\sigma = \uparrow, \downarrow$; the fact that the incoming and outgoing 
legs end at the same point on the vertex
represents the antisymmetry of this vertex with respect to the exchange of the corresponding external labels.
(c) Induced mixed fermion-boson interaction.
(d)
 Induced two-body interaction between superfluid fluctuations; again, the symmetry of this vertex with respect to the exchange of 
 the labels associated with the two incoming or outgoing legs is represented by the attachment of the legs to the same point on the vertex.
}
\label{fig:induced4}
\end{figure}

We consider a system of electrons with quadratic energy dispersion
$\epsilon_{\bd{k}} = \bd{k}^2 /(2m)$
which are coupled by a short-range attractive two-body interaction
with strength $g_0 > 0$. The coupling $g_0$ represents some effective
interaction in the spin-singlet particle-particle channel. Since we neglect long-range Coulomb
interactions and do not consider the coupling to external electromagnetic fields,
we do not distinguish between superfluidity and superconductivity.
At finite temperature $T$ and chemical potential $\mu$
the Euclidean action of the system is
 \begin{equation}
S [ \bar{c} , c ] =  \int_{K } \sum_{\sigma}
 ( - i \omega + \epsilon_{\bd{k}} - \mu ) \bar{c}_{K \sigma} c_{K \sigma}
- g_0 \int_P \bar{C}_P C_P,
 \label{eq:Sbare}
\end{equation}
where $c_{K \sigma}$ and $\bar{c}_{K \sigma}$ are Grassmann fields
labeled by momentum $\bd{k}$, Matsubara frequencies $i \omega$, and spin projection
$\sigma = \uparrow, \downarrow$ (we introduce collective labels $K = ( \bd{k} , i \omega )$ and use units where $\hbar$ and the 
Boltzmann constant can be set equal to unity),
and the collective fields $C_P$ and $\bar{C}_P$ are defined by
 \begin{subequations}
 \begin{eqnarray}
  C_P & = & \int_K c_{ -K \downarrow} c_{K+P \uparrow},
 \label{eq:Cdef}
 \\
  \bar{C}_P & = & \int_K \bar{c}_{ K+P \uparrow} \bar{c}_{-K \downarrow}.
 \label{eq:Cbardef}
 \end{eqnarray}
 \end{subequations}
Here $P = ( \bd{p}, i \bar{\omega} )$ represents the total (bosonic) Matsubara frequency
$i \bar{\omega}$ and the total momentum $\bd{p}$ 
of a pair of electrons with opposite spin, and the
integration symbols are defined by
$\int_K = T \sum_{\omega} \int d^D k /( 2 \pi )^D$ and
$\int_P = T \sum_{\bar{\omega}} \int d^D p /( 2 \pi )^D$.
Although we are eventually interested in $D=3$ dimensions,
we will keep $D$ arbitrary before we explicitly start evaluating momentum integrals.
We represent the bare interaction of our model defined in Eq.~(\ref{eq:Sbare})
by the graphical element shown in Fig.~\ref{fig:bareint} (a).
The physics in the vicinity
of the superfluid transition is dominated by
the effective interaction in the particle-particle channel.
It is then natural to decouple the two-body interaction 
in Eq.~(\ref{eq:Sbare})  
by means of a complex
bosonic Hubbard-Stratonovich field $\psi$
such that the composite particle-particle fields defined
in Eqs.~(\ref{eq:Cdef}) and (\ref{eq:Cbardef}) 
appear in the decoupled action, which
then assumes the form
 \begin{eqnarray}
 S [ \bar{c} , c , \bar{\psi} , \psi ] & = &  - \int_{K } \sum_{\sigma}
  G_0^{-1} ( K )  \bar{c}_{K \sigma} c_{K \sigma}
+ \int_P g_0^{-1} \bar{\psi}_P \psi_P
 \nonumber
 \\
 &  & + \int_P \left[ \bar{C}_P \psi_P + {C}_P \bar{\psi}_P \right],
 \label{eq:Sbare2}
 \end{eqnarray}
where we have introduced the bare fermion propagator
 \begin{equation}
 G_0 ( K ) = \frac{1}{ i \omega - \epsilon_{\bd{k}} + \mu }.
 \label{eq:G0def}
 \end{equation}
The interaction in the last term of Eq.~(\ref{eq:Sbare2}) involves three-legged (Yukawa)
vertices with one bosonic and two fermionic external legs, 
as shown in Fig.~\ref{fig:bareint} (b).
We shall refer to $\psi$ as the order parameter field, because a finite expectation
value of this field signals the existence of superfluidity in the system.
In this work, we shall focus on the temperature regime
above the superfluid critical temperature. In this case the
exact fermionic propagator is given by
 \begin{equation}
 G ( K ) = \frac{1}{G_0 ( K )^{-1} - \Sigma ( K ) } ,
 \label{eq:Dysonsigma}
 \end{equation}
where $\Sigma ( K )$ is the exact fermionic self-energy in the normal state.
Similarly, the exact propagator of our order parameter field is of the form
 \begin{equation}
F ( P ) = \frac{1}{ g_0^{-1} - \Phi ( P )},
 \label{eq:Dysonphi}
 \end{equation}
where the function $\Phi ( P)$ can be identified with the one-interaction-line irreducible
bosonic self-energy. Graphical representations of the two Dyson equations
(\ref{eq:Dysonsigma}) and (\ref{eq:Dysonphi}) are shown in Fig.~\ref{fig:Dyson}.
Note that in lowest order perturbation theory $\Phi ( P ) \approx \Phi_0 ( P )$ can be identified with the particle-particle bubble with bare fermionic propagators,
 \begin{equation}
 \Phi_0 ( P )  =   \int_K  {G}_0 ( K ) G_0 ( P-K ).
 \label{eq:Phi0def}
 \end{equation}
The transition temperature to the superfluid state can be determined from
 the condition that the order-parameter field for $P =0$ becomes gapless at $T = T_c$, i.e.,
 \begin{equation}
 F^{-1} ( P=0) = g_0^{-1} - \Phi (P=0) = 0,
 \label{eq:FTc}
 \end{equation}
which is equivalent with the statement that the corresponding
uniform susceptibility diverges. To determine the critical temperature $T_c$ for superfluidity, we should calculate the 
function $\Phi ( 0 ) = \Phi (\bd{p} =0 , i \bar{\omega} =0 )$
to a certain approximation and then tune the temperature $T$ such that
Eq.~(\ref{eq:FTc}) is satisfied. The corrections to the non-interacting bubble given
in Eq.~(\ref{eq:Phi0def}) can be expressed in terms of the induced interactions, 
which take 
scattering processes in all channels into account.
Although the Hubbard-Stratonovich transformed bare action (\ref{eq:Sbare2})
does not contain two-body and higher order interaction vertices, these vertices
will appear in the effective low-energy theory when we integrate out high-energy
degrees of freedom. In particular,
two types of fermionic two-body interaction vertices will be generated, which we denote by
$\Gamma^{ \bar{c}_{\uparrow} \bar{c}_{\downarrow} c_{\downarrow} c_{\uparrow} } ( K_1^{\prime} , K_2^{\prime} ; K_2 , K_1 )$,
and
$\Gamma^{ \bar{c}_{\sigma} \bar{c}_{\sigma} c_{\sigma} c_{\sigma} } 
( K_1^{\prime} , K_2^{\prime} ; K_2 , K_1 )$, where $\sigma = \uparrow, \downarrow$ 
and the supercripts denote the
field types associated with the external legs and the energy-momentum labels
refer to the corresponding superscripts.
Moreover,
the two-body interactions between the superfluid order parameter are encoded in
the bosonic interaction vertex
$\Gamma^{ \bar{\psi} \bar{\psi} \psi \psi } ( P_1^{\prime} , P_2^{\prime} ; P_2 , P_1 )$.
Finally, symmetry allows also mixed four-point vertices
$\Gamma^{ \bar{c}_{\sigma} {c}_{\sigma} \bar{\psi} \psi  } ( K^{\prime} ; 
K ; P^{\prime}  ; P ) $ with two fermionic and two bosonic external legs.
Graphical representations of these different types of induced interaction vertices
are shown in Fig.~\ref{fig:induced4}.

\subsection{Skeleton equations}

\begin{figure}[tb]
  \centering
 \includegraphics[width=0.45\textwidth]{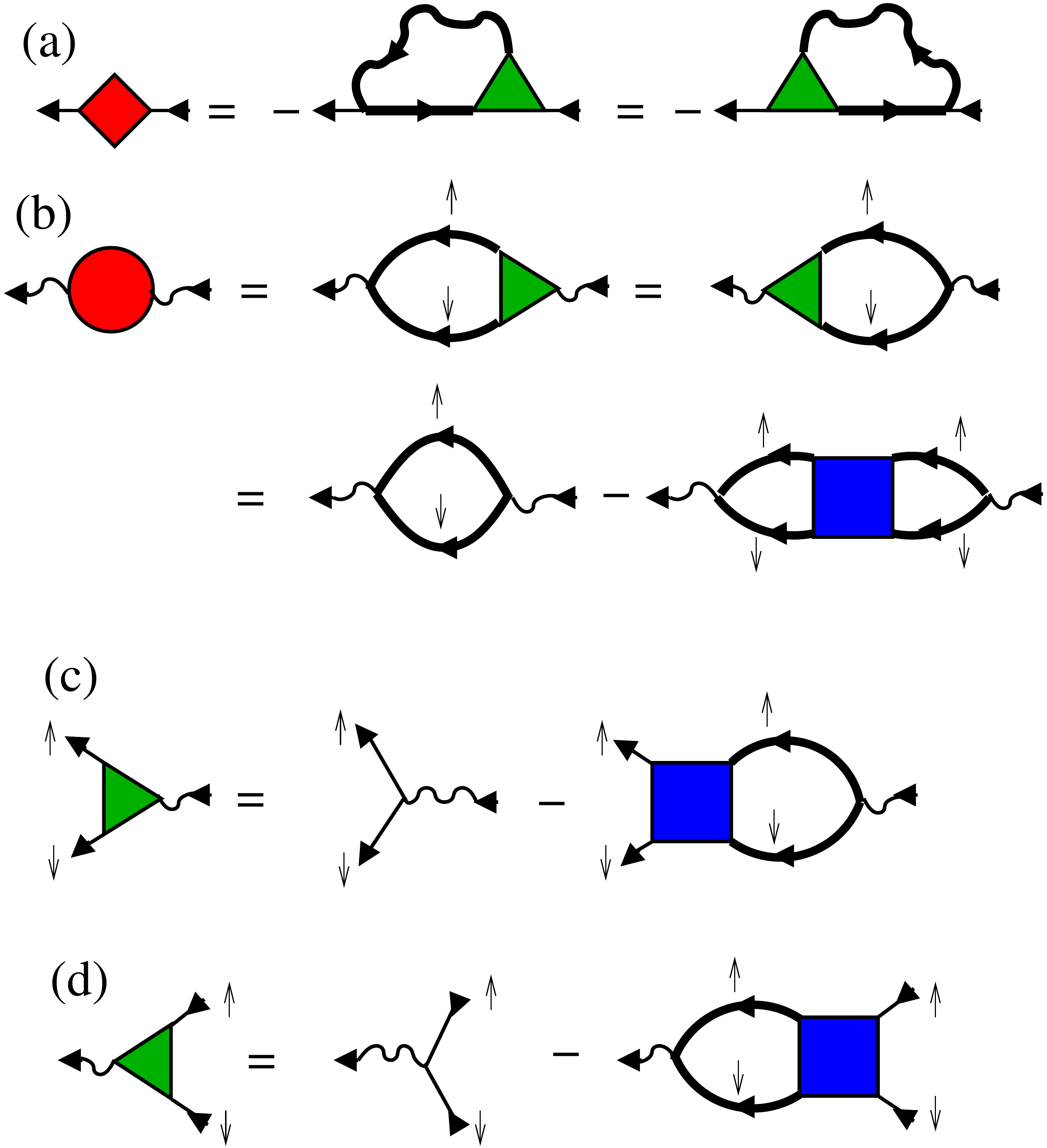}%
\vspace{5mm}
   \vspace{-4mm}
  \caption{%
The skeleton equation (a) expresses the exact fermionic self-energy
$\Sigma ( K )$ in terms
of the the exact three-point vertices
$\Gamma^{\bar{c}_{\uparrow} \bar{c}_{\downarrow} \psi }
 ( K_1^{\prime} , K_2^{\prime} ; P )$ 
and
$\Gamma^{{c}_{\downarrow} {c}_{\uparrow} \bar{\psi} }
 ( K_1^{\prime} , K_2^{\prime} ; P )$ which are represented by green
shaded triangles. In (b) we show three different ways of expressing the
exact self-energy $\Phi ( P )$ associated with the superfluid
order parameter in terms of the three-point vertices or in terms of
the exact effective interaction $\Gamma^{ \bar{c}_{\uparrow} \bar{c}_{\downarrow} c_{\downarrow} c_{\uparrow} } ( K_1^{\prime} , K_2^{\prime} ; K_2 , K_1 )$
between two fermions with opposite spin.
(c) and (d) represent skeleton equations relating the three-point vertices
in terms of the effective interaction.  
}
\label{fig:skeleton}
\end{figure}

Before calculating the 
fermionic and bosonic irreducible self-energies
$\Sigma ( K )$ and $\Phi ( P )$ using the FRG, 
it is instructive to rederive the GM result for the critical temperature 
using the effective field theory derived above.
Therefore it is convenient to start from
formally exact skeleton equations (also called Dyson-Schwinger equations),
which allow us to express
 the self-energies in terms of the induced interaction 
$\Gamma^{ \bar{c}_{\uparrow} \bar{c}_{\downarrow} c_{\downarrow} c_{\uparrow} } ( K_1^{\prime} , K_2^{\prime} ; K_2 , K_1 )$
between fermions with opposite spin. Graphically, the skeleton equations for the self-energies
$\Sigma ( K )$ and $\Phi ( P )$
and for the irreducible three-point vertices 
$\Gamma^{\bar{c}_{\uparrow} \bar{c}_{\downarrow} \psi }
 ( K_1^{\prime} , K_2^{\prime} ; P )$ 
and
$\Gamma^{{c}_{\downarrow} {c}_{\uparrow} \bar{\psi} }
 ( K_1^{\prime} , K_2^{\prime} ; P )$ 
are shown in in Fig.~\ref{fig:skeleton}.
Formally, these equations can be derived by using
the invariance of the functional integral representing the
generating functional of the irreducible vertices
under shift transformations of the fields \cite{ZinnJustin02,Kopietz10}.
Explicitly, the skeleton equations relating the fermionic and bosonic 
self-energies to the three-point vertices are
 \begin{eqnarray}
  \Sigma ( K ) & = & - \int_P  F ( P ) G ( P-K )
 \Gamma^{\bar{c}_{\uparrow} \bar{c}_{\downarrow} \psi}
 ( K, P-K ; P ),
 \nonumber
 \\
 &=   &  - \int_P    \Gamma^{{c}_{\downarrow} {c}_{\uparrow} \bar{\psi}} ( P-K , K; P ) 
     F ( P ) G ( P-K ),
 \nonumber
 \\
 & &
 \label{eq:DSself}
 \\
  \Phi ( P ) & = &  \int_K  {G} ( K ) G ( P-K )
 \Gamma^{{c}_{\downarrow} {c}_{\uparrow} \bar{\psi}} ( P-K , K; P ) 
 \nonumber
 \\
 & =  &
  \int_K   
 \Gamma^{\bar{c}_{\uparrow} \bar{c}_{\downarrow} \psi} ( K, P-K ; P )
   {G} ( K ) G ( P-K ),
 \nonumber
 \\
 & & 
 \label{eq:DysonSchwingerPhi}
 \end{eqnarray}
while the skeleton equations (c) and (d)
in Fig.~\ref{fig:skeleton} relating the three-point vertex
to the effective interaction between two fermions with opposite
spin are
 \begin{eqnarray}
 & &  \Gamma^{\bar{c}_{\uparrow} \bar{c}_{\downarrow} \psi } ( K_1 , K_2 ; P ) =  
 1 -  \int_K {G} ( K ) G ( P-K ) 
 \nonumber
 \\
 & & \times \Gamma^{\bar{c}_{\uparrow} \bar{c}_{\downarrow} c_{\downarrow} c_{\uparrow} }
 ( K_1 , K_2 ; K, P-K ),
 \\
 & & \Gamma^{{c}_{\downarrow} {c}_{\uparrow} \bar{\psi} } ( K_1 , K_2 ; P )
 = 1   -  \int_K {G} ( K ) G ( P-K ) 
 \nonumber
 \\
 & & \times \Gamma^{\bar{c}_{\uparrow} \bar{c}_{\downarrow} c_{\downarrow} c_{\uparrow} }
 ( K, P-K ; K_1 , K_2).
\end{eqnarray}
Substituting these expressions into Eqs.~(\ref{eq:DysonSchwingerPhi}), we obtain the skeleton equation for the bosonic self-energy shown in the second line of
Fig.~\ref{fig:skeleton} (b),
 \begin{eqnarray}
  \Phi ( P ) & = &  \int_K G ( K ) G ( P-K) 
  \nonumber
 \\
 & - &
  \int_K \int_{K^{\prime}} G ( K ) G ( P-K) G ( K^{\prime} )
  G ( P - K^{\prime} )
 \nonumber
 \\
 & & \hspace{8mm} \times  
\Gamma^{ \bar{c}_{\uparrow} \bar{c}_{\downarrow} c_{\downarrow} c_{\uparrow} }
 ( K_ , P-K ; K^{\prime} ,P-K^{\prime} ).
 \hspace{7mm}
 \label{eq:phi2}
 \end{eqnarray}

\subsection{Perturbative expansion in powers of the scattering length}

\begin{figure}[tb]
  \centering
 \includegraphics[width=0.4\textwidth]{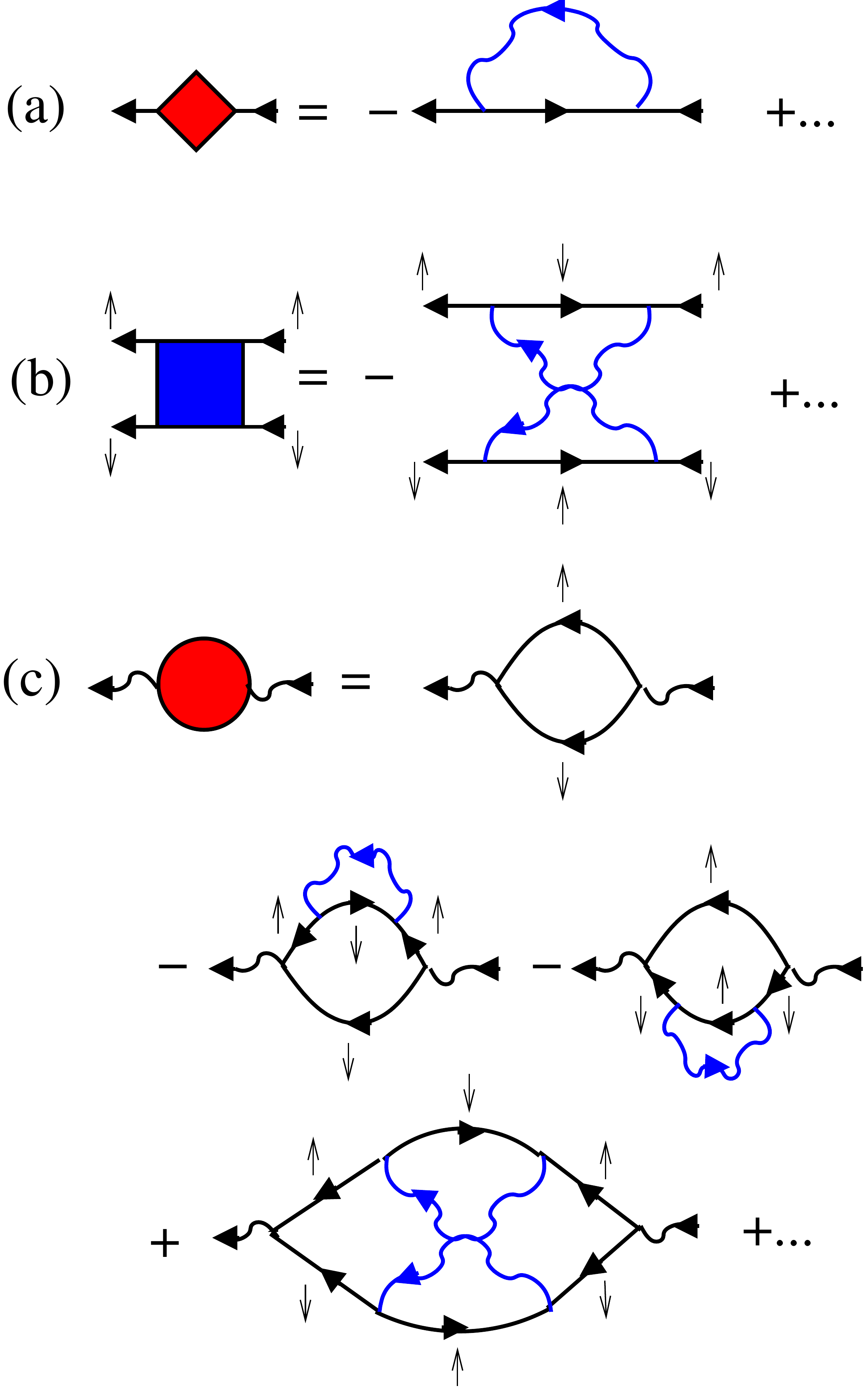}
\vspace{5mm}
  \vspace{-4mm}
  \caption{%
Perturbative expansion in powers
of the $T$-matrix $g$ which is represented by a blue wavy arrow.
(a) Fermionic self-energy, (b) induced interaction  between two fermions
with opposite spin, and (c) bosonic self-energy.
}
\label{fig:pert}
\end{figure}

The GM correction to the critical temperature can now be obtained
by expanding  the induced  interaction between electrons with opposite spin
appearing in the skeleton equation
(\ref{eq:phi2}) to second order in the scattering length.
Recall that in three dimensions
the $s$-wave scattering length $a_s$ is defined by
 \begin{equation}
 g =  \frac{4 \pi   a_s}{m},
 \end{equation}
where the so-called $T$-matrix in vacuum at vanishing total momentum is related to
the bare interaction via
 \begin{equation}
 g^{-1}  =  g_0^{-1}  - \Phi_0^{\rm vac} (0),
 \end{equation}
and the  particle-particle bubble at vanishing temperature and chemical
potential is in three dimensions  given by
 \begin{equation}
 \Phi_0^{\rm vac} (0)  =   \int_{\bd{k}} \frac{\Theta ( \Lambda_0 - | \bd{k} | )}{ 2
 \epsilon_{\bd{k}}} =  \nu \frac{\Lambda_0}{k_F}.
 \label{eq:bandwidth}
 \end{equation}
Here $\Lambda_0$ is an ultraviolet cutoff in momentum space  and
 \begin{equation}
\nu =  m k_F /(2 \pi^2)
 \label{eq:nu3def} 
 \end{equation}
  is the density of states (per spin projection) 
at the Fermi energy, where $k_F$ is the Fermi momentum.
To generate an expansion in powers of $g$,
let us write the propagator of the
pairing field in Gaussian approximation
(where the bosonic self-energy
is approximated by $\Phi ( P ) \approx \Phi_0 ( P )$, see
Eq.~(\ref{eq:Phi0def}))
in the following form
 \begin{eqnarray}
 F_0 ( P ) & = &  \frac{1}{g_0^{-1} - \Phi_0 ( P ) }
 =
  \frac{1}{g^{-1} - \Phi_0^{\rm reg} ( P ) },
 \label{eq:F0def}
 \end{eqnarray}
where the regularized particle-particle bubble is
 \begin{equation}
\Phi_0^{\rm reg} (P )  =  \Phi_0 (P) - \Phi_0^{\rm vac} (0).
 \label{eq:phi0reg}
 \end{equation}
Due to the subtraction this expression is ultraviolet convergent
so that we may take the limit $\Lambda_{0} \rightarrow \infty$.
Assuming $  | g | \ll 1$ and that the relevant momenta in loop
integrations are such that $ |g \Phi_0^{\rm reg} (P ) | \ll 1$
we may approximate
\begin{eqnarray}
 F_0 ( P ) &  \approx &  g  + {\cal{O}} ( g^2 ).
 \end{eqnarray}
The leading terms in the expansion of the fermionic self-energy
and the effective interaction between fermions with opposite spin
are shown in Fig.~\ref{fig:pert} (a) and (b).
Explicitly, the first order self-energy correction in Fig.~\ref{fig:pert} (a)
is
 \begin{equation}
 \Sigma_1 = - g \int_K G_0 ( K ) = - g \rho_0,
 \end{equation}
where $\rho_0 =  \int_K G_0 ( K ) $ is the density (per spin projection) in
the non-interacting limit. The induced interaction between fermions with opposite spin
to order $g^2$ shown in Fig.~\ref{fig:pert} (b) can be written as
\begin{equation}
\Gamma^{ \bar{c}_{\uparrow} \bar{c}_{\downarrow} c_{\downarrow} c_{\uparrow} }
 ( K_1^{\prime} , K_2^{\prime} ; K_2 , K_1 )
     \approx -  g^2 \Pi_0 ( K_1^{\prime} - K_2 ) ,
 \label{eq:Upert}
 \end{equation}
where
 \begin{equation}
 \Pi_0 ( Q  ) = \int_K G_{0} ( K ) G_0 ( K-Q )
 \label{eq:Pi0def}
 \end{equation}
is the non-interacting particle-hole bubble.
Substituting these expansions into the skeleton equation (\ref{eq:phi2})
for the bosonic self-energy we obtain the
expansion shown in Fig.~\ref{fig:pert} (c), which can be written as
 \begin{equation}
 \Phi ( P )  =  \Phi_0 ( P ) + \Phi_1 ( P ) + \Phi_2 ( P ) + \ldots,
 \end{equation}
where the non-interacting particle-particle bubble $\Phi_0 ( P )$
is given in Eq.~(\ref{eq:Phi0def}), the first order correction is
\begin{eqnarray}
  \Phi_1 ( P ) & = & - 2  g \rho_0 \int_K G_0^2 ( K ) G_0 ( P-K ),
 \label{eq:phi1exp}
 \hspace{7mm}
 \end{eqnarray}
while the leading correction due to the induced interaction is
 \begin{eqnarray}
   \Phi_2 ( P )
 & =  &  g^2 \int_K \int_{K^{\prime}} G_0 ( K ) G_0 ( P-K) 
 \nonumber
 \\
 & & \times  \Pi_0 ( K  - K^{\prime} ) G_0 ( K^{\prime} )  G_0 ( P - K^{\prime} )    .
 \label{eq:phi2integral}
 \end{eqnarray} 
If we follow GM\cite{Gorkov61} and work at constant density, the self-energy $\Sigma_1$ is exactly
canceled by a shift in the chemical potential which is necessary to keep the
density fixed; in this case we should  ignore the first order correction  $\Phi_1 ( P )$, so that
the leading interaction correction to $\Phi ( P)$ is given by the second order term $\Phi_2 ( P )$.
On the other hand, at constant chemical potential
the term $\Phi_1 ( P )$ modifies the GM result.

According to Eq.~(\ref{eq:FTc}), the critical temperature is determined by
 \begin{equation}
 0= g_0^{-1} - \Phi (0) \approx  g^{-1} - \Phi_0^{\rm reg} (0) - \Phi_1 ( 0 ) - \Phi_2 ( 0 ).
 \label{eq:tc2}
 \end{equation}
An explicit evaluation of the three contributions on the right-hand side of this equation 
in the BCS regime (where $\mu \approx E_{F}$) 
is given in Appendix~A. Here we briefly summarize the main results.
First of all, for temperatures $T \ll E_F$ the regularized particle-particle bubble is given by
\begin{equation}
   \Phi_0^{\rm reg} ( 0 )  = \nu \left[ \ln ( A / \tau )  + {\cal{O}}( \tau ) \right],
 \label{eq:phiregres}
 \end{equation}
where $\tau = T / E_F$ and
 \begin{equation}
 A = \frac{ 8}{\pi e^{ 2 - \gamma_E }}.
 \label{eq:Adef}
 \end{equation}
Here $\gamma_E = 0.577...$ is the Euler-Mascheroni constant.
If we ignore the terms $\Phi_1 ( 0 )$ and $\Phi_2 ( 0)$
on the right-hand side of Eq.~(\ref{eq:tc2}) we obtain the mean-field critical temperature
\begin{equation}
  \frac{T_{c0}}{ E_F}  = \tau_{c0} = A e^{-1 /\tilde{g}}  =  \frac{8 e^{\gamma_E}}{\pi e^2}   e^{ -1 / \tilde{g}}, 
 \label{eq:tcweak}
 \end{equation}
with the dimensionless interaction constant
 \begin{equation}
 \tilde{g} = \nu g = \frac{2}{\pi} k_F a_s.
 \end{equation}
As shown in Appendix A, for $P=0$ the first order correction
to the particle-particle bubble (\ref{eq:phi1exp}) is at low temperatures
given by
 \begin{equation}
 \Phi_1 ( 0 ) = \alpha_1 g \nu^2 [ \ln ( A / \tau )  + \lambda_0 ], 
 \label{eq:phi1res}
 \end{equation}
where
 \begin{equation}
 \alpha_1 = 1/3,
 \end{equation}
and  $\lambda_0 = \Lambda_0 / k_F$ is a dimensionless ultraviolet cutoff which is necessary
to regularize the relevant momentum integral.
As discussed above, the contribution $\Phi_1 ( 0 )$ should be 
omitted if we work at constant density.
The second order correction to the particle-particle bubble is
 \begin{eqnarray}
 \Phi_2 (0) 
 & = &    \alpha_2 g^2 \nu^3  \left[ \ln ( A / \tau ) + \lambda_0 \right]^2,
 \label{eq:phi2res}
 \end{eqnarray}
with
\begin{equation}
\alpha_2 = - \frac{ 1 + \ln 4 }{ 3}.
 \label{eq:alpha2}
 \end{equation}
Consider first the case of constant density, where the contribution from $\Phi_1 (0)$
should be omitted.
Substituting Eqs.~(\ref{eq:phiregres}) and (\ref{eq:phi2res})
into Eq.~(\ref{eq:tc2}) we obtain the following estimate of the dimensionless
critical temperature,
 \begin{equation}
 \tau_c = A e^{ \alpha_2} e^{ - 1 / \tilde{g}} \times \left[
 1 + {\cal{O}} ( \tilde{g} \lambda_0  ) \right].
 \label{eq:GMB}
 \end{equation}
In the asymptotic weak coupling limit  $\tilde{g}  \lambda_0  \ll 1$
we may neglect the cutoff-dependent correction  and find that
the induced interaction due to particle-hole fluctuations reduces the
critical temperature for superfluidity by a factor of
\begin{equation}
 \frac{T_c }{ T_{c0}}  =    e^{  \alpha_2}   = \frac{1}{( 4 e )^{1/3}}  \approx 0.451 ,
 \label{eq:tcalpha1}
 \end{equation}
in agreement with  GM\cite{Gorkov61}.
Note that according to Eq.~(\ref{eq:GMB}) the cutoff-dependent correction
to the GM result is 
of the order $\tilde{g} \Lambda_0  = \nu g \Lambda_0 / k_F$
which depends linearly  on the ultraviolet cutoff $\Lambda_0$.
We show in Appendix A that this linear cutoff dependence
is an artifact of neglecting the momentum- and frequency dependence
of the particle-hole bubble  $\Pi_0 ( K - K^{\prime} )$
in the evaluation of Eq.~(\ref{eq:phi2integral}).
In a more accurate calculation taking
the momentum or the frequency dependence of $\Pi_0 ( K - K^{\prime} )$ into account
the correction depends only  logarithmically on the cutoff.
If the chemical potential is held constant, then the term $\Phi_1 (0)$ is not canceled 
and we obtain 
\begin{equation}
 \frac{T_c }{ T_{c0}}  =    e^{  \alpha_1 + \alpha_2}   = \frac{1}{ 4^{1/3} }  \approx 0.630
 \; \; \; \mbox{for constant $\mu$,}
 \label{eq:tcalpha2}
 \end{equation}
which is larger than the GM result in  Eq.~(\ref{eq:tcalpha1}).
The discrepancy to the GM result for $T_c$ found in a recent renormalization group
calculation by Tanizaki {\it{et al.}} \cite{Tanizaki14} seems to be due to the fact that 
these authors did not fix the density
in their calculation.

\section{Induced interactions and vertex corrections from the FRG}
\label{sec:FRG}

\begin{figure}[tb]
  \centering
 \includegraphics[width=0.45\textwidth]{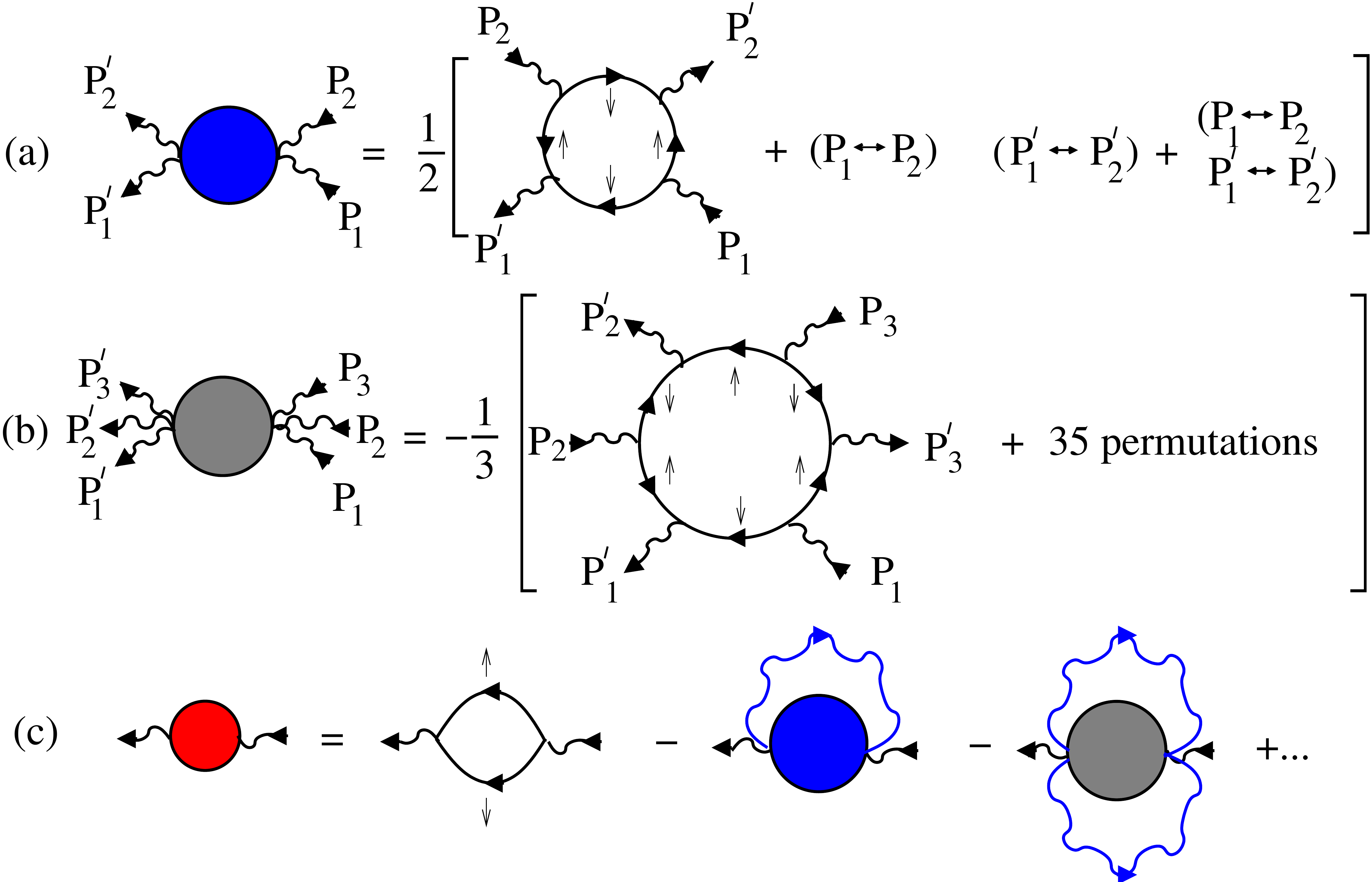}
\vspace{5mm}
  \vspace{-4mm}
  \caption{%
(a) and (b) represent the vertices with four and six bosonic external legs
at the initial cutoff $\Lambda = \Lambda_0$ in our cutoff scheme where only the bosonic propagator
is regularized. 
In (c) we show the perturbative expansion of the bosonic self-energy
in powers of the $T$-matrix $g$ in vacuum (blue wavy arrows), which 
is obtained by approximating the Gaussian propagators of the order parameter field by
 $F_0 ( P ) \approx g$.
The one-loop contraction of the four-point vertex
contains the self-energy corrections to the particle-particle bubble
shown in Fig.~\ref{fig:pert} (c), while the
two-loop contraction of the six-point vertex contains the GM correction
shown in the last line of Fig.~\ref{fig:pert} (c).
The minus signs in (c) are due to the fact that in Eq.~(\ref{eq:Gammatrunc})
there is a relative minus sign between $\Phi_{\Lambda} ( P )$ and the other vertices.
}
\label{fig:loops}
\end{figure}

In order to understand the origin of the GM correction from the renormalization  group point of view and to
set up a machinery which allows us to calculate the fermionic self-energy non-perturbatively,
we develop in this section a general FRG approach for our model 
with bare action given by Eq.~(\ref{eq:Sbare2}).
To derive formally exact FRG flow equations for the irreducible vertices of our model, we
 introduce an additional cutoff $\Lambda$ such that
for large $\Lambda$ fluctuations are suppressed while for $\Lambda \rightarrow 0$
we obtain our original model \cite{Kopietz10}. The evolution of the  generating functional
of the one-line irreducible vertices under changes of the cutoff 
is described by the Wetterich equation \cite{Wetterich93}.
By expanding this equation in powers of the fields, we obtain a formally exact
hierarchy of FRG flow equations for all one-line irreducible vertices 
of our theory.
For the implementation of this procedure 
there is considerable freedom in the choice of the
cutoff scheme. For our purpose 
it is most convenient to use the particle-particle
version of the momentum-transfer cutoff scheme proposed in 
Refs.~[\onlinecite{Schuetz05,Schuetz06}],
which has been shown to be useful in several other contexts \cite{Kopietz10,Drukier12,Sharma16}. In this {\it{interaction-momentum cutoff scheme}},
we replace the inverse bare coupling $g_0^{-1}$ of our model
by the cutoff- and momentum-dependent coupling
  \begin{equation}
 g^{-1}_{0, \Lambda} ( \bd{p} ) = g_0^{-1} + R_{\Lambda} ( \bd{p} ),
 \end{equation}
there the regulator function vanishes for $\Lambda \rightarrow 0$ and
approaches some large value for $\Lambda \rightarrow \Lambda_0$,
where $\Lambda_0$ is some large initial value of the cutoff.
Below we will work with a sharp momentum  cutoff  which amounts
to setting
 \begin{equation}
  g_{0, \Lambda} ( \bd{p} ) = g_0 \Theta ( | \bd{p} | - \Lambda ) .
 \end{equation}
For  $\Lambda < \Lambda_0$,
the generating functional of the cutoff-dependent one-particle  irreducible vertices
of our model can be expanded in powers of the fields as follows
 \begin{widetext} 
\begin{eqnarray}
 \Gamma_{\Lambda} [ \bar{c}, c , \bar{\psi} , \psi ] & = & 
 \int_K \sum_{\sigma} \Sigma_{\Lambda} ( K ) 
 \bar{c}_{K \sigma} c_{K \sigma}
-  \int_P  \Phi_{\Lambda} ( P )  \bar{\psi}_P \psi_P
 \nonumber
 \\
 &    + & \int_K \int_P \Bigl[ \Gamma_{\Lambda}^{\bar{c}_{\uparrow} \bar{c}_{\downarrow} \psi}
 ( K+P, -K ; P ) \,  \bar{c}_{K+P \uparrow} \bar{c}_{-K \downarrow} \psi_P
+ \Gamma_{\Lambda}^{{c}_{\downarrow} c_{\uparrow} \bar{\psi}}
 ( -K , K+P;  P )\, {c}_{-K \downarrow} {c}_{K + P  \uparrow} \bar{\psi}_P
 \Bigr]
 \nonumber
 \\
 & + & \int_{K_1^{\prime}} \int_{K_2^{\prime}} \int_{K_2} \int_{K_1}
 \delta_{ K_1^{\prime} + K_2^{\prime} , K_2 + K_1 }
 \Gamma^{\bar{c}_{\uparrow} \bar{c}_{\downarrow} c_{\downarrow}
 c_{\uparrow}}_{\Lambda} ( K_1^{\prime} , K_2^{\prime} ; K_2 , K_1 ) \,
 \bar{c}_{ K_1^{\prime} \uparrow} \bar{c}_{ K_2^{\prime} \downarrow} c_{ K_2 \downarrow}
 c_{ K_1 \uparrow}
 \nonumber
\\
 &  + &  \frac{1}{(2!)^2} \int_{K_1^{\prime}} \int_{K_2^{\prime}} \int_{K_2} \int_{K_1}
  \sum_{\sigma}   \delta_{ K_1^{\prime} + K_2^{\prime} , K_2 + K_1 } 
 \Gamma^{\bar{c}_{\sigma} \bar{c}_{\sigma} c_{\sigma}
 c_{\sigma}}_{\Lambda} ( K_1^{\prime} , K_2^{\prime} ; K_2 , K_1 ) \,
 \bar{c}_{ K_1^{\prime} \sigma} \bar{c}_{ K_2^{\prime} \sigma} c_{ K_2 \sigma}
 c_{ K_1 \sigma}
 \nonumber
 \\
 & + &  \int_{K^{\prime}}  \int_{K}    \int_{P^{\prime}} \int_{P} \sum_{\sigma}
 \delta_{ K^{\prime} + P^{\prime} ,  K + P }
 \Gamma^{\bar{c}_{\sigma} {c}_{\sigma}  \bar{\psi} \psi }_{\Lambda} 
( K^{\prime} , K  ; P^{\prime} , P ) \,
 \bar{c}_{ K^{\prime} \sigma} {c}_{ K  \sigma} \bar{\psi}_{ P^{\prime}}
 \psi_P
 \nonumber
 \\
 &  + &  \frac{1}{ (2!)^2} \int_{P_1^{\prime}} \int_{P_2^{\prime}} \int_{P_2} \int_{P_1}
 \delta_{ P_1^{\prime} + P_2^{\prime} , P_2 + P_1 }
 \Gamma^{\bar{\psi} \bar{\psi} \psi \psi }_{\Lambda} ( P_1^{\prime} , P_2^{\prime} ; 
 P_2 , P_1 ) \,
 \bar{\psi}_{ P_1^{\prime}} \bar{\psi}_{ P_2^{\prime} } \psi_{ P_2} \psi_{ P_1 } 
 + \ldots ,
 \label{eq:Gammatrunc}
 \end{eqnarray}
 \end{widetext}
where the ellipsis represents terms  involving five and more powers of the fields and all
vertices are assumed to be properly symmetrized with respect to
permutations of the labels associated with fields of the same type.
$\Sigma_{\Lambda} (K)$ and 
$\Phi_{\Lambda} (P) $ are the cutoff-dependent fermionic and bosonic
irreducible self-energies.
The corresponding cutoff-dependent inverse propagators are 
 \begin{eqnarray}
 G^{-1}_{\Lambda} ( K ) & = & G^{-1}_{0 , \Lambda} ( K ) - \Sigma_{\Lambda} ( K ),
 \label{eq:Dyson1}
  \\
 F^{-1}_{\Lambda} ( P ) & = & g^{-1}_{0 , \Lambda} ( \bd{p} ) - \Phi_{\Lambda} ( P ).
 \label{eq:Dyson2}
 \end{eqnarray}
The last four lines in Eq.~(\ref{eq:Gammatrunc})
represent the various induced interactions shown graphically in
Fig.~\ref{fig:induced4}. Although these interactions do not appear in our bare
action given in Eq.~(\ref{eq:Sbare2}), they are generated by the FRG flow.
Since we do not introduce the regulator into the fermionic sector of our model,
we have to start the FRG flow at some large initial scale $\Lambda = \Lambda_0$ with
a non-trivial initial condition, as explained in Refs.~\onlinecite{Kopietz10,Schuetz06}.
Hence, apart from the initial values of the three-legged vertices
which appear in the bare action (\ref{eq:Sbare2}),
 \begin{equation}
 \Gamma_{\Lambda_0}^{\bar{c}_{\uparrow} \bar{c}_{\downarrow} \psi}
 ( K+P, -K ; P ) = 
\Gamma_{\Lambda_0}^{{c}_{\downarrow} c_{\uparrow} \bar{\psi}}
 ( -K , K+P;  P ) = 1,
 \label{eq:gamma3initial}
 \end{equation}
all purely bosonic $2n$-point vertices with $n$ incoming and $n$ outgoing 
boson lines are finite at the initial cutoff $\Lambda_0$. 
Diagrammatically, these vertices can be identified with the symmetrized closed fermion loops
with $n$ incoming and $n$ outgoing external bosonic legs,
as shown in Fig.~\ref{fig:loops}.
Specifically,  the initial value of the bosonic self-energy is the non-interacting particle-particle bubble,
 \begin{eqnarray}
 \Phi_{\Lambda_0} ( P ) & = & \Phi_0 ( P ) = \int_K G_0 ( K ) G_0 ( P-K ),
 \end{eqnarray}
while the bosonic four-point vertex at the initial cutoff scale is
 \begin{eqnarray}
  & & \Gamma^{\bar{\psi} \bar{\psi} \psi \psi }_{\Lambda_0} ( P_1^{\prime} , P_2^{\prime} ; P_2 , P_1 )  =  \frac{1}{2} \int_K
 \Bigl[ G_0 ( K ) G_0 ( - K + P_1^{\prime} ) 
 \nonumber \\
 & & \times G_0 ( K - P_1^{\prime} + P_2 ) 
 G_0 ( - K + P_1^{\prime} - P_2 + P_2^{\prime} )
 \nonumber
  \\
 & & + ( P_1 \leftrightarrow P_2 ) + ( P_1^{\prime} \leftrightarrow P_2^{\prime} )
 +   ( P_1 \leftrightarrow P_2 , P_1^{\prime} \leftrightarrow P_2^{\prime} )
 \Bigr].
 \nonumber
 \\
 & &
 \label{eq:4loop}
 \end{eqnarray}
Note that this vertex 
is symmetric  with respect to the independent exchange $P_1^{\prime} \leftrightarrow
 P_2^{\prime}$ and $P_1 \leftrightarrow P_2$. 
Setting all external momenta and frequencies equal to zero we obtain
 \begin{equation}
 \Gamma^{\bar{\psi} \bar{\psi} \psi \psi }_{\Lambda_0} (0 ) = \frac{7  \zeta ( 3 ) }{4} \frac{\nu}{(\pi T )^2 }.
 \label{eq:u0def}
 \end{equation}
The contribution of the higher order bosonic vertices 
to the initial value of the generating functional (\ref{eq:Gammatrunc}) is
 \begin{eqnarray}
 \Gamma_{\Lambda_0}^{\rm n > 3} [ \bar{\psi} , \psi ] & = & \sum_{ n =3}^{\infty} 
 \frac{1}{ (n!)^2} \int_{P_1^{\prime}} \ldots 
 \int_{P_n^{\prime}} \int_{P_n}\ldots \int_{P_1}
 \nonumber
 \\
 & & \hspace{-20mm} \times 
 \delta_{ P_1^{\prime} + \ldots + P_n^{\prime}, P_n + \ldots+ P_1 }
 \Gamma^{(2n) }_{\Lambda_0} 
 ( P_1^{\prime} , \ldots  , P_n^{\prime}; 
 P_n , \ldots  , P_1 ) 
 \nonumber
 \\
 & & \hspace{-20mm} \times
 \bar{\psi}_{ P_1^{\prime}} \ldots  \bar{\psi}_{P_n^{\prime}}
 \psi_{ P_n} \ldots \psi_{ P_1 } .
 \end{eqnarray}
In our approach the GM correction to $T_c$ is determined 
by the initial value of the 
bosonic six-point vertex, which 
after symmetrization can be written as
 \begin{eqnarray}
& & \Gamma^{(6) }_{\Lambda_0} ( P_1^{\prime} , P_2^{\prime} , P_3^{\prime} ; 
P_3, P_2 , P_1 )  =
 \nonumber
 \\
 &  &  - \frac{1}{3} \int_K
 \Bigl[ G_0 ( K ) G_0 ( - K + P_1^{\prime} )   G_0 ( K - P_1^{\prime} + P_2 ) 
 \nonumber \\
 & & \times
 G_0 ( - K + P_1^{\prime} - P_2 + P_2^{\prime} )
 G_0 ( K - P_1^{\prime} + P_2 - P_2^{\prime} + P_3) 
 \nonumber
  \\
 & & \times 
 G_0 ( - K + P_1^{\prime} - P_2 + P_2^{\prime} - P_3 + P_3^{\prime} )
 \nonumber
 \\
 & &
+  \mbox{ \small  $(3 !)^2-1$ permutations of  
 $( P_1^{\prime} , P_2^{\prime} , P_3^{\prime} )$ and $(P_1, P_2 , P_3)$ } 
 \Bigr]. 
 \nonumber
 \\
 & &
 \label{eq:gamma6init}
 \end{eqnarray}
All other vertices vanish at the initial scale, but all vertices which are compatible with the $U(1)$-symmetry of the
bare action are generated by the FRG flow, in particular the
induced interactions shown in Fig.~\ref{fig:induced4}.
From the bosonic sector of our initial action $\Gamma_{\Lambda_0} [ \bar{c} , c ,
 \bar{\psi} , \psi ]$ it is easy to reproduce the
perturbation series for the renormalized particle-particle bubble
shown graphically in Fig.~\ref{fig:pert} (c).
The non-interacting particle-particle bubble is contained in the
Gaussian propagator $F_0 ( P ) ] = [ g_0^{-1} - \Phi_0 ( P ) ]^{-1}$, see Eq.~(\ref{eq:F0def}).
The first order corrections shown in the second line of Fig.~\ref{fig:pert} (c) 
can be recovered from the one-loop contraction of the four-point vertex shown 
in Fig.~\ref{fig:loops}~(c), while the
last diagram in  Fig.~\ref{fig:pert} (c) which gives the GM correction
is contained in the two-loop contraction of the six-point vertex
shown in Fig.~\ref{fig:loops} (c).

\begin{figure}[tb]
  \centering
 \includegraphics[width=0.4\textwidth]{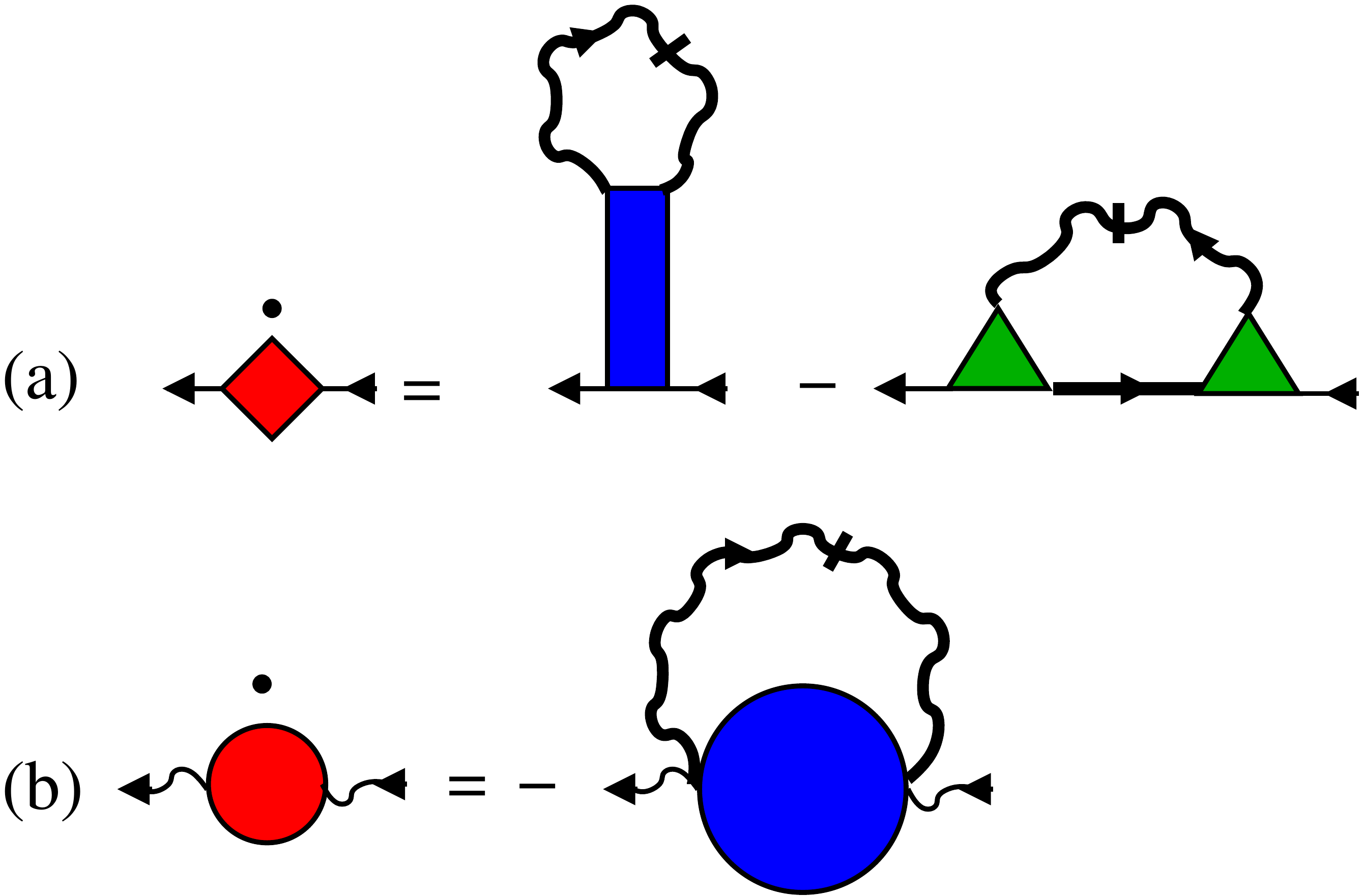}
\vspace{5mm}
  \vspace{-4mm}
  \caption{%
Graphical representation  of the exact FRG flow equations for (a) the  fermionic and (b) the 
bosonic self-energy of our model
using the interaction-momentum cutoff scheme, see Eqs.~(\ref{eq:flowself}) and
(\ref{eq:flowselfbos}).
The notations for the vertices and propagators are the same
as in Figs.~\ref{fig:Dyson}, \ref{fig:induced4} and \ref{fig:skeleton}.
A dot over a vertex denotes the cutoff derivative and
wavy arrows with an additional slash represent the bosonic single-scale
propagator defined in Eq.~(\ref{eq:singlescaledef}).
}
\label{fig:flowself}
\end{figure}

Let us now write down exact FRG flow equations for the self-energies of our model
in our interaction-momentum cutoff scheme.
The derivation of these flow equations is straightforward following the general procedure
outlined in Ref.~[\onlinecite{Kopietz10}].
The cutoff-dependent fermionic self-energy satisfies
 \begin{eqnarray}
  \partial_{\Lambda} \Sigma_{\Lambda} ( K ) & = &
 \int_{P } \dot{F}_{\Lambda} ( P )
\Gamma^{\bar{c}_{\sigma} {c}_{\sigma} \bar{\psi} \psi}_{\Lambda} ( K; K ;  P ; P )
 \nonumber
 \\
& - &  \int_P  \dot{F}_{\Lambda} ( P )  G_{\Lambda} ( P-K ) 
 \Gamma_{\Lambda}^{\bar{c}_{\uparrow} \bar{c}_{\downarrow}  \psi}
 ( K, P-K ; P )
 \nonumber
\\
  & & \hspace{3mm} \times
\Gamma_{\Lambda}^{{c}_{\downarrow} c_{\uparrow} \bar{\psi}}
 ( P-K , K ;  P ) ,
 \label{eq:flowself}
 \end{eqnarray}
while the flow of the bosonic self-energy (which can be identified with the
renormalized particle-particle bubble) is given by
 \begin{equation}
  \partial_{\Lambda} \Phi_{\Lambda} ( P )   =  
 - \int_{P^{\prime} }  \dot{F}_{\Lambda} ( P^{\prime} )
\Gamma^{\bar{\psi} \bar{\psi} \psi  \psi}_{\Lambda} ( P,P^{\prime}; P^{\prime} , P ).
 \label{eq:flowselfbos}
 \end{equation}
Here  $\dot{F}_{\Lambda} ( P )$ is the bosonic single-scale propagator,
which for our sharp interaction-momentum cutoff scheme is simply given by
 \begin{equation}
  \dot{F}_{\Lambda} ( P ) = - \frac{ \delta ( p - \Lambda )}{ g_0^{-1} - \Phi_{\Lambda} ( P ) }.
 \label{eq:singlescaledef}
 \end{equation}
A graphical representation of Eqs.~(\ref{eq:flowself}) and (\ref{eq:flowselfbos})
is shown in Fig.~\ref{fig:flowself}. 

Note that in our 
interaction-momentum cutoff scheme only diagrams with bosonic single-scale propagators
appear in the flow equations. For this simplification we pay the price that we have to start the FRG flow with a non-trivial initial condition, as explained above.

\begin{figure}[b!]
  \centering
 \includegraphics[width=0.45\textwidth]{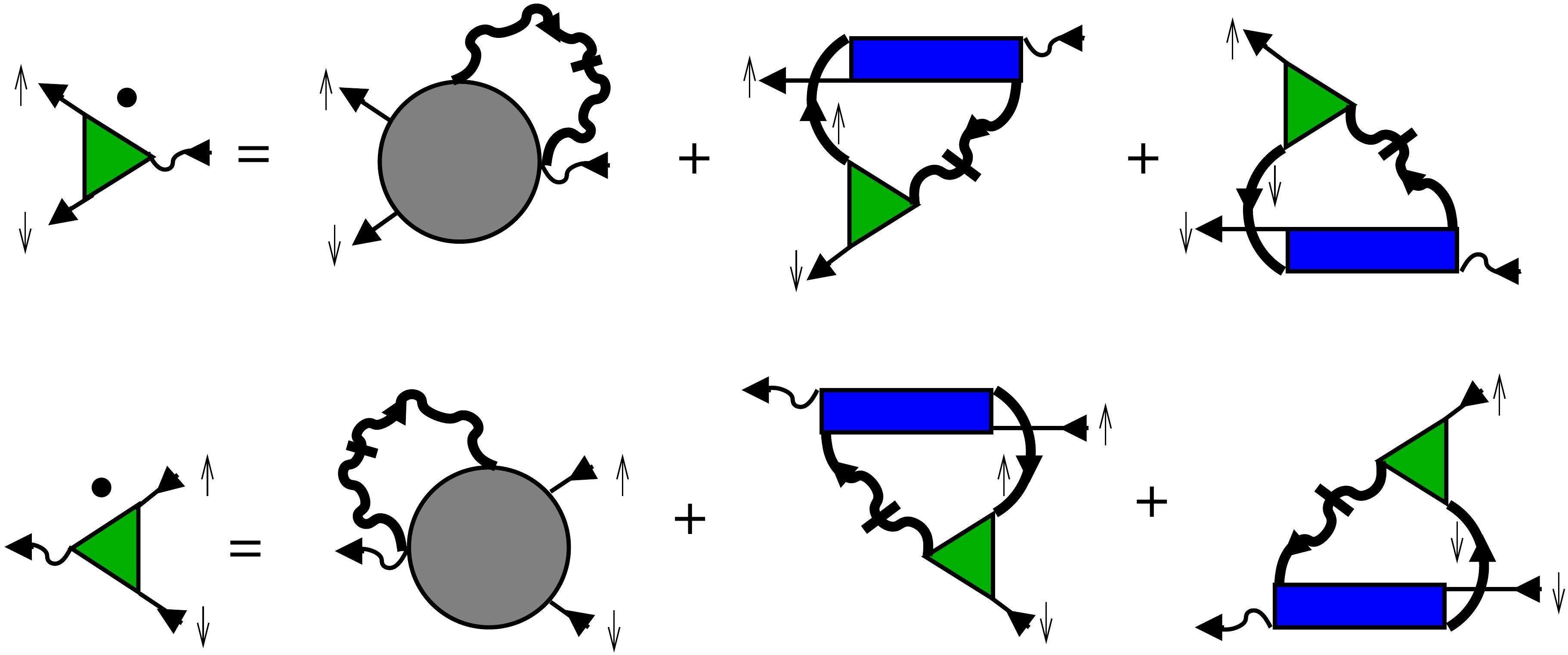}
\vspace{5mm}
  \vspace{-4mm}
  \caption{%
Graphical representation of the exact FRG flow equations (\ref{eq:vertex3flow1}) and
(\ref{eq:vertex3flow2})  for the three-point vertices.
}
\label{fig:flow3}
\end{figure}

The right-hand sides of the flow equations (\ref{eq:flowself}) and (\ref{eq:flowselfbos}) for the self-energies
depend  on higher order vertices with three and four external legs 
for which we can derive again exact FRG  flow equations.
The  flow equations for the three-point vertices are
 \begin{widetext}
 \begin{eqnarray}
 \partial_{\Lambda} \Gamma_{\Lambda}^{\bar{c}_{\uparrow} \bar{c}_{\downarrow} \psi } 
 ( K_1 , K_2 ; P ) & =&
 \int_{P^{\prime}} \dot{F}_{\Lambda} ( P^{\prime} ) \Gamma^{ \bar{c}_{\uparrow} \bar{c}_{\downarrow}
 \bar{\psi} {\psi} \psi }_{\Lambda} ( K_1 , K_2; P^{\prime}; P^{\prime} , P )
 \nonumber
 \\
  &+ & \int_{P^{\prime}} 
 \dot{F}_{\Lambda} ( P^{\prime} ) G_{\Lambda} ( P^{\prime} - K_1 )
 \Gamma^{\bar{c}_{\uparrow} \bar{c}_{\downarrow} {\psi}}
 ( P^{\prime} - K_1, K_1 ; P^{\prime} )
 \Gamma_{\Lambda}^{ \bar{c}_{\uparrow} c_{\uparrow} \bar{\psi} \psi }
 ( K_2; P^{\prime} - K_1 ; P^{\prime}; P  )
 \nonumber
 \\
 &+ & 
\int_{P^{\prime}} 
 \dot{F}_{\Lambda} ( P^{\prime} ) G_{\Lambda} ( P^{\prime} - K_2 )
 \Gamma^{\bar{c}_{\uparrow} \bar{c}_{\downarrow} {\psi}}
 ( K_2, P^{\prime} - K_2 ; P^{\prime} )
 \Gamma_{\Lambda}^{ \bar{c}_{\downarrow} c_{\downarrow} \bar{\psi} \psi }
 ( K_1 ; P^{\prime} - K_2 ; P^{\prime}; P  ) ,
 \label{eq:vertex3flow1}
 \\
 \partial_{\Lambda} \Gamma_{\Lambda}^{{c}_{\downarrow} {c}_{\uparrow} \bar{\psi} } 
 ( K_1 , K_2 ; P ) & = &
 \int_{P^{\prime}} \dot{F}_{\Lambda} ( P^{\prime} ) \Gamma^{ {c}_{\downarrow} {c}_{\uparrow}
 \bar{\psi} \bar{\psi} \psi }_{\Lambda} ( K_1 , K_2; P, P^{\prime}, P^{\prime}  )
 \nonumber 
\\
&+ & \int_{P^{\prime}} 
 \dot{F}_{\Lambda} ( P^{\prime} ) G_{\Lambda} ( P^{\prime} - K_1 )
 \Gamma^{{c}_{\downarrow} {c}_{\uparrow} \bar{\psi}}
 ( K_1, P^{\prime} - K_1 ; P^{\prime} )
 \Gamma_{\Lambda}^{ \bar{c}_{\uparrow} c_{\uparrow} \bar{\psi} \psi }
 ( P^{\prime} - K_1 ; K_2 ; P^{\prime}; P  )
 \nonumber
 \\
 &+ & 
\int_{P^{\prime}} 
 \dot{F}_{\Lambda} ( P^{\prime} ) G_{\Lambda} ( P^{\prime} - K_2 )
 \Gamma^{{c}_{\downarrow} {c}_{\uparrow} \bar{\psi}}
 ( P^{\prime} - K_2, K_2 ; P^{\prime} )
 \Gamma_{\Lambda}^{ \bar{c}_{\downarrow} c_{\downarrow} \bar{\psi} \psi }
 ( P^{\prime} - K_2; K_1 ; P^{\prime}; P  ) .
 \label{eq:vertex3flow2}
 \end{eqnarray}
\end{widetext}
A graphical representation of Eqs.~(\ref{eq:vertex3flow1}) and (\ref{eq:vertex3flow2}) is  shown 
in Fig.~\ref{fig:flow3}.

\begin{figure}[b!]
  \centering
 \includegraphics[width=0.45\textwidth]{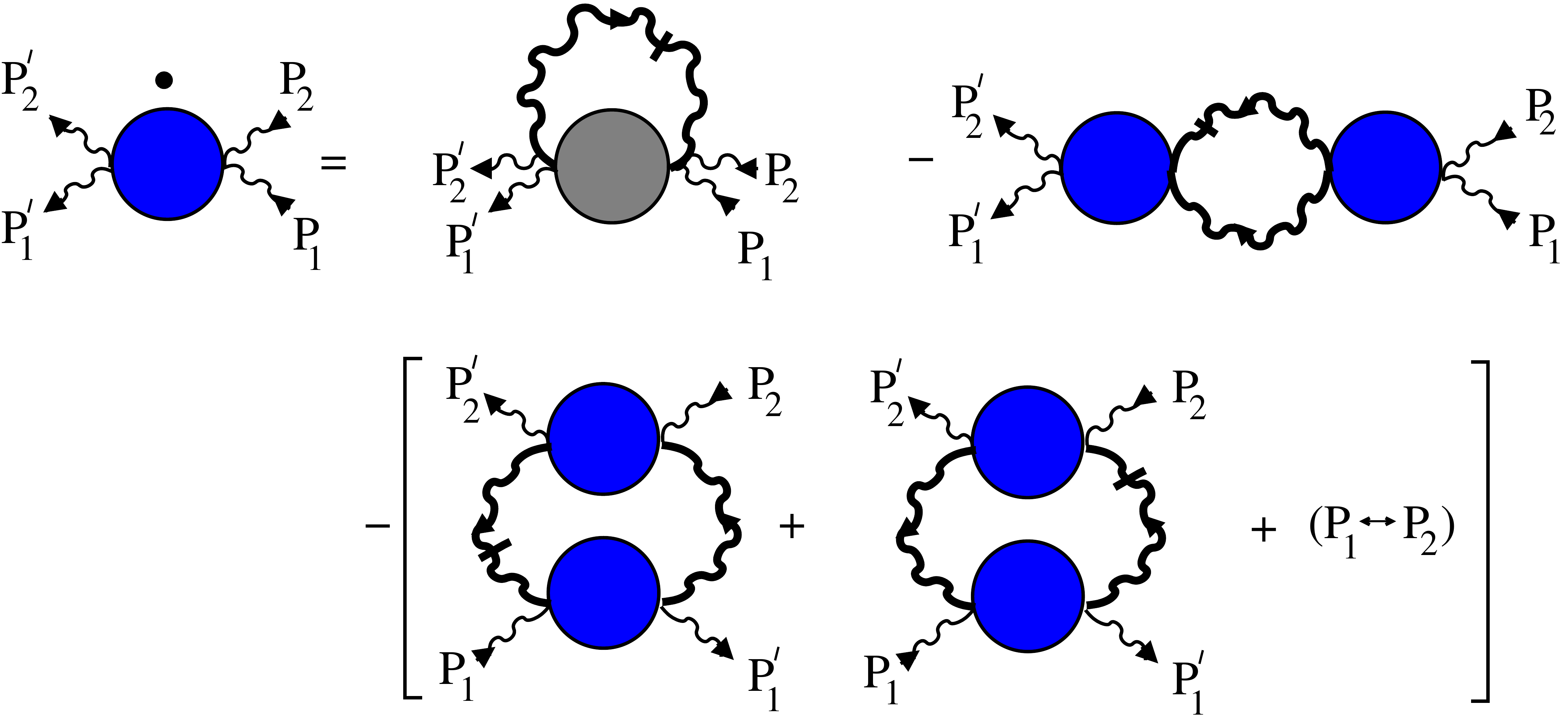}
\vspace{5mm}
  \vspace{-4mm}
  \caption{%
Graphical representation of the exact FRG flow equation ~(\ref{eq:flowgamma4})  for the
induced  interaction between pairing fluctuations.
}
\label{fig:flowgamma}
\end{figure}

Next, consider the bosonic four-point  vertex
 $\Gamma^{\bar{\psi} \bar{\psi} \psi \psi }_{\Lambda} ( P_1^{\prime} , P_2^{\prime} ; P_2 , P_1 ) $ which controls the FRG flow of the
bosonic self-energy in Eq.~(\ref{eq:flowselfbos}).
Recall that
in our interaction-momentum cutoff scheme this vertex,
which describes the induced interaction  between fluctuations of the
superfluid order parameter,
has a finite initial value  at $\Lambda = \Lambda_0$
given by the symmetrized fermion loop in Eq.~(\ref{eq:4loop}).
The FRG flow equation for the bosonic four-point vertex is
(see Fig.~\ref{fig:flowgamma})
 \begin{widetext}
 \begin{eqnarray}
& &\partial_{\Lambda}   \Gamma^{\bar{\psi} \bar{\psi} \psi \psi }_{\Lambda} ( P_1^{\prime} , P_2^{\prime} ; P_2 , P_1 )   =  
  \int_{P }  \dot{F}_{\Lambda} ( P )
\Gamma^{(6)}_{\Lambda} ( P_1^{\prime} , P_2^{\prime} , P^{\prime}; P^{\prime} , P_2 , P_1 ) 
 \nonumber
 \\
&  &  - \int_{P }  \dot{F}_{\Lambda} ( P ) F_{\Lambda} ( P_1 + P_2 - P )
 \Gamma^{\bar{\psi} \bar{\psi} \psi \psi }_{\Lambda} ( P_1^{\prime} , P_2^{\prime} ; 
 P_1 + P_2  - P , P ) 
 \Gamma^{\bar{\psi} \bar{\psi} \psi \psi }_{\Lambda} ( P, P_1 +  P_2 - P ; 
P_2   , P_1 ) 
 \nonumber
 \\
 & & - \int_{P }  \left[ {F}_{\Lambda} ( P ) F_{\Lambda} ( P + P_1 - P_1^{\prime} )  \right]^{\bullet}
 \Gamma^{\bar{\psi} \bar{\psi} \psi \psi }_{\Lambda} ( P_1^{\prime} , P + P_1 - P_1^{\prime}  ; 
 P , P_1 ) 
 \Gamma^{\bar{\psi} \bar{\psi} \psi \psi }_{\Lambda} ( P_2^{\prime} , P ; 
P + P_1 - P_1^{\prime}   , P_2 ) 
 \nonumber
 \\
& & - \int_{P }  \left[ {F}_{\Lambda} ( P ) F_{\Lambda} ( P + P_2 - P_1^{\prime} )
 \right]^{\bullet}
 \Gamma^{\bar{\psi} \bar{\psi} \psi \psi }_{\Lambda} ( P_1^{\prime} , P + P_2 - P_1^{\prime}  ; 
 P , P_2 ) 
 \Gamma^{\bar{\psi} \bar{\psi} \psi \psi }_{\Lambda} ( P_2^{\prime} , P ; 
P + P_2 - P_1^{\prime}   , P_1 ),
 \label{eq:flowgamma4}
 \end{eqnarray}
\end{widetext}
where we have introduced the following product rule notation,
 \begin{equation}
 [ F_{\Lambda} (P ) F_{\Lambda} ( P^{\prime} ) ]^{\bullet} =
  \dot{F}_{\Lambda} (P ) F_{\Lambda} ( P^{\prime} ) + F_{\Lambda} (P ) \dot{F}_{\Lambda} ( P^{\prime} ).
  \label{eq:product_rule}
 \end{equation}

\begin{figure}[tb]
  \centering
 \includegraphics[width=0.45\textwidth]{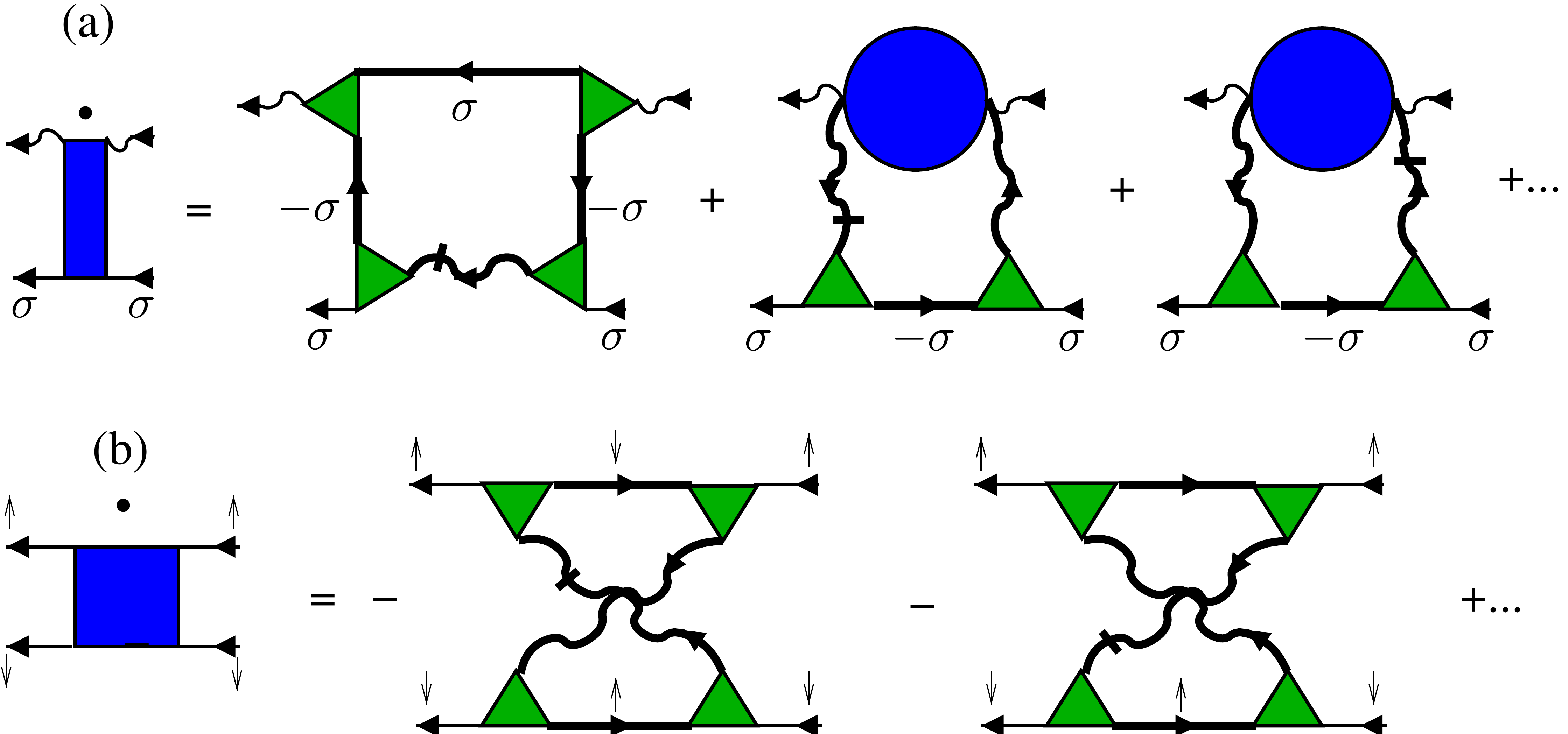}
\vspace{5mm}
  \vspace{-4mm}
  \caption{%
Approximate FRG flow equations for the induced interactions in a truncation where only
those vertices are retained on the right-hand side which are finite at the initial cutoff scale:
(a) mixed fermion-boson vertex; (b) interaction  vertex between two fermions with opposite spin.
}
\label{fig:flowinduced}
\end{figure}

To conclude this section, let us briefly discuss the flow equations of the induced interactions
which vanish at the initial cutoff scale:
the mixed fermion-boson vertex
$\Gamma^{\bar{c}_{\sigma} {c}_{\sigma} \bar{\psi} \psi}_{\Lambda}$
and the induced fermionic interactions
$\Gamma_{\Lambda}^{ \bar{c}_{\uparrow} \bar{c}_{\downarrow} c_{\downarrow} c_{\uparrow} }$ and
$\Gamma_{\Lambda}^{\bar{c}_{\sigma} \bar{c}_{\sigma} c_{\sigma} c_{\sigma} }$
which appear in the vertex expansion ~(\ref{eq:Gammatrunc}) and are represented by the
symbols defined in Fig.~\ref{fig:induced4} (a) and (b).
The exact FRG flow equations for these vertices are rather complicated and are given in Appendix~B.
Because the right-hand sides of the flow equations 
for the mixed fermion-boson vertex $\Gamma^{\bar{c}_{\sigma} {c}_{\sigma} \bar{\psi} \psi}_{\Lambda}$
and for the fermionic interaction vertex
$\Gamma_{\Lambda}^{ \bar{c}_{\uparrow} \bar{c}_{\downarrow} c_{\downarrow} c_{\uparrow} }$
are finite even if
the above four-point vertices are neglected, 
the FRG flow generates finite values of these induced interactions.
An approximate method to take these induced interactions into account 
is to retain only those vertices in the FRG flow equations which
are finite at the initial scale. 
In Ref.~\onlinecite{Sharma16} we have obtained reasonable results using a  similar strategy 
to truncate the
hierarchy of FRG flow equations for the vertices in a low-energy model for graphene.
Following this strategy, 
we arrive to the simplified FRG flow equations for the induced interactions shown graphically
in Fig.~\ref{fig:flowinduced}.
Note that the induced interaction between fermions with parallel spin still vanishes
within this approximation.

In Appendix~E we present an approximate evaluation of the flow equation 
for the mixed four-point vertex shown in Fig.~\ref{fig:flowinduced} (a)
and study the  effect of this vertex on the fermionic self-energy.
At this point, let us make three comments on the truncated flow equations
of the induced interactions shown in Fig.~\ref{fig:flowinduced}. 
First of all, if we replace the boson propagators in the approximate flow equation 
for the fermionic interaction vertex shown in Fig.~\ref{fig:flowinduced} (b) by the
$T$-matrix $g$ and neglect all self-energy corrections to the fermionic
propagators, we recover the leading term in the perturbative
expansion of this interaction vertex shown in Fig.~\ref{fig:pert} (b).
Note, however, that within our interaction-momentum cutoff scheme this vertex
does not directly couple to the FRG flow of the bosonic self-energy.
The corresponding renormalization of the critical temperature is
taken into account via the bosonic six-point vertex, as explained in the text after
Eq.~(\ref{eq:gamma6init}).

Next, we note that for small values of the scattering matrix $g$ we can use these flow equations
to calculate higher order vertex corrections to various physical quantities.
For example, from the truncated flow equation for the mixed
fermion-boson vertex shown in Fig.~\ref{fig:flowinduced} (a) it is obvious that this vertex is at least of 
order $g$. From the exact flow equations for the three-point vertices shown in Fig.~\ref{fig:flow3}
we then see that the latter are at least of order $g^2$. 

Finally, let us point out that the induced interactions can exhibit some rather complicated momentum- and
frequency dependence. 
Only in cases where this can be  neglected, one can try to avoid the appearance 
of the  induced interactions by redefining the
bosonic Hubbard-Stratonovich fields $\psi$ and $\bar{\psi}$.
This strategy, which has been called dynamical re-bosonization \cite{Gies02},
was adopted by Floerchinger {\it{et al.}} \cite{Floerchinger08} who attempted to  reproduce the 
GM correction to the critical temperature using this strategy.
However, the numerical value of the GM correction is determined by the
full momentum dependence of the 
induced interaction in (\ref{eq:Uindflow}), so that it is not surprising that
Floerchinger {\it{et al.}} \cite{Floerchinger08} could not reproduce the precise
numerical value of the GM correction given in Eq.~(\ref{eq:tcalpha1}).

\section{Density of states and quasiparticle damping  
in the normal state close to $T_c$}
\label{sec:selfGauss}

In this section, we shall consider the effect of superfluid fluctuations on the
electronic self-energy in the normal state at and slightly above the critical temperature.
This effect is usually neglected \cite{Larkin05}, which is only correct for temperatures
not too close to $T_c$. Surprisingly, a quantitatively accurate calculation 
of the electronic self-energy in this regime cannot be found in the literature. Although such a theory is currently needed in other contexts, 
e.g. temporal development of an order parameter following a sudden quench in the field of 
out-of-equilibrium dynamics\cite{Yuzbashyan06,Yuzbashyan09,Straeter12}.
To begin with, we analyze this problem in  Sec.~\ref{subsec:gauss} 
within the Gaussian approximation for the propagator of the 
superfluid order parameter field. 
However,  the critical behavior of the superfluid order parameter
belongs to the XY-universality class, which below four dimensions
is controlled by the Wilson-Fisher fixed point.
In Sec.~\ref{subsec:FRGdamp} we shall therefore present a more accurate analysis of this
problem using the FRG approach developed in Sec.~\ref{sec:FRG}.

\subsection{Gaussian approximation}
\label{subsec:gauss}

To begin with, let us calculate the electronic self-energy within the
Gaussian approximation, which is equivalent to calculating the effective
interaction in ladder approximation. 
In the normal state the self-energy is then given by
 \begin{equation}
 \Sigma_1 ( K )  =  - \int_P  F_0 ( P ) G_0 ( P-K ),
 \label{eq:sigmak}
 \end{equation}
where the Gaussian propagator of the pairing field
is given in Eq.~(\ref{eq:F0def}).
Since we  are interested in the effect of long-wavelength and low-energy
order parameter fluctuations on the
fermionic self-energy, we may expand the inverse Gaussian propagator
to leading order in momenta and frequencies,
\begin{eqnarray}
 F_0^{-1} ( \bd{p} ,  i \bar{\omega} )  &  = &  g^{-1} - \Phi_0^{\rm reg} ( 
 \bd{p} ,  i \bar{\omega} ) 
 \nonumber
 \\
 & \approx & \nu [ t_0 + {p}^2 /p_0^2 +  
 | \bar{\omega} | / \omega_0  ].
 \label{eq:Finvexp}
 \end{eqnarray}
In the BCS regime and for $| T - T_{c0} | \ll T_{c0}$ 
the dimensionless parameter $t_0$ can be identified with the 
reduced temperature
 \begin{equation}
 t_0 = \frac{ T - T_{c0}}{T_{c0} } ,
 \end{equation}
while the momentum scale $p_0$ and the energy scale $\omega_0$ are both 
proportional to the temperature \cite{Larkin05}
\begin{eqnarray}
  p_0 &  = &  \sqrt{ \frac{ 48 }{ 7 \zeta (3) }}  \frac{  \pi T}{v_F },
  \\
 \omega_0  & = &  \frac{8 T }{ \pi}.
 \end{eqnarray}
Note that $1/p_0 = \xi_0$ can be identified with the coherence length 
of a clean three-dimensional superconductor with isotropic Fermi surface \cite{Larkin05}.
The Ginzburg-Levanyuk number $Gi$ introduced in Eq.~(\ref{eq:Gi})
can be written as \cite{Larkin05}
 \begin{equation}
 Gi = \left( \frac{ 7 \zeta (3) p_0^3 }{64 \pi^3 \nu T_c} \right)^2 =
 \frac{ 27}{28 \zeta ( 3 )}  \left( \frac{ \pi T_c }{E_F } \right)^4.
 \label{eq:Gi2}
 \end{equation}
On the other hand, in the strong coupling regime where $\nu g$ is not small 
the coefficients in the long-wavelength expansion of $F_0^{-1} ( \bd{p} , i \bar{\omega} )$
have a more complicated dependence  in $T$ and $\mu$, as discussed in Appendix~C.
In particular, at the unitary point $g^{-1} =0$ the momentum scale
$p_0$ is of the order of $k_F$ while $\omega_0$
is of order  $E_F$.
Note that the corresponding expressions given by Larkin and Varlamov~\cite{Larkin05} are only valid 
in the BCS limit $\nu g \ll 1$.

\begin{figure}[tb]
	\centering
	\includegraphics[width=0.48\textwidth]{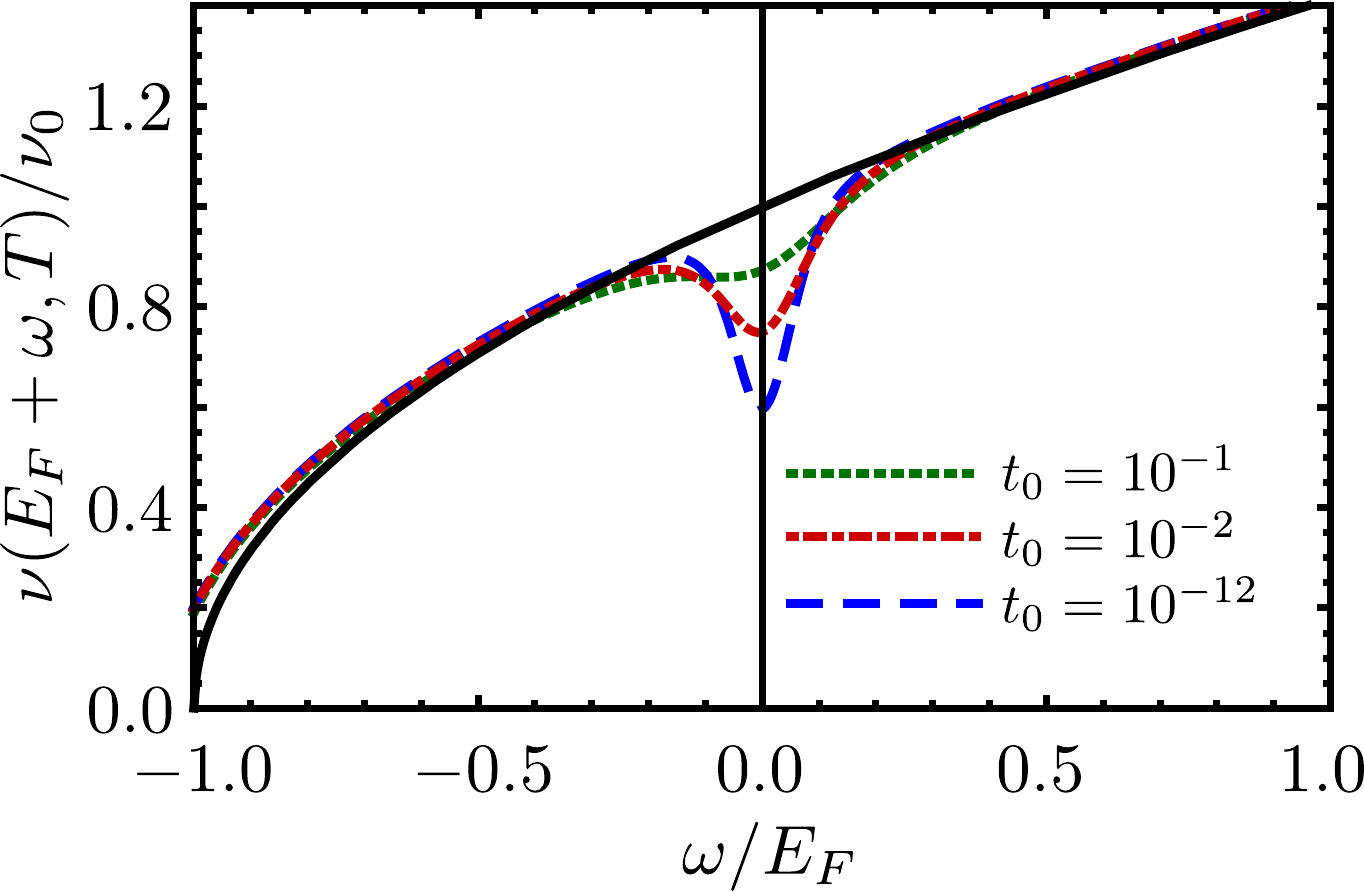}
	\vspace{5mm}
	\vspace{-4mm}
	\caption{%
		Frequency-dependence of the modification of the density of states
		due to classical order parameter fluctuations obtained from the numerical
		evaluation of Eq.~(\ref{eq:dos}).
		The curves are for $ T/ E_F = 0.1$,  $\mu =E_F$, and 
		$ t_0 = 10^{-1}$ (green dotted line), $ t_0 = 10^{-2}$ (red dashed-dotted line),
		and $ t_0 = 10^{-12}$ (blue dashed line). The solid black line is the non-interacting
		density of states $\nu(\epsilon) / \nu_0  = \sqrt{\epsilon / E_F}$.
	}
	\label{fig:dos}
\end{figure}

Let us now focus on the effect of classical
long-wavelength fluctuations of the superfluid order parameter on the
fermionic self-energy.
Because in the vicinity of the critical temperature the dynamics of the order parameter is slow compared with the
electron dynamics, it is then sufficient to retain only the
contribution from the zeroth Matsubara frequency in Eq.~(\ref{eq:sigmak}).
In Appendix~D we present a formal justification of this approximation.
The resulting  critical contribution to the fermionic self-energy
is 
\begin{equation}
 \Sigma_{\rm crit} ( \bd{k} , i \omega ) =
  \frac{T}{\nu} \int_{\bd{p}} \frac{\Theta ( p_0 - | \bd{p} |) }{ t_0 +  \bd{p}^2 / p_0^2 }
 \frac{1}{  i \omega + \xi_{\bd{p} - \bd{k}} },
 \label{eq:sigmasing}
 \end{equation}
where the cutoff $\Theta ( p_0 - | \bd{p} |)$ takes into account the range of validity of
our long-wavelength expansion (\ref{eq:Finvexp}).
Eq.~(\ref{eq:sigmasing}) can be evaluated analytically without further approximation,
but the result is very complicated so that we do not present it here.
In Fig.~\ref{fig:dos} we plot the corresponding renormalized density of states
\begin{equation}
 \nu_{\rm crit} (E_F +  \omega ) = - \frac{1}{\pi}  {\rm Im} \int_{\bd{k}}
  \frac{1}{ \omega - \xi_{\bd{k}} - \Sigma_{\rm crit} ( \bd{k} , \omega  + i 0^{+} ) } .
  \label{eq:dos}
\end{equation}

\begin{figure}[tb]
  \centering
 \includegraphics[width=0.45\textwidth]{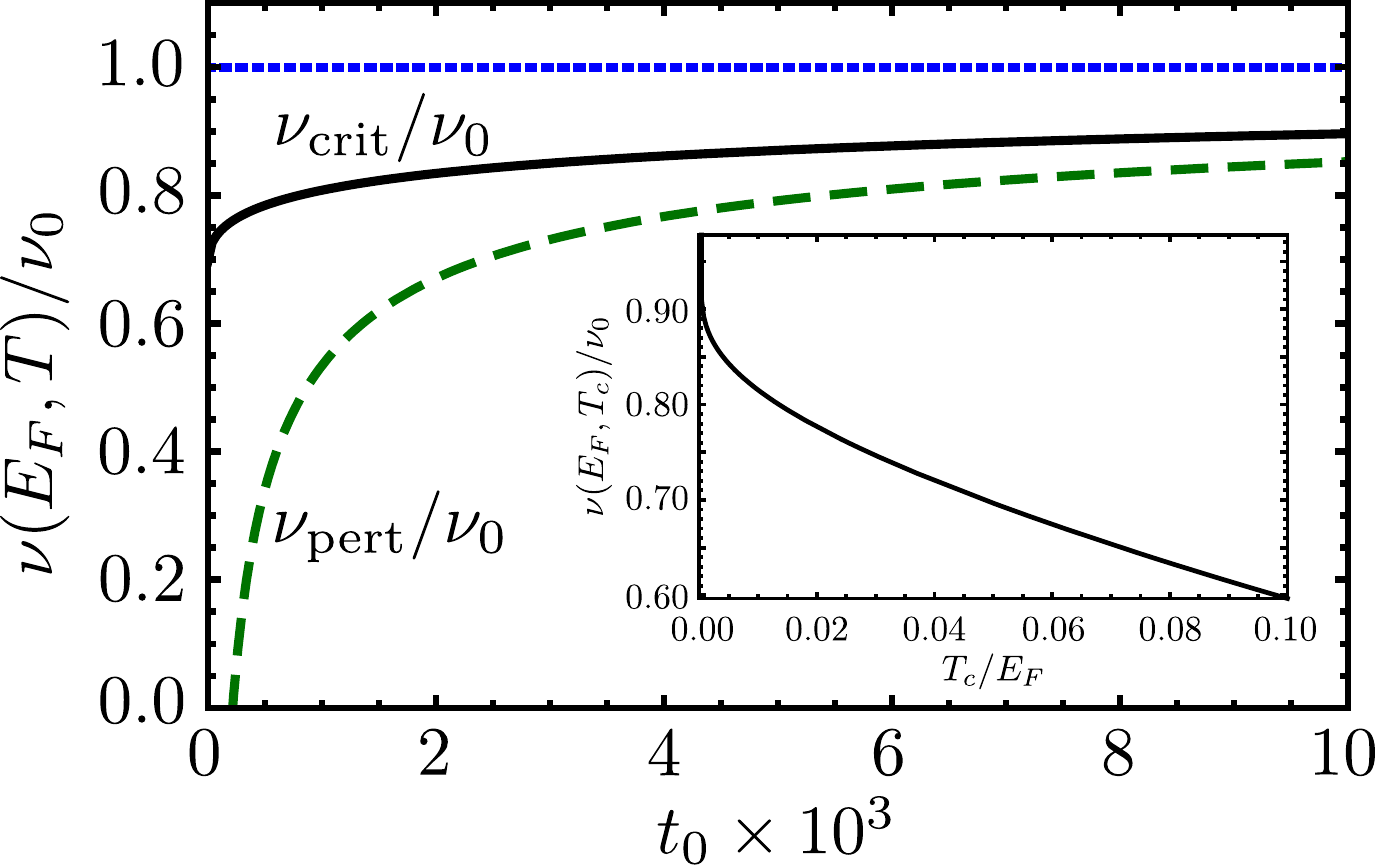}
\vspace{5mm}
  \vspace{-4mm}
  \caption{%
The solid line represents our result (\ref{eq:dos}) for the density of
states at the Fermi energy $\nu_{\rm crit} ( E_F ,T )$
as a function of the reduced temperature $t_0 = ( T - T_{c0}) / T_{c0}$  for $T_c / E_F=0.05$.
Note that for $t_0 \rightarrow 0$ the density of states has a finite limit.
The green dashed line represents  the perturbative result (\ref{eq:nu_dc})
derived by
Di Castro {\it{et al.}}\cite{DiCastro90} which diverges at the critical temperature as $ -1/\sqrt{t_0}$. 
In the inset we show the behavior of $\nu_{\rm crit} ( E_F ,T_{c0} )$ as a function of $T_{c0}$.
}
\label{fig:dosvsT}
\end{figure}

Obviously, for $T \rightarrow T_c$ classical pairing fluctuations
give rise to a pronounced pseudogap in the density of states at the Fermi energy.
This has already been noticed by
Di Castro \textit{et al.} in Ref.~[\onlinecite{DiCastro90}] within a perturbative approach
which amounts to expanding the right-hand side of
Eq.~(\ref{eq:dos}) to first order in the self-energy.
With this approximation  Di Castro \textit{et al.} obtained
for the density of states at the Fermi-energy \cite{DiCastro90}
\begin{equation}
\nu_{\rm pert}(E_F) = \nu_0 \left[ 1-  \sqrt{\frac{3}{7\zeta(3)}} \frac{(\pi T/E_F)^2}{\sqrt{t_0}} \right].
\label{eq:nu_dc}
\end{equation}
This suppression of the density of states has been observed experimentally in the
fluctuation regime above the superfluid transition of a strongly interacting
Fermi gas\cite{Gaebler10}. However, for $t_0 \rightarrow 0$ the correction 
in Eq.~(\ref{eq:nu_dc})  diverges.
Clearly, this divergence is unphysical and 
signals the breakdown of perturbation theory
for temperatures close to $T_c$. 
In contrast to the perturbative result (\ref{eq:nu_dc}) our
expression  (\ref{eq:dos})
obtained within the Gaussian approximation predicts a finite suppression of the density of states
for all $t_0 \geq 0$.
To show this, we have evaluated  Eq.~(\ref{eq:dos}) numerically for different temperatures.
In  Fig.~\ref{fig:dosvsT}.
and show our numerical result for $\nu_{\rm crit} ( E_F )$ as a function of $t_0$.
Note that at the critical point $t_0 =0$ the Gaussian approximation (\ref{eq:dos}) predicts a finite
suppression of the density of states at the Fermi energy.

The phenomenon that within perturbation theory
superconducting fluctuations 
above $T_c$ give rise to singular corrections
to various physical quantities has first been noticed
by  Aslamazov and Larkin \cite{Aslamazov69}, who
discovered a $1/t_0$ singularity in the conductivity 
of normal metals due to virtually formed Cooper pairs above $T_c$.
Moreover, Maki~\cite{Maki68} and Thompson\cite{Thompson69} have shown
that Cooper pair formation along diffusive paths in a disordered conductor also lead
to singularities in the 
the transport coefficients. Although the Maki-Thompson correction to the conductivity
has generally a weaker functional dependence on the reduced temperature $t_0$,  
in certain regimes it can be larger than the Aslamazov-Larkin correction. 
However, similar to the singularity in the density of states given in
Eq.~(\ref{eq:nu_dc}), the perturbatively generated
singularities at $T = T_c$
should be regularized by some higher order process.
The only systematic way of introducing a cutoff at $T=T_c$ 
so far is an external pair-breaking mechanism,
such as magnetic impurities  or electron-phonon interactions.\cite{Keller71}
There were a few attempts to identify a cutoff at $T_c$ within the microscopic theory itself: 
by accounting for some subclasses of higher order diagrams\cite{Patton71} and by accounting for non-linear effects 
of the fluctuations through Gorkov equation\cite{Larkin72}, 
at least in dirty superconductors. But the results are still inconclusive.
Here we focus on the singularity in the density of states and 
propose a new strategy to solve this long-standing problem
 using renormalization
group methods. In fact, from Fig.~\ref{fig:dosvsT} it is clear that
the singularity in the perturbative result (\ref{eq:nu_dc}) can be removed
if we do not expand the density of the states in powers of the self-energy but 
insert the perturbative self-energy into the Dyson equation and use  
Eq.~(\ref{eq:dos}) to calculate the density of states.

To gain a better analytical understanding of the low-energy behavior of the self-energy, 
let us simplify the integrand in Eq.~(\ref{eq:sigmasing}) by
setting $\bd{k} = \bd{k}_F + \bd{q} $ and assuming $ | \bd{q} | \ll k_F$.
We may then approximate $\xi_{ \bd{p} - \bd{k}} \approx - \bd{v}_F \cdot ( \bd{p} - \bd{q} )
 = \xi_{\bd{k}} - \bd{v}_F \cdot \bd{p}$. 
We have verified numerically that this approximation correctly reproduces the main
low-energy features of the self-energy.
After analytic continuation ($ i \omega \rightarrow \omega + i 0^+$)
we obtain from Eq.~(\ref{eq:sigmasing}) for the imaginary part
for $t_0 \ll 1$ and $ | \omega + \xi_{\bd{k}} | \ll v_F p_0$,
 \begin{equation}
 {\rm Im} \Sigma_{\rm crit} ( \bd{k} , \omega + i 0^+ ) = T  \frac{ p_0^2}{ \nu v_F}
 \frac{\pi}{2}
 \ln \left[ 
t_0 + \left( \frac{  \omega + \xi_{\bd{k}} }{ v_F p_0 } \right)^2
 \right].
 \label{eq:imres}
 \end{equation}
To calculate the real part of the self-energy for $t_0 \ll 1$ and $ | \omega + \xi_{\bd{k}} | \ll v_F p_0$
we first perform the angular integration in Eq.~(\ref{eq:sigmasing}) and obtain
 \begin{eqnarray}
 {\rm Re} \Sigma_{\rm crit} ( \bd{k} , \omega + i 0^+ ) & = &  T   \frac{ p_0^2}{ \nu v_F}
  {\rm sgn} ( \omega + \xi_{\bd{k}} )
 \nonumber
 \\
 & & \hspace{-15mm} \times 
 \int_0^1  \frac{dx x }{t_0 + x^2 }
 \ln \left| \frac{ x +   \frac{  \left| \omega + \xi_{\bd{k}}  \right| }{v_F p_0 }   }{ 
     x -   \frac{  \left| \omega + \xi_{\bd{k}}  \right| }{v_F p_0 }       }
 \right|.
 \end{eqnarray}
In the regime $ \sqrt{t_0 } \ll | \omega + \xi_{\bd{k}} | / (v_F p_0 ) \ll 1$
we may set $t_0=0$ and move the upper limit of the $x$- integral to infinity.
Using the fact that for $a > 0$,
 \begin{equation}
 \int_0^{\infty} \frac{dx}{x} \ln \left| \frac{ x+a}{x-a} \right| = \frac{ \pi^2}{2}
 \end{equation}
we obtain for $ \sqrt{t_0 } \ll | \omega + \xi_{\bd{k}} | / (v_F p_0 ) \ll 1$,
 \begin{equation}
 {\rm Re} \Sigma_{\rm crit} ( \bd{k} , \omega + i 0^+ )   \approx  T   \frac{ p_0^2}{ \nu v_F}
  \frac{\pi^2}{2} {\rm sgn} ( \omega + \xi_{\bd{k}} ).
 \end{equation}
In the opposite regime
$ | \omega + \xi_{\bd{k}} | / (v_F p_0 ) \ll \sqrt{t_0} \ll 1$ we may expand
the logarithm for $x \gg | \omega + \xi_{\bd{k}} | / (v_F p_0 )$,
 \begin{equation}
\ln \left| \frac{ x +   \frac{  \left| \omega + \xi_{\bd{k}}  \right| }{v_F p_0 }   }{ 
     x -   \frac{  \left| \omega + \xi_{\bd{k}}  \right| }{v_F p_0 }       }
 \right| \approx \frac{2}{x}  \frac{  \left| \omega + \xi_{\bd{k}}  \right| }{v_F p_0 } .
 \end{equation}
Then we obtain to leading order
 \begin{equation}
  {\rm Re} \Sigma_{\rm crit} ( \bd{k} , \omega + i 0^+ )  \approx   T   \frac{ p_0^2}{ \nu v_F}
  \frac{\pi}{\sqrt{t_0}} \frac{ \omega + \xi_{\bd{k}} }{  v_F p_0 }.
 \end{equation}
Note that in both regimes the imaginary part of the self-energy is parametrically larger
than the real part, so that from now on we shall simply neglect the real part of the self-energy.

Our result (\ref{eq:imres}) for the imaginary part of the self-energy due to classical fluctuations
of the superfluid order parameter
implies that for $T \rightarrow T_{c0} $ the
damping of quasiparticles on the Fermi surface diverges as
 \begin{equation}
 \gamma_{\rm crit} = - {\rm Im} \Sigma_{\rm crit} ( \bd{k}_F , i 0^+ ) = \frac{\pi T }{2} \frac{ p_0^2}{\nu v_F } \ln \left( \frac{ T_{c0}}{ T - T_{c0} } \right).
\label{eq:gammacrit}
 \end{equation}
While within the Gaussian approximation the density of states is finite, 
the quasiparticle damping exhibits an unphysical logarithmic singularity 
for $T \rightarrow T_c$. 
In the BCS regime the logarithm is multiplied by a small 
prefactor $ T p_0^2 / \nu v_F \propto T^3 / E_F^2$, 
while in the vicinity of the unitary point where $p_0 \propto k_F$
the prefactor is linear in the temperature, such that
\begin{equation}
 \gamma_{\text{crit}} \propto T.
\end{equation}
Comparing the above  $\gamma_{\rm crit}$ with the generic form
of the quasiparticle damping in a three-dimensional Fermi liquid,
 \begin{equation}
 \gamma_{\rm FL} =  C_{\rm FL} T^2 / E_F,
 \label{eq:dampFL}
 \end{equation}
where the numerical constant $C_{\rm FL}$ is usually of the order of unity \cite{Kittel66,Thouless77},
we conclude that for $ \ln (1/ t_0 ) \gtrsim E_F / T $ the contribution from classical
superconducting fluctuations to the
quasiparticle damping dominates.

It turns out, however, that the logarithmic divergence
in Eq.~(\ref{eq:gammacrit}) is an artifact of the Gaussian approximation. 
Physically, it is clear that both the damping of the intermediate states as well as the existence  of an
anomalous dimension $\eta$ of the superfluid order parameter field will  smooth out this singularity.
For example, to take into account the usual Fermi liquid damping (\ref{eq:dampFL}) 
we should replace the free propagator in Eq.~(\ref{eq:sigmak}) by
 \begin{equation}
 G_{1} ( K ) = \frac{1}{ i \omega - \xi_{\bd{k}} + i \gamma_{\rm FL}   {\rm sgn} \omega  }.
 \end{equation}
Then we obtain for $T \rightarrow T_{c0}$ 
 instead of Eq.~(\ref{eq:gammacrit}),
 \begin{equation}
 \gamma_{\rm crit} = \frac{\pi T_{c0} }{2} \frac{ p_0^2}{\nu v_F } \ln \left( 
 \frac{  v_F p_0 }{\gamma_{\rm FL}} \right),
\label{eq:gammacrit_fl}
 \end{equation}
which is proportional to $T_{c0}^3 \ln (E_F /T_{c0} )$ in the BCS regime.
A similar sub-leading
non-analytic correction to the self-energy of
three-dimensional Fermi liquids is also generated by 
short-range interactions \cite{Chubukov06}.
Note, however, that Eq.~(\ref{eq:gammacrit_fl}) does not take into account
that the  anomalous dimension $\eta$ of
superfluid fluctuations at the critical point.
Recall that critical behavior of the superconduting transition
belong to the universality class of the classical XY-model
which is characterized by a finite critical exponent (anomalous dimension) $\eta \approx 0.038$ 
in three dimensions [\onlinecite{Pelissetto02}]. The true static propagator of the order-parameter field
at $T = T_c$ is therefore for small momenta $\bd{p}$ of the form
 \begin{equation}
 F_{\ast} ( \bd{p} , 0 ) \sim \frac{A_\ast}{\nu} \left(
 \frac{p_0}{ p } \right)^{ 2 - \eta} ,
 \label{eq:Fcrit}
 \end{equation}
where $A_{\ast}$ is a dimensionless  constant.
If we replace the Gaussian propagator in Eq.~(\ref{eq:sigmasing})
by Eq.~(\ref{eq:Fcrit})  we obtain for the
self-energy at the critical point,
\begin{equation}
 \Sigma_{\rm crit} ( \bd{k} , i \omega ) \approx
  \frac{T }{\nu} \int_{\bd{p}}
 \left( \frac{p_0}{ p } \right)^{ 2 - \eta}
 \frac{  A_{\ast}  \Theta ( p_0 - | \bd{p} |)   }{  i \omega + \xi_{\bd{p} - \bd{k}} }.
 \label{eq:sigmaeta}
 \end{equation}
From this expression it is easy to show that
 \begin{equation}
 \gamma_{\rm crit} \propto \frac{T}{\eta} \frac{p_0^2}{\pi v_F } \propto 
\frac{1}{\eta} \frac{T^3}{ E_F^2}.
 \label{eq:gammaeta}
 \end{equation}
Due to the small value of $\eta$, the prefactor of the leading $T^3 $-behavior
is unusually large.
Of course, the above procedure is not satisfactory because  it does not self-consistently
take the interplay between critical fluctuations and quasiparticle damping
of intermediate states into account.
We shall address this problem below using
the FRG.  This allows us to
consistently take into account the feedback of non-Gaussian critical
order parameter fluctuations on the electronic properties,
which provide an intrinsic cutoff
of the  logarithmic singularity in the quasiparticle damping
encountered in Gaussian approximation, see Eq.~(\ref{eq:gammacrit}).

\subsection{FRG calculation of the quasiparticle damping}
\label{subsec:FRGdamp}

The exact FRG flow equation of the fermionic self-energy $\Sigma_{\Lambda} ( K )$
 is given
in Eq.~(\ref{eq:flowself}) and is shown graphically in Fig.~\ref{fig:flowself} (a).
This flow equation depends on the cutoff-dependent mixed fermion-boson interaction 
$\Gamma^{\bar{c}_{\sigma} {c}_{\sigma} \bar{\psi} \psi}_{\Lambda}$
and on the three-point vertices
$\Gamma_{\Lambda}^{ \bar{c}_{\uparrow} \bar{c}_{\downarrow}  \psi }$ and
$\Gamma_{\Lambda}^{c_{\downarrow} c_{\uparrow} \bar{\psi} }$.
In this subsection we shall neglect all vertices which
vanish at the initial cutoff scale within our cutoff scheme. In particular, we
set the mixed fermion-boson interaction vertex equal to zero,
 \begin{equation}
\Gamma^{\bar{c}_{\sigma} {c}_{\sigma} \bar{\psi} \psi}_{\Lambda} \approx 0.
 \end{equation}
From the exact FRG flow equations (\ref{eq:vertex3flow1}, \ref{eq:vertex3flow2})
for the three-point vertices
shown graphically in Fig.~\ref{fig:flow3} it is obvious that in  our interaction-momentum cutoff scheme 
this truncation is consistent with approximating the
three-point vertices by their initial values,
 \begin{equation}
 \Gamma_{\Lambda}^{ \bar{c}_{\uparrow} \bar{c}_{\downarrow}  \psi } =
 \Gamma_{\Lambda}^{c_{\downarrow} c_{\uparrow} \bar{\psi} } \approx 1,
 \end{equation}
see Eq.~(\ref{eq:gamma3initial}).
In Appendix~E we shall use a more elaborate truncation 
strategy where the RG flow of the three-point vertices
$\Gamma_{\Lambda}^{ \bar{c}_{\uparrow} \bar{c}_{\downarrow}  \psi }$,
$\Gamma_{\Lambda}^{c_{\downarrow} c_{\uparrow} \bar{\psi} }$
and the mixed four-point vertex
$\Gamma^{\bar{c}_{\sigma} {c}_{\sigma} \bar{\psi} \psi}_{\Lambda}$
is regained. 
However, our main result for the quasiparticle 
damping derived in this subsection is not qualitatively modified by the
higher order vertex corrections represented by the RG flow of
$\Gamma_{\Lambda}^{ \bar{c}_{\uparrow} \bar{c}_{\downarrow}  \psi }$,
$\Gamma_{\Lambda}^{c_{\downarrow} c_{\uparrow} \bar{\psi} }$, and
$\Gamma^{\bar{c}_{\sigma} {c}_{\sigma} \bar{\psi} \psi}_{\Lambda}$.

Since we are interested in the effect of classical critical fluctuations, we retain
only the contribution from the zeroth Matsubara frequency
to the right-hand side of the flow equation (\ref{eq:flowself}).
After analytic continuation to the real frequencies we obtain the
following FRG flow equation for
the fermionic self-energy,
 \begin{eqnarray}
 & & \partial_{\Lambda} \Sigma_{\Lambda} ( \bd{k} , \omega + i 0^+ ) =
 \nonumber
 \\
 &   & T \int_{\bd{p}}
 \frac{ \dot{F}_{\Lambda} ( \bd{p}  )     }{ \omega + \xi_{ \bd{p} - \bd{k}} + \Sigma_{\Lambda} ( \bd{p} - \bd{k} , - \omega - i 0^+ ) }.
 \label{eq:selftrunc}
 \end{eqnarray}
We approximate the flowing static single-scale propagator by its 
long wavelength limit
 \begin{equation}
 \dot{F}_{\Lambda} ( \bd{p}  )  \approx \frac{ - \delta ( p - \Lambda )}{
 r_{\Lambda} + c_{\Lambda} \Lambda^2 }.
 \end{equation}
The parameters $r_{\Lambda}$ and $c_{\Lambda}$
are  determined by the FRG flow equation (\ref{eq:flowselfbos})
for the bosonic self-energy $\Phi_{\Lambda} ( P )$ shown graphically in
Fig.~\ref{fig:flowself}~(b). Since we are only interested in classical fluctuations
we may set all Matsubara frequencies equal to zero in these equations and obtain
 \begin{eqnarray}
 \partial_{\Lambda} r_{\Lambda} & = & T \int_{\bd{p}} \dot{F}_{\Lambda} ( \bd{p} )
 \Gamma_{\Lambda}^{\bar{\psi} \bar{\psi} \psi \psi } (0, \bd{p}  ;  \bd{p} ,0 ),
 \label{eq:rcl}
 \\
 \partial_{\Lambda} c_{\Lambda} & = & T \int_{\bd{p}} 
 \dot{F}_{\Lambda} ( \bd{p}  )
 \lim_{ \bd{q} \rightarrow 0 }
 \frac{ \partial }{\partial q^2 }
 \Gamma_{\Lambda}^{\bar{\psi} \bar{\psi} \psi \psi } ( \bd{q} , \bd{p}  ;  
 \bd{p} , \bd{q}  ). \hspace{7mm}
 \label{eq:ccl}
 \end{eqnarray}
Note that the parameter $c_{\Lambda}$ is related to the scale-dependent anomalous
dimension
$\eta_\Lambda$ of the superfluid order parameter field as follows \cite{Kopietz10}
 \begin{equation}
 \eta_\Lambda 
 = \Lambda \partial_{\Lambda}  \ln \left( \frac{ c_0}{c_{\Lambda}} \right)   =
 - \frac{  \Lambda \partial_{\Lambda} c_{\Lambda}}{c_{\Lambda} }
 .
 \label{eq:etadef}
 \end{equation}
Our truncated FRG flow equation (\ref{eq:selftrunc}) therefore contains
both the effect of  the anomalous dimension of the superfluid order parameter and
the damping of intermediate states.
In fact, our evaluation of the self-energy in Gaussian approximation presented in 
Sec.~\ref{subsec:gauss} shows that
critical fluctuations mainly renormalize the imaginary part of the self-energy.
We therefore ignore the real part of the self-energy in Eq.~(\ref{eq:selftrunc})
and focus on the FRG flow of its imaginary part  on the Fermi surface,
 \begin{equation}
 \gamma_{\Lambda} = - {\rm Im} \Sigma_{\Lambda} ( \bd{k}_F , i 0^+ ).\label{eq:gamma_FRG}
 \end{equation}
This quasiparticle damping is determined by the flow equation
 \begin{equation}
 \partial_{\Lambda} \gamma_{\Lambda} = - T \int_{\bd{p}} \frac{\delta ( p - \Lambda ) }{ r_{\Lambda} + c_{\Lambda} \Lambda^2 }
 \frac{ \gamma_{\Lambda}}{ \gamma_{\Lambda}^2 + \xi_{ \bd{p} - \bd{k}_F }^2 }.
 \end{equation}
Assuming $\Lambda \ll k_F$ we may linearize the energy dispersion around the
Fermi surface, $\xi_{\bd{p} - \bd{k}_F } \approx - \bd{v}_F \cdot \bd{p}$.
In three dimensions, the angular integration is then elementary and we obtain
for the flow of the quasiparticle damping
on the Fermi surface,
 \begin{equation}
 \partial_\Lambda \gamma_{\Lambda} = - K_3 \frac{T \Lambda }{v_F}
 \frac{ \arctan ( v_F \Lambda / \gamma_{\Lambda} ) }{
  r_{\Lambda}  + c_{\Lambda} \Lambda^2 },
 \label{eq:flowdamp}
 \end{equation} 
where 
 \begin{equation} 
 K_3 = 1 /(2 \pi^2 ).
 \end{equation}

To obtain the self-consistent
quasiparticle damping from Eq.~(\ref{eq:flowdamp}), we need 
additional  RG flow equations for the two parameters $r_{\Lambda}$ and $c_{\Lambda}$.
Within our classical approximation this flow is  determined by
Eqs.~(\ref{eq:rcl}) and (\ref{eq:ccl}) which depend on the
induced interaction
$\Gamma_{\Lambda}^{\bar{\psi} \bar{\psi} \psi \psi } ( \bd{p}_1^{\prime},
\bd{p}_2^{\prime}; \bd{p}_2 , \bd{p}_1)$  between classical
order parameter fluctuations.
Note that in our interaction-momentum cutoff scheme the FRG flow of all vertices 
without fermionic external legs 
is completely decoupled from the
FRG flow of the other vertices with fermionic legs so that we may 
use the strategy developed
in Refs.~[\onlinecite{Ledowski04,Hasselmann04}] 
to obtain a closed systems of RG flow equations for $r_{\Lambda}$ and $c_{\Lambda}$.
In a first step, we define 
 \begin{equation}
 u_{\Lambda} = \Gamma_{\Lambda}^{\bar{\psi} \bar{\psi} \psi \psi } ( 0, 0; 0 , 0),
 \end{equation}
and neglect the momentum-dependence
of $\Gamma_{\Lambda}^{\bar{\psi} \bar{\psi} \psi \psi } ( \bd{q},
\bd{p} ; \bd{p} , \bd{q})$ on the right-hand sides of the flow equations
(\ref{eq:rcl}) and (\ref{eq:ccl}). In this approximation $c_{\Lambda}$ does not flow and
the RG flow of $r_{\Lambda}$ is
\begin{equation}
 \partial_{\Lambda} r_{\Lambda} = - K_3 T \frac{ u_{\Lambda} \Lambda^2}{ 
r_{\Lambda} 
 + c_{\Lambda} \Lambda^2 }.
 \end{equation}
To obtain the RG flow of the interaction $u_{\Lambda}$,
we neglect again the momentum-dependence of the
four-point vertices on the right-hand side of the
flow equation (\ref{eq:flowgamma4}) for the induced interaction
between order parameter fluctuations and obtain
 \begin{equation}
 \partial_{\Lambda} u_{\Lambda} = \frac{5}{2} K_3 T \frac{u_{\Lambda}^2 \Lambda^2}{ 
  [ r_{\Lambda}  + c_{\Lambda} \Lambda^2]^2},
 \end{equation} 
where we have also neglected the flow of the six-point vertex. Actually, 
as discussed in Sec.~\ref{sec:FRG}, within our cutoff scheme the 
GM correction to the critical temperature can be obtained by calculating the
effect of the initial value of the six-point vertex on the bosonic self-energy
to second order in the Gaussian propagator of the order parameter field, see
Fig.~\ref{fig:loops} (c).  In our FRG approach this contribution can be simply taken into
account via the initial condition $r_0 \propto T - T_c$, where
the value of $T_c$ includes the GM correction.
Finally, to obtain the RG of $c_{\Lambda}$
and the associated  flowing anomalous dimension $\eta_\Lambda$ from 
Eq.~(\ref{eq:etadef}), 
we need the momentum-dependence of the
induced interaction  $\Gamma_{\Lambda}^{\bar{\psi} \bar{\psi} \psi \psi } ( \bd{p}_1^{\prime},
\bd{p}_2^{\prime}; \bd{p}_2 , \bd{p}_1)$, which is determined by the exact FRG
flow equation (\ref{eq:flowgamma4}) shown graphically in Fig.~\ref{fig:flowgamma}.
 Following
Refs.~\onlinecite{Ledowski04,Hasselmann04}, we obtain an approximate solution
of this flow equation by neglecting the flowing six-point vertex
as well as the momentum-dependence of the
four-point vertices on the right-hand side.
Moreover, since we are  interested in classical order parameter fluctuations,
we only need the classical component of the interaction
which can be obtained by setting all external Matsubara frequencies 
in our exact flow equation~(\ref{eq:flowgamma4}) equal to zero.
With these approximations we obtain for the momentum-dependent
induced interaction between order parameter fluctuations,
 \begin{eqnarray}
 \partial_{\Lambda}   \Gamma^{\bar{\psi} \bar{\psi} \psi \psi }_{\Lambda} 
 ( \bd{p}_1^{\prime} , \bd{p}_2^{\prime} ; \bd{p}_2 , \bd{p}_1 )   & 
 \approx & - u_{\Lambda}^2   \Bigl[
 \frac{1}{2} {I}_{\Lambda} ( \bd{p}_1 + \bd{p}_2 ) 
 \nonumber
 \\
 & & \hspace{-17mm}
 + I_{\Lambda} ( \bd{p}_1 - \bd{p}_1^{\prime} )
 + I_{\Lambda} ( \bd{p}_2 - \bd{p}_1^{\prime} ) \Bigr],
 \hspace{7mm}
 \label{eq:intflow2}
 \end{eqnarray}
where
 \begin{eqnarray}
 I_{\Lambda} ( \bd{p} ) = 2 T \int_{\bd{q}}  \dot{F}_{\Lambda} ( \bd{q} )
 F_{\Lambda} ( \bd{q} + \bd{p} ) .
 \end{eqnarray}
Integrating Eq.~(\ref{eq:intflow2}) over the flow parameter $\Lambda$ we find for
the induced two-body interaction between classical superfluid fluctuations 
 \begin{eqnarray}
  \Gamma^{\bar{\psi} \bar{\psi} \psi \psi }_{\Lambda} 
 ( \bd{p}_1^{\prime} , \bd{p}_2^{\prime} ; \bd{p}_2 , \bd{p}_1 ) &  = & 
  \Gamma^{\bar{\psi} \bar{\psi} \psi \psi }_{\Lambda_0} 
 ( \bd{p}_1^{\prime} , \bd{p}_2^{\prime} ; \bd{p}_2 , \bd{p}_1 ) 
 \nonumber
 \\
 &   & \hspace{-20mm} +\int_{\Lambda}^{\Lambda_0} d \Lambda^{\prime}  u_{\Lambda^{\prime}}^2
 \Bigl[\frac{1}{2} {I}_{\Lambda^{\prime}} ( \bd{p}_1 + \bd{p}_2 ) 
 \nonumber
 \\
 & & \hspace{-15mm} + I_{\Lambda^{\prime}} ( \bd{p}_1 - \bd{p}_1^{\prime} )   + I_{\Lambda^{\prime}} ( \bd{p}_2 - \bd{p}_1^{\prime} ) \Bigr].
 \hspace{7mm}
 \label{eq:gamma4int}
 \end{eqnarray}
Recall that in our cutoff scheme the initial value 
  $\Gamma^{\bar{\psi} \bar{\psi} \psi \psi }_{\Lambda_0} 
 ( \bd{p}_1^{\prime} , \bd{p}_2^{\prime} ; \bd{p}_2 , \bd{p}_1 ) $ of the induced
interaction 
is given by the symmetrized closed fermion loop defined in 
Eq.~(\ref{eq:4loop}) (see also  Fig.~\ref{fig:loops}~(a)),
which is momentum-dependent.
Substituting Eq.~(\ref{eq:gamma4int})
 into our flow equation (\ref{eq:ccl}) for the coupling  $c_{\Lambda}$
we find for the flowing anomalous dimension defined in Eq.~(\ref{eq:etadef}),
 \begin{eqnarray}
 \eta_{\Lambda} & = & - \frac{\Lambda T}{c_{\Lambda}}
 \int_{\bd{p}} \dot{F}_{\Lambda} ( \bd{p} )
\lim_{ \bd{q} \rightarrow 0 }
 \frac{ \partial }{\partial q^2 }
 \Gamma_{\Lambda_0}^{\bar{\psi} \bar{\psi} \psi \psi } ( \bd{q} , \bd{p}  ;  
 \bd{p} , \bd{q}  )
 \nonumber
 \\
& -  & \frac{3}{2} \frac{\Lambda T}{c_{\Lambda}}
 \int_{\bd{p}} \dot{F}_{\Lambda} ( \bd{p} ) \int_{\Lambda}^{\Lambda_0} d \Lambda^{\prime}  u_{\Lambda^{\prime}}^2
 \lim_{\bd{q} \rightarrow 0} \frac{ \partial}{\partial q^2 }
  I_{\Lambda^{\prime}} ( \bd{p} + \bd{q} ).
 \nonumber
 \\
 & &
 \label{eq:etares1}
 \end{eqnarray}
From the explicit expression for the initial interaction
 $\Gamma^{\bar{\psi} \bar{\psi} \psi \psi }_{\Lambda_0} 
 ( \bd{p}_1^{\prime} , \bd{p}_2^{\prime} ; \bd{p}_2 , \bd{p}_1 ) $ in
Eq.~(\ref{eq:4loop}) we obtain for $ p \lesssim p_0$ the estimate
 \begin{eqnarray}
& & \lim_{ \bd{q} \rightarrow 0 }
 \frac{ \partial }{\partial q^2 }
 \Gamma_{\Lambda_0}^{\bar{\psi} \bar{\psi} \psi \psi } ( \bd{q} , \bd{p}  ;  
 \bd{p} , \bd{q}  )
 \nonumber
 \\
 & \approx & \lim_{ \bd{q} \rightarrow 0 }
 \frac{ \partial }{\partial q^2 }
 \Gamma_{\Lambda_0}^{\bar{\psi} \bar{\psi} \psi \psi } ( \bd{q} , 0  ;  
  0  , \bd{q}  ) = - A_0  u_0 / p_0^2,
 \end{eqnarray}
where $A_0$ is a numerical constant of the order of unity and
$u_0 = 
 \Gamma^{\bar{\psi} \bar{\psi} \psi \psi }_{\Lambda_0} (0,0; 0,0 )$ 
is given in
Eq.~(\ref{eq:u0def}). It is then easy to see that the first term
in Eq.~(\ref{eq:etares1}) cannot modify the  fixed point limit of
$\eta_{\Lambda}$ for $\Lambda \rightarrow 0$, so that from now on we shall omit this term.  The resulting system of coupled RG flow equations for the three
couplings $r_{\Lambda}$, $c_{\Lambda}$ and $u_{\Lambda}$ is then formally identical
to the system discussed in Refs.~\onlinecite{Ledowski04,Hasselmann04}. 
Introducing the logarithmic flow parameter
$l =  \ln ( \Lambda_0 / \Lambda )$, the RG flow of the dimensionless rescaled couplings
 \begin{eqnarray}
 \tilde{r}_l  & = &  \frac{ r_{\Lambda}}{ c_{\Lambda} \Lambda^2 },
 \label{eq:rldef}
 \\
 \tilde{u}_l & = &  \frac{ K_3 T u_{\Lambda}}{c_{\Lambda}^2 \Lambda }
 \label{eq:uldef}
 \end{eqnarray}
is given by
 \begin{eqnarray}
 \partial_l \tilde{r}_l & = & ( 2 - \eta_l ) \tilde{r}_l + \frac{ \tilde{u}_l }{ 1 + \tilde{r}_l },
 \label{eq:rlflow}
  \\
 \partial_l \tilde{u}_l & = & ( 1 - 2 \eta_l ) \tilde{u}_l - \frac{5}{2} \frac{ \tilde{u}^2_l }{ 
(1 + \tilde{r}_l )^2 }.
 \label{eq:ulflow}
 \end{eqnarray}
The scale-dependent anomalous dimension satisfies the integral equation
  \begin{equation}
 \eta_l = \int_0^l dt K ( l , t ) u^2_{ l-t} e^{ - 2 \int_{ l-t}^l d \tau \eta_{\tau}},
 \label{eq:etaflow}
 \end{equation}
where the kernel $K ( l, t )$ can be expressed in terms of the
dimensionless function
  \begin{equation}
 f_l  ( p / \Lambda ) = - \frac{ \Lambda^2 c_{\Lambda}^2}{ K_3 T } 
 I_{\Lambda} ( \bd{p} )
 \end{equation}
as follows,
 \begin{equation}
 K ( l , t ) =  \frac{1}{4(1 + \tilde{r}_l) }
 \left[ 2 f^{\prime}_{l-t} ( e^{-t} ) + e^{-t} f^{\prime \prime}_{l-1} ( e^{-t } ) \right].
 \end{equation}
Here $f_l^{\prime} (  x )$ and $f_l^{\prime \prime} ( x )$ denote the 
first and the second derivative of $f_l ( x )$.

\begin{figure}[tb]
  \centering
 \includegraphics[width=0.5\textwidth]{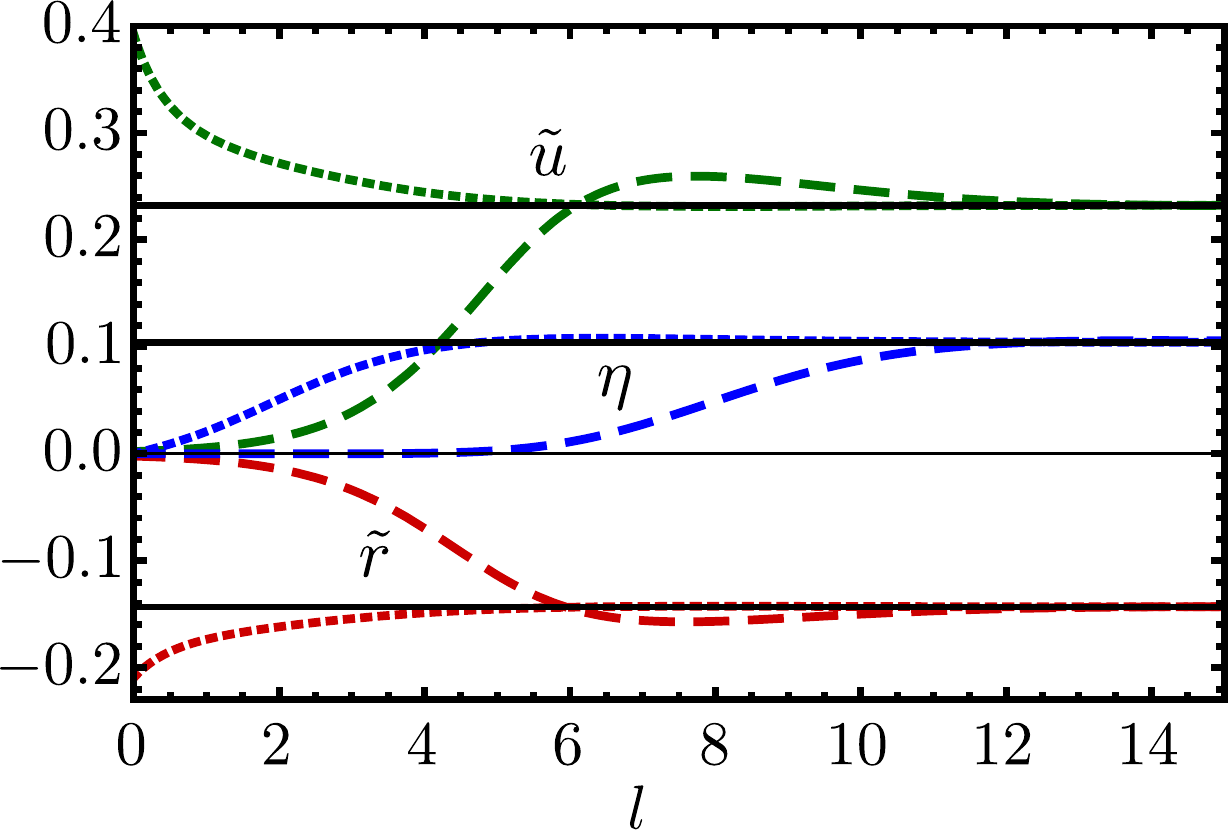}
\vspace{5mm}
  \vspace{-4mm}
  \caption{%
RG flow of the flowing anomalous dimension $\eta_l$ and dimensionless couplings 
$\tilde{r}_l$ and $\tilde{u}_l$ obtained from the 
numerical solution of the coupled integro-differential 
equations (\ref{eq:rlflow}--\ref{eq:etaflow})
for two different critical temperatures:  
$T=T_c = 0.13 E_F$ (dotted lines),
$ T= T_c=0.01E_F$ (dashed lines).
The black solid lines mark the fixed point values given in 
Eqs.~(\ref{eq:rstar}--\ref{eq:etastar}).
}
\label{fig:eta_r_u}
\end{figure}

At the critical temperature the rescaled couplings $\tilde{r}_l$, 
$\tilde{u}_l$, and $\eta_l$ approach finite  limits  for $l \rightarrow \infty$.
In Fig.~\ref{fig:eta_r_u} we plot the flow of $\tilde{r}_l$,
$\tilde{u}_l$, and $\eta_l$ for two different values
of the critical temperature as an example.
Within our simple truncation the fixed point values are \cite{Ledowski04,Hasselmann04}
 \begin{subequations}
 \begin{eqnarray}
 \tilde{r}_{\star} & = & -0.143,
 \label{eq:rstar}
 \\
 \tilde{u}_{\star} & = & 0.232,
 \label{eq:ustar}
 \\
\eta_\star  & = &  0.104.
 \label{eq:etastar}
 \end{eqnarray}
 \end{subequations}
Note that the
fixed point value of the anomalous dimension $\eta_{\star}$ is larger
than the accepted value $\eta = 0.038$ for the XY-universality class in three 
dimensions \cite{Pelissetto02}, this discrepancy can be significantly reduced  using
more sophisticated truncation strategies \cite{Hasselmann04,Berges02} of the FRG flow equations. For our purpose, the simple truncation strategy 
described above is sufficient.

Given the RG flow of the rescaled quantities $\tilde{r}_l$, $\tilde{u}_l$, and $\eta_l$,
we can reconstruct the flow of
the
dimensionful relevant coupling  $r_{\Lambda} = c_{\Lambda} \Lambda^2 \tilde{r}_l$ and
of the marginal coupling
 \begin{equation}
 c_{ \Lambda}  = c_0 \exp \left[  \int_0^{\ln ( \Lambda_0 / \Lambda)} d t \eta_t  \right], 
 \end{equation}
which we need for calculating the quasiparticle damping $\gamma_{\Lambda}$
from the flow equation (\ref{eq:flowdamp}).

\begin{figure}[tb]
  \centering
 \includegraphics[width=0.49\textwidth]{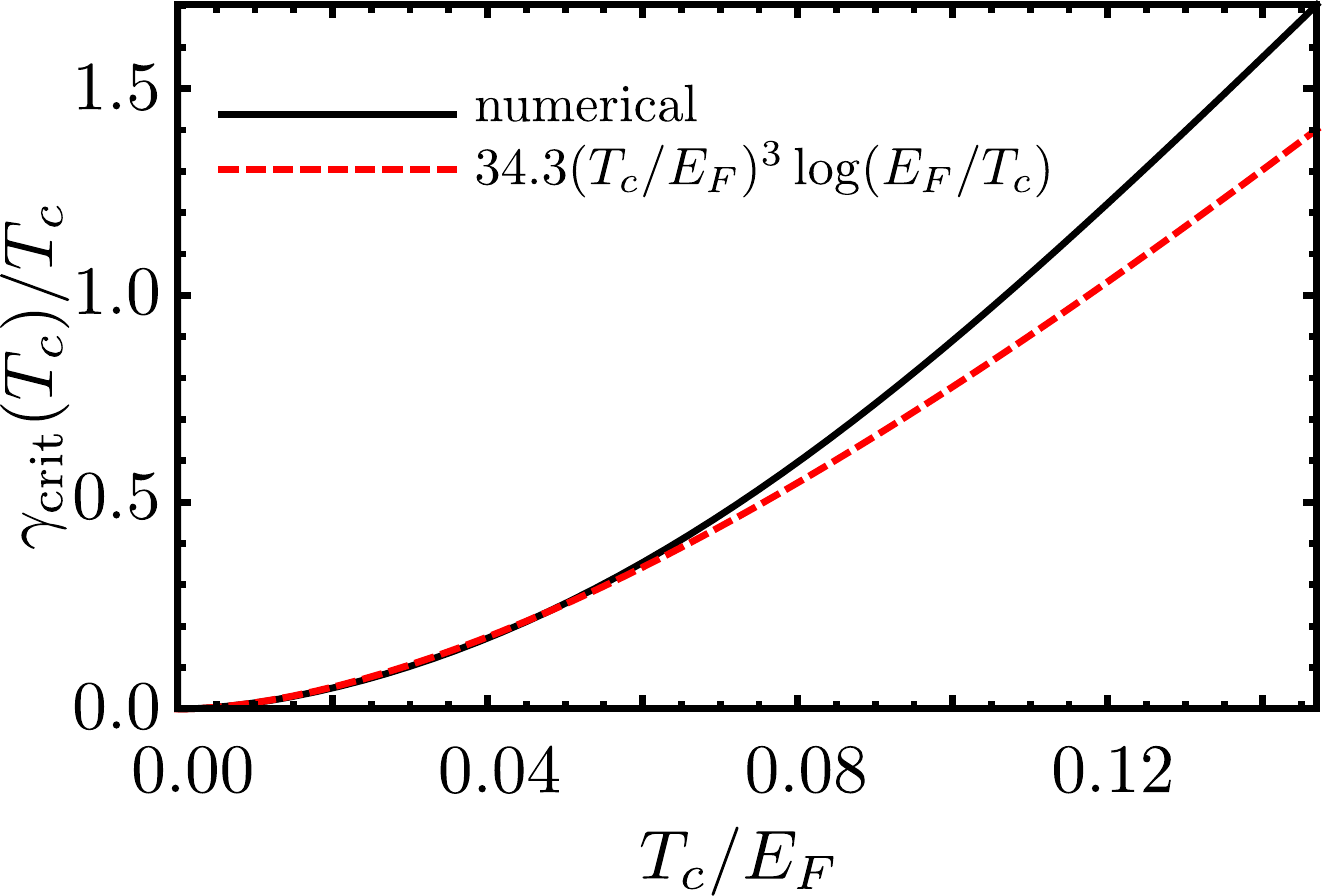}%
\vspace{5mm}
  \vspace{-4mm}
  \caption{%
Quasiparticle damping $\gamma ( T_c)$ as a function of the 
critical temperature $T_c$.
The black solid line represents the numerical solution
of the  FRG flow equations
 (\ref{eq:rlflow}), (\ref{eq:ulflow}), (\ref{eq:etaflow})  and
\eqref{eq:flowdamp}.
The red dashed line is a fit to the interpolation formula (\ref{eq:gammaint})
with $C = 34.3$.
}
\label{fig:gamma_wo_vertex}
\end{figure}

By solving the coupled flow equations \eqref{eq:rlflow}, \eqref{eq:ulflow}, and 
\eqref{eq:etaflow} 
for various temperatures and using the result as an input for the flow
equation \eqref{eq:flowdamp} for the quasiparticle damping we obtain the 
quasiparticle damping $\gamma ( T )$ as a function of the temperature.
Our numerical result for the damping $\gamma ( T_c)$ 
as a function of  the critical temperature $T_c$ 
is plotted in Fig.~\ref{fig:gamma_wo_vertex}.
In the weak coupling regime  $T_c \ll E_F$ 
the quasiparticle damping due to classical critical fluctuations 
is described by the interpolation formula
 \begin{equation}
 \gamma_{\rm crit} ( T_c ) \approx C \; \frac{T_c^3}{  E^2_F} \ln \left( \frac{E_F }{ T_c} \right) ,
 \label{eq:gammaint}
 \end{equation}
where the numerical value of the prefactor is
 \begin{equation}
 C \approx 34.3.
 \label{eq:Cres}
 \end{equation}
Using  Eq.~(\ref{eq:Gi2}) to express the the logarithm $\ln ( E_F / T_c )$
in Eq.~(\ref{eq:gammaint}) in terms of the Ginzburg-Levanyuk number $Gi$
we can express the quasiparticle damping due to critical fluctuations
in the form (\ref{eq:gammares}) given in the introduction.

The appearance of the logarithm in Eq.~(\ref{eq:gammaint} 
is related to the logarithmic divergence of the quasi-particle damping
encountered in Gaussian approximation, see Eq.~(\ref{eq:gammacrit}).
Note that the numerical value of the prefactor $C$ is rather large.
Although the precise numerical value of $C$ given above is an artifact of
our truncation scheme, we show in Appendix E that a more sophisticated truncation
including the RG flow of the three-point and mixed four-point vertices 
confirms the validity of Eq.~(\ref{eq:gammaint}) with a prefactor
$C \approx 18$ which is still large compared with unity.

\begin{figure}[tb]
  \centering
 \includegraphics[width=0.48\textwidth]{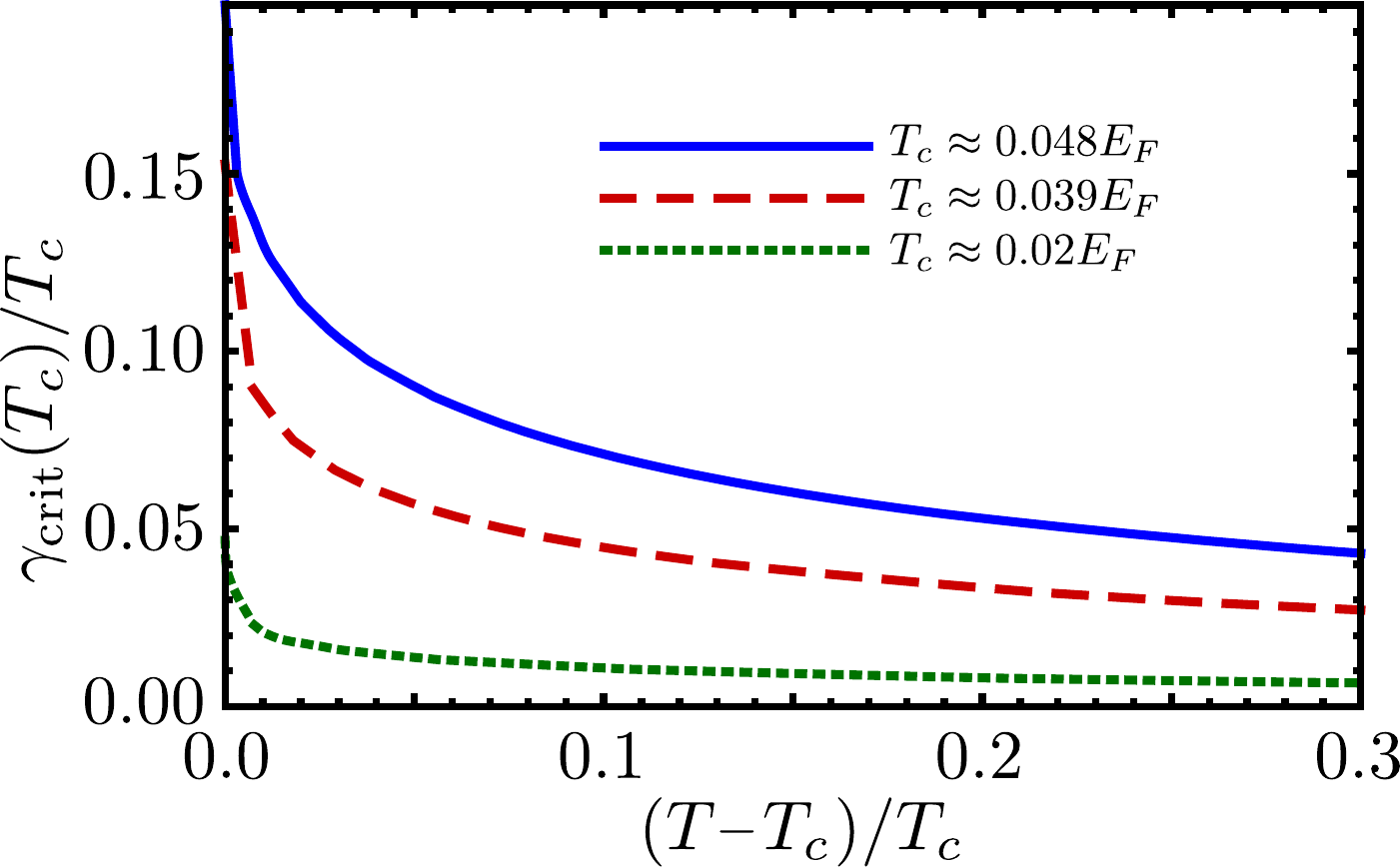}
 \vspace{5mm}
  \vspace{-4mm}
  \caption{%
FRG result for the
temperature dependence of the quasiparticle damping $\gamma ( T )$
as a function of $t$ for different values of $T_c$.  
}
\label{fig:damping_v_T}
\end{figure}

\begin{figure}[t]
\centering
\includegraphics[width=0.42\textwidth]{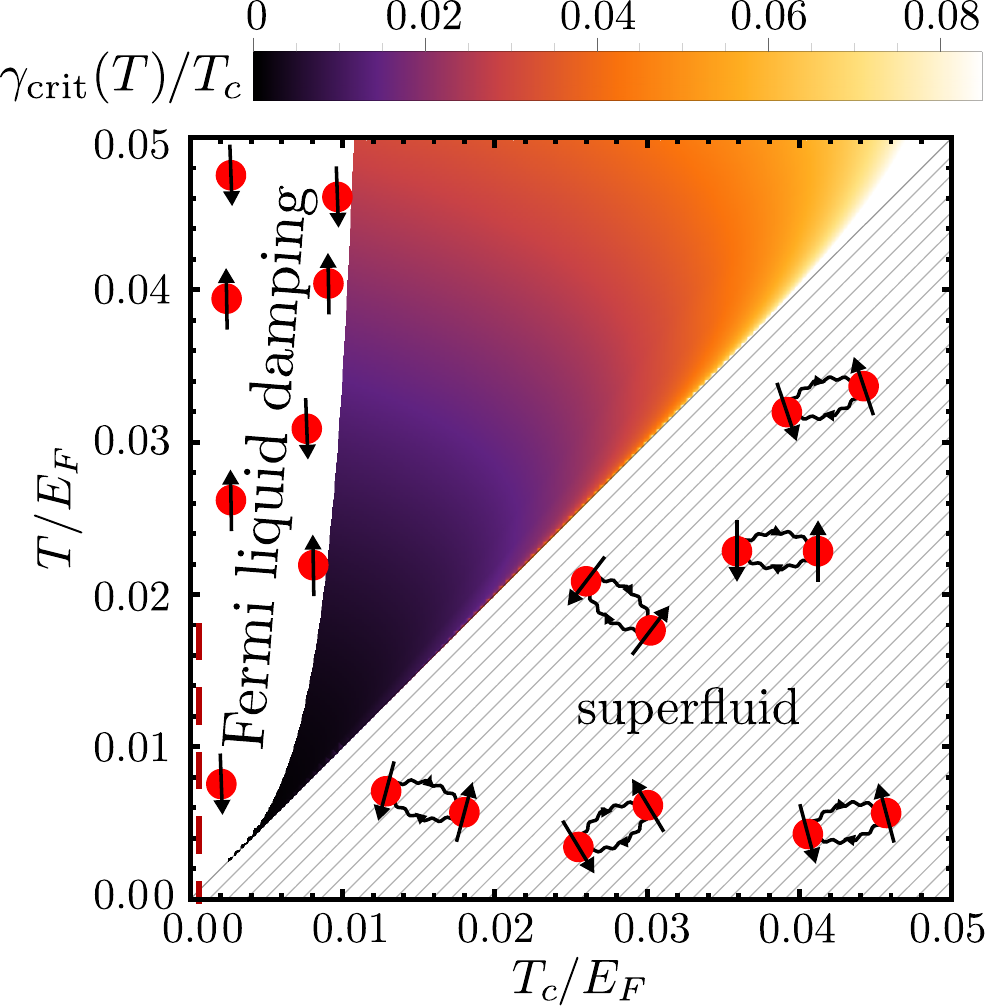}%
\caption{%
The colorful area represents the regime in the
plane spanned by the interaction length (parametrized by the dimensionless parameter $T_c / E_F $) and temperature
where the quasi-particle damping due to classical pairing fluctuations $\gamma ( T)$
obtained from our FRG approach 
is larger than  the Fermi liquid result $\gamma_{\rm FL} \approx T^2 / E_F$.
In the white region $\gamma_{\rm FL}$ is still larger than $\gamma_{\rm crit} ( T )$.
}
\label{fig:surface_damping}
\end{figure}

The above results should be compared with the
well known quadratic low-temperature behavior of the
quasi-particle damping in a three-dimensional Fermi liquid, see Eq.~(\ref{eq:dampFL}).
At  $T = T_c$ the Fermi liquid damping is
 \begin{equation}
 \gamma_{\rm FL} ( T_c ) = C_{\rm FL} T_c^2 / E_F, 
 \end{equation}
where the numerical value of $C_{\rm FL}$ depends on the strength of the screened 
interaction but is usually close to unity \cite{Kittel66}.
Although for sufficiently small $T_c$ the Fermi liquid damping is 
always larger than the damping due to critical superconducting fluctuations discussed above,
due to the large prefactor in 
Eq.~(\ref{eq:gammaint}) 
there is a substantial temperature regime where the
damping due to critical fluctuations dominates.
Note also that short-range interactions in Fermi liquids
give rise to a non-analytic correction of the form (\ref{eq:gammaint}), see
Ref.~[\onlinecite{Chubukov06}]. However, the corresponding prefactor $C_{\rm FL}$
is or order unity, so that the numerical value of the prefactor of the non-analytic
$T^3 \ln T$-contribution to the quasi-particle damping at $T_c$
is dominated by classical critical fluctuations.
Moreover, for 
for $T > T_c$ the quasi-particle damping $\gamma ( T )$ 
due to classical critical fluctuations is a 
decreasing function of temperature,  as shown in
Fig.~\ref{fig:damping_v_T}.
This is very different from any perturbative correction to the quasi-particle damping, which usually increases with temperature.
The fact that the contribution from classical critical fluctuations to $\gamma ( T )$
grows as the temperature is lowered is closely  
related to the logarithmic divergence of the damping
for $T \rightarrow T_c$
encountered within the Gaussian approximation, see Eq.~(\ref{eq:gammacrit}).
Note that, in contrast to the result for the Gaussian approximation,
our  FRG result for the damping approaches a finite limit  for $ T \rightarrow T_c$,
as given in Eqs.~(\ref{eq:gammaint}). 
The decrease of  relaxation rates with 
temperature as one moves away from the critical point 
has also been observed for disordered metals above the superconducting 
transition\cite{Larkin05}. 
In  Fig.~\ref{fig:surface_damping} we illustrate the regime in the 
plane spanned by the interaction length (which we parametrize by $T_c$)
and the temperature where the damping $\gamma ( T )$ due to classical pairing fluctuations 
obtained from our FRG approach is larger than the Fermi liquid damping
$\gamma_{\rm FL} ( T ) \approx T^2 / E_F$, see Eq.~(\ref{eq:dampFL}).
Obviously, the colorful area where this condition is fulfilled is
is sizable even for rather small values of the interaction.

Finally, let us point out that
our result (\ref{eq:gammaint}) 
for the quasiparticle damping due to classical critical fluctuations
is not qualitatively modified by vertex corrections. 
In Appendix~E we present an improved   truncation of the FRG flow equations
where we retain, in
addition to the purely bosonic vertices in Eqs. (\ref{eq:rlflow}-\ref{eq:etaflow}) and 
the fermionic self-energy in Eq. (\ref{eq:flowdamp}), 
the three-legged vertices $\Gamma_{\Lambda}^{ \bar{c}_{\uparrow} \bar{c}_{\downarrow}  \psi }$,
$\Gamma_{\Lambda}^{c_{\downarrow} c_{\uparrow} \bar{\psi} }$, as well as
the mixed four-legged vertex   
$\Gamma^{\bar{c}_{\sigma} {c}_{\sigma} \bar{\psi} \psi}_{\Lambda}$.
 From the numerical 
solution of the extended set of the flow equations we can confirm  the validity of
Eq. (\ref{eq:gammaint}) with a modified prefactor $C \approx 18$, which is still
large compared with unity.

\section{Summary and conclusions}
\label{sec:conclusions}

In this work we have used functional renormalization group
methods to study the effect of static pairing fluctuations on the 
electronic properties of metals in the critical (Ginzburg) region above the superconducting transition temperature. 
Our approach is based on partial bosonization of the
electron-electron interaction 
in the particle-particle channel and the application
of FRG methods to the resulting mixed
Bose-Fermi model within a special cutoff scheme
where a regulator is introduced only in 
the bosonic sector (interaction-momentum cutoff scheme).
To illustrate the efficiency of our approach, 
we have re-derived in a simplified way
the correction to $T_c$ obtained by Gorkov and 
Melik-Barkhudarov\cite{Gorkov61} by identifying 
$T_c$ with the temperature where the gap of the inverse bosonic propagator vanishes.
Moreover, we have shown, within our more streamlined approach, that this correction 
to $T_c$ is changed by a numerical factor when the chemical potential is fixed instead of the particle density.
Another advantage of our approach is that it allows us to understand the emergence of various types of induced interaction vertices
involving pairing fluctuations from the renormalization group point of view.

We have then used our powerful method to study the  effect of critical pairing
fluctuations on the electronic single-particle excitations  in the normal state, especially on the
electronic density of states and on the damping $\gamma$ of quasiparticles with momenta on the Fermi surface.
Within the Gaussian approximation (which corresponds to the  ladder or $T$-matrix approximation for the 
effective two-body interaction) we have found an (up to now unnoticed) logarithmic
divergence of the quasiparticle damping $\gamma \propto T_c^3 \ln [ T_c /( T- T_c )] $
for $T \rightarrow T_c$, while the density of states
exhibits a finite pseudogap.
At this level of approximation a finite value of $\gamma$
can only be introduced by invoking other interaction processes,
for instance inelastic electron-electron 
collisions within Fermi liquid theory\cite{Kittel66}  that are further 
enhanced by disorder present in real materials due to the weak localization effect\cite{Altshuler82,Altshuler85,Altshuler85_2}.
The logarithmic divergence of $\gamma$ encountered in Gaussian approximation implies that
Gaussian pairing fluctuations
completely destroy the Fermi liquid behavior of the single-particle Green function 
at and slightly above the critical temperature.
In view of the fact that in three dimensions the critical fluctuations of the
pairing field are not controlled by the Gaussian  fixed point
this is perhaps not so surprising.
Note also that  for $T < T_c$, i.e. in the superfluid phase,
order parameter fluctuations are known to have a strong effect on the
single-particle properties. For example, in Ref.~\onlinecite{Lerch08}
it has been shown  that
at $T =0$ Gaussian fluctuations of the pairing  field give rise to a logarithmic suppression
of the quasiparticle residue and the density of states.

Given the fact the Gaussian approximation is not sufficient, we
have used the FRG to take the non-Gaussian nature
of critical pairing fluctuations into account, which is the main technical
part of our work.
Let us point out that this approach should also be useful 
for a systematic evaluation  of corrections to the Gaussian approximation in other
cases where the dominant scattering channel between electrons can be uniquely identified
on physical grounds.
It is then convenient to treat the dominant  channel non-perturbatively using a suitable
Hubbard-Stratonovich field, so that the Gaussian approximation for this field 
amounts  to solving a single-channel Bethe-Salpeter equation for the effective interaction.
Other scattering channels and the corresponding vertex corrections
can then be taken into account approximately
via the induced interaction vertices which are generated as we integrate the
FRG flow equations. A similar strategy is adopted by the
dynamical re-bosonization method \cite{Gies02}, which is, however, restricted to
situations where the momentum- and frequency dependence of the induced
interactions can be neglected.

In the fluctuation regime just above $T_c$ our FRG approach  
gives a finite relaxation rate of the fermionic 
quasiparticles, which increases down to the transition temperature but remains 
finite at the transition point, $\gamma_{\rm crit} = C T_c^3/E_F^2\log(E_F/T_c)$, where the numerical constant $C$
is large compared with unity.
Physically, the
corresponding finite lifetime $\tau_{\rm crit} = 1 / \gamma_{\rm crit}$ 
of quasiparticles on the Fermi surface is due to
collisions between the unpaired fermions and virtually 
formed Cooper pairs associated with critical pairing fluctuations.
This is similar to the effect of a disorder potential on the phase coherence of quasi-particles in dirty systems\cite{Altshuler98,Aleiner99,Narozhny02}, 
where limited applicability of the ergodicity hypothesis makes direct 
observation of some of the coherence effects harder\cite{Tsyplyatyev03,Falko04}. 

It is tempting to associate $\tau_\textrm{crit}$ with the phase breaking time $\tau_\varphi$ due to the Anderson's theorem\cite{Anderson59}, 
which is applicable to the s-type superconductor studied in the present work. However, its manifestation in the particle-particle (fluctuation) 
propagator has only been studied within the ladder approximation, see Ref. [\onlinecite{Abrikosov59}] 
and a comprehensive book by Larkin and Varlamov\cite{Larkin05}. The finite quasi-particle lifetime obtained in Eq.~(\ref{eq:gammaint}) 
requires essentially a beyond ladder approach, i.e. renormalization of the bosonic line in the second term in Fig.~\ref{fig:flowself}(a) 
corresponds to a sum over the ladder diagrams in Eq. (4.1) but renormalization of the fermionic line accounts for 
more diagrams of  a different type; here we refer to the analysis in Subsec. IIIB where the first term in Fig.~\ref{fig:flowself}(a) 
is neglected. Thus, the two-particle correlation function would need to be calculated using the approach developed in this paper in 
order to put such an interpretation on a solid ground, which could be a subject of a future work.

At higher temperatures above $T_c$ the 
Fermi liquid  damping \cite{Kittel66,Thouless77} $\gamma_\textrm{FL}\simeq T^2/E_F$ becomes 
larger, but close to the transition temperature 
there is a finite region where the damping in clean systems is dominated by critical fluctuations, 
as shown in Fig.~\ref{fig:surface_damping}.
To be specific, we estimate that the effect of critical  pairing fluctuations
can be seen if the critical temperature is larger than $T^*_c\approx 2\times 10^{-4} E_F$.
With  $E_F=2$~eV this gives $T^*_c\approx5$~K, so that 
the contribution of critical pairing fluctuations 
to the quasiparticle damping 
should be observable in superconductors with $T_c  \gtrsim 5$~K.

Our approach also provides a microscopic and fully consistent theory for the pseudogap  
in a clean electronic system originating  from the superconducting fluctuations only. 
The density of states in Eq. (\ref{eq:dos}) evaluated using the result of the 
FRG in Eq. (\ref{eq:gamma_FRG}) is significantly different from the Gaussian 
approximation in Eq. (\ref{eq:sigmasing}). The former result predicts a partial suppression of the density of states at 
the Fermi energy and a finite quasi-particle relaxation rate down to the point of the superconducting 
transition. Fitting the result of our numerical integration of the FRG equations in subsection \ref{subsec:FRGdamp2},
we find for the leading order behavior for the density of states at the Fermi level in the weak coupling regime,
\begin{equation}
 \frac{\nu_0-\nu}{\nu_0} \propto  \left(\frac{T_c}{E_F} \right)^2\propto\sqrt{Gi}.
\end{equation}
This functional dependence of the pseudogap strength on $T_c$ should be observable 
in clean superconductors with higher $T_c$, which transition temperatures exceeding 
our estimate for $T^*_c$ given above.

Generally, the effect of critical pairing fluctuations on the electronic spectrum is
most pronounced in strongly coupled superconductors with small coherence length 
and broad fluctuation regimes, corresponding to Ginzburg-Levanyuk numbers $Gi$ of
the order of unity. For instance, this regime should be relevant for the normal state of the
cuprate superconductors,\cite{Dagatto94,Basov05,Devereaux07,Fischer07,Alloul09,Armitage10} which exhibits a pseudogap and a linear temperature dependence of the 
quasiparticle damping, in agreement with our prediction.
Another class of fermionic superfluids where
fluctuation effects $T_c$ can be studied experimentally are 
ultracold gases of fermionic
atoms or molecules. In these systems the effective two-body interaction can be controlled 
using the Feshbach-resonance technique \cite{Bloch08}.  
In particular, in the vicinity of the unitary point where the
scattering length diverges, fluctuation effects above $T_c$ are expected to be most 
pronounced. It should be interesting to extend the calculations for the
quasi-particle damping and the pseudogap presented in this work  
to the unitary point and discuss the
BCS-BEC crossover of these quantities.

\section*{acknowledgements}
We thank Andrey Varlamov for many useful discussions and for his 
insightful suggestions and comments on the manuscript.
We also thank Casper Drukier for his contributions at initial states of this work and
acknowledge financial support by the DFG through SFB/TRR 49 and FOR 723.

\begin{appendix}

\section*{APPENDIX A: Interaction corrections to the particle-particle 
bubble}
\setcounter{equation}{0}
\renewcommand{\theequation}{A\arabic{equation}}
 \label{sec:bubble}

The self-energy $\Phi ( P)$ of the pairing field introduced in Eq.~(\ref{eq:Dysonphi})
can be identified with the
renormalized particle-particle bubble. In this appendix we 
will explicitly evaluate the regularized non-interacting bubble $\Phi_0^{\rm reg} ( 0)$
defined in Eq.~(\ref{eq:phi0reg}) and the 
first two interaction
corrections $\Phi_1 (0)$ and $\Phi_2 ( 0 )$ 
given in Eqs.~(\ref{eq:phi1exp}) and (\ref{eq:phi2integral}) 
for vanishing total momentum and energy.

Consider first the non-interacting particle-particle bubble $\Phi_0 ( P )$ defined
in Eq.~(\ref{eq:Phi0def}).  
After performing the Matsubara sum and setting $P = ( \bd{p}, i \bar{\omega} )$ we obtain
 \begin{eqnarray}
 \Phi_0 ( \bd{p} , i \bar{\omega} ) & = & 
  \int_K G_0 ( K ) G_0 (P-K)
 \nonumber
 \\
& = &
 \int_{\bd{k}} \Theta ( \Lambda_0 - | \bd{k} | )  
 \frac{1 - f ( \xi_{\bd{k}} ) - f ( \xi_{\bd{p} - \bd{k}} ) }{
 \xi_{\bd{k}} + \xi_{\bd{p} - \bd{k} }  - i \bar{\omega} }
 \nonumber
 \\
 & = &  \int_{\bd{k}}  \Theta ( \Lambda_0 - | \bd{k} | )     \frac{\tanh ( \beta \xi_{\bd{k}} /2 )  }{
 \xi_{\bd{k}} + \xi_{\bd{p} - \bd{k} }  - i \bar{\omega} },
 \label{eq:phi0int}
 \end{eqnarray}
where $\xi_{\bd{k}} = \epsilon_{\bd{k}} - \mu$,  $f ( \xi_{\bd{k} } ) = 1 / (e^{\beta \xi_{\bd{k}} } +1 )$ is the Fermi function, and we have defined 
the integration symbol  $\int_{\bd{k}} =
 \int \frac{ d^3 k}{ (2 \pi )^3}$.
The ultraviolet cutoff $\Lambda_0 \gg k_F$ restricts
the momentum integration to the regime $ | \bd{k} | \leq \Lambda_0$.
We assume that the external momentum satisfies
 $ \Lambda_0 \gg | \bd{p} |$ so that  the shift in
the integration variable in the third line of Eq.~(\ref{eq:phi0int})
does not affect the cutoff.
At $P=0$ the integral in the last line of Eq.~(\ref{eq:phi0int})
can be transformed to a dimensionless form by substituting $x = \epsilon_{\bd{k}} / E_F$,
\begin{equation}
 \Phi_0 (0) = \nu \int_0^{\lambda_0^2} dx \sqrt{x} 
 \frac{ \tanh \left( \frac{ x- {\mu} / E_F  }{2 \tau } \right)}{2 (x- {\mu}/E_F ) },
 \end{equation}
where  $\lambda_0 =   \Lambda_0 / k_F$, $\tau = T / E_F$, and
$\nu = m k_F / ( 2 \pi^2 )$ is the density of states at the Fermi energy per spin projection.
We focus on the BCS regime where  $\mu \approx E_F$. The asymptotic behavior of
this integral for $\tau \ll 1$ can then be 
extracted following the procedure outlined by GM \cite{Gorkov61} and we finally obtain
\begin{eqnarray}
 \Phi_0 (0) =  \nu \left[ \ln ( A / \tau ) + \lambda_0 + {\cal{O}}( \tau, \lambda_0^{-1} ) \right],
 \label{eq:Phi00}
 \end{eqnarray}
with the numerical constant $A = 8 /( \pi e^{ 2 - \gamma_E } )$, see Eq.~(\ref{eq:Adef}).
If we subtract from $\Phi_0 ( P )$ the vacuum bubble defined in Eq.~(\ref{eq:bandwidth})
the cutoff-dependent term $\nu \lambda_0$
on the right-hand side of (\ref{eq:Phi00}) is canceled so that we may take the limit
$\lambda_0 \rightarrow \infty$ and obtain the low-temperature
asymptotics of the regularized particle-particle bubble given in Eq.~(\ref{eq:phiregres}).

Next, let us evaluate the second order correction $\Phi_2 ( 0 )$
to the particle-particle bubble arising from the induced interaction in the particle-hole channel,
which according to Eq.~(\ref{eq:phi2integral}) can we written as
 \begin{eqnarray}
   \Phi_2 ( 0 )
 &\approx  & g^2 \int_K \int_{K^{\prime}} G_0 ( K ) G_0 (-K) 
 \nonumber
 \\
 & & \times  \Pi_0 ( K  - K^{\prime} ) G_0 ( K^{\prime} )  G_0 ( - K^{\prime} ) ,
 \label{eq:phi2null}
 \end{eqnarray} 
where the non-interacting particle-hole bubble $\Pi_0 ( Q )$
is defined in Eq.~(\ref{eq:Pi0def}). 
It turns out that this integral is still ultraviolet divergent so that
we  introduce again an ultraviolet cutoff $\Lambda_0 \gg k_F$ as in Eq.~(\ref{eq:phi0int}).
Following GM, we simplify the integrand in Eq.~(\ref{eq:phi2null}) as follows:
\begin{enumerate}
\item Neglect the frequency dependence of the particle hole bubble,
 \begin{equation}
   \Pi_0 ( K - K^{\prime} ) \approx   \Pi_0 ( \bd{k} - \bd{k}^{\prime} , 0 ).
 \end{equation}
\item Project the momentum dependence of the particle-hole bubble
onto the Fermi surface, 
 \begin{equation}
  \Pi_0 ( \bd{k} - \bd{k}^{\prime} , 0 ) \approx
   \Pi_0 ( \bd{k}_F -  \bd{k}^{\prime}_F , 0 ) ,
 \label{eq:pimom}
\end{equation}
where $\bd{k}_F$ is the point on the Fermi surface closest to $\bd{k}$.
\end{enumerate}
By numerically evaluating  Eq.~(\ref{eq:phi2null}) we have explicitly verified that
the above approximations do not modify the prefactor  of the leading 
$ \ln^2 ( 1/ \tau )$ dependence of $\Phi_2 ( 0 )$ given in  Eq.~(\ref{eq:phi2res}),
which determines the fluctuation correction to $T_c$ in the weak coupling limit.
With these approximations the second order correction (\ref{eq:phi2null})
to the particle-particle bubble reduces to
 \begin{eqnarray}
   \Phi_2 ( 0 )
 &\approx  &  g^2 \int_K \int_{K^{\prime}} G_0 ( K ) G_0 (-K)  \Pi_0 (  \bd{k}_F -  \bd{k}^{\prime}_F, 0   )
 \nonumber
 \\
 & & \hspace{13mm}  \times  G_0 ( K^{\prime} )  G_0 ( - K^{\prime} )    .
 \label{eq:phi2null2}
 \end{eqnarray} 
The particle-hole bubble is given by 
 \begin{eqnarray}
 \Pi_0 ( Q ) & = & \int_K G_0 ( K ) G_0 (K-Q)
 \nonumber
 \\
 & = &
\int \frac{ d^3 k}{(2 \pi )^3} \frac{ f ( \xi_{\bd{k}} ) - f ( \xi_{\bd{k} - \bd{q}} )}{
 \xi_{\bd{k} } - \xi_{\bd{k} - \bd{q}} - i \bar{\omega} }.
 \end{eqnarray}
At zero temperature and in the static limit $( \bar{\omega} =0)$ this reduces to
  \begin{eqnarray}
 \Pi_0 ( \bd{q} , 0 ) & = & - \nu \left[ \frac{1}{2} + \frac{ 1- \tilde{q}^2 }{4 \tilde{q} }
 \ln \left| \frac{ 1 + \tilde{q} }{1 - \tilde{q} } \right| \right],
 \label{eq:Pi0q}
 \end{eqnarray}
where $\tilde{q} = | \bd{q} |  / (2k_F)$ and $\nu$ is the density of states at the 
Fermi energy. Setting
 \begin{equation}
 | {\bd{k}}_{F} - \bd{k}_{F}^{\prime} | = k_F \sqrt{2 - 2 \cos \vartheta} ,
\end{equation}
where  $\vartheta$ is the angle between $\bd{k}_{F}$ and $\bd{k}_{F}^{\prime}$,
we may expand $\Pi_0 (  \bd{k}_F -  \bd{k}^{\prime}_F, 0   )$
 in Legendre polynomials $P_l ( \cos \vartheta )$,
 \begin{eqnarray}
& &
\Pi_0 (  \bd{k}_F -  \bd{k}^{\prime}_F, 0   ) 
=  \sum_{l =0}^{\infty}
  a_l P_l ( \cos \vartheta ),
 \end{eqnarray}
where
 \begin{equation}
  a_l =  \frac{ 2 l+1}{2} \int_{-1}^1 dx   \Pi_0 ( k_F \sqrt{2 - 2 x}      , 0 )     P_l ( x ).
 \end{equation} 
Actually,  the integration 
in Eq.~(\ref{eq:phi2null2}) projects out the $l =0$ component
so that under the integral sign we may replace
$\Pi_0 (  \bd{k}_F -  \bd{k}^{\prime}_F, 0   ) $ by its angular average
 \begin{eqnarray}
  a_0 & = &  \frac{1}{2} \int_{-1}^1 dx \Pi_0 ( k_F \sqrt{ 2 - 2 x } , 0 )
 \nonumber
 \\
 & = & - \nu \int_0^1 dy 
 \left[ \frac{1}{2} + \frac{ 1 -y}{4 \sqrt{y}} \ln \left| \frac{ 1 + \sqrt{y}}{1 - \sqrt{y}} \right|
 \right] =  \nu \alpha_2, \hspace{7mm}
 \end{eqnarray}
where the numerical constant $\alpha_2 < 0 $ is given in Eq.~(\ref{eq:alpha2}).
The second order correction to the particle-particle bubble
then reduces to
 \begin{eqnarray}
 \Phi_2 (0) & = & g^2 a_0 \left[ \int_K G_0 ( K ) G_0 ( - K ) \right]^2 =
  g^2 \nu \alpha_2 [ \Phi_0 (0 ) ]^2 
 \nonumber
 \\
 & = &       g^2 \nu^3 \alpha_2 \left[ \ln ( A / \tau ) + \lambda_0 \right]^2,
 \label{eq:phi2res2}
 \end{eqnarray}
as given in Eq.~(\ref{eq:phi2res}) of the main text.
The cutoff-dependence in Eq.~(\ref{eq:phi2res2}) is
an artifact of the approximation  (\ref{eq:pimom}); if we do not project the momenta onto the
Fermi surface, the resulting  integral in Eq.~(\ref{eq:phi2null})
depends only logarithmically on the
ultraviolet cutoff, which follows from the fact that for large $| \bd{q} | $ the static polarization
$\Pi_0 ( \bd{q} , 0)$ vanishes as $1/ \bd{q}^2$.

Finally, let us evaluate the term $\Phi_1 (P=0)$ defined in
Eq.~(\ref{eq:phi1exp}), which contributes to the shift of $T_c$ if we fix the chemical potential
instead of the density.
Therefore we manipulate the right-hand side of Eq.~(\ref{eq:phi1exp}) for $P=0$ as follows,
 \begin{eqnarray}
 \Phi_1 ( 0 ) & = & - 2 g \rho_0 \int_K G_0^2 ( K ) G_0 ( -K )  
 \nonumber
 \\
& = &   - 2 g \rho_0 \int_K
 \frac{1}{( i \omega - \xi_{\bd{k}}  )^2 ( - i \omega - \xi_{\bd{k}}  ) }
\nonumber
 \\
 & =  & - g \rho_0 \int_K \frac{\partial}{\partial \xi_{\bd{k}}}
\frac{1}{( i \omega - \xi_{\bd{k}}  ) ( - i \omega - \xi_{\bd{k}}  ) }
 \nonumber
 \\
 & = &  - g \rho_0 \int_{\bd{k}}  \frac{\partial}{\partial \xi_{\bd{k}}}
 \frac{ \tanh ( \frac{\beta}{2} \xi_{\bd{k}} ) }{ 2 \xi_{\bd{k}} }
 \nonumber
 \\
 & = &   g \rho_0 \int_0^{\infty} d \epsilon \frac{ \partial \nu ( \epsilon ) }{\partial
 \epsilon}     \frac{ \tanh ( \frac{\beta}{2} (\epsilon - \mu ) ) }{ 2 ( \epsilon - \mu ) },
 \end{eqnarray}
where we have integrated by parts to express the integral in terms of the derivative 
of the energy-dependent density of states
$\nu ( \epsilon )$.
Using the fact that in $D$ dimensions the density (per spin projection)
can be related to the density of states at the Fermi energy as 
$\rho_0 = ( 2 / D) \nu / E_F$,
we obtain in three dimensions to leading logarithmic order
 \begin{eqnarray}
 \Phi_1 ( 0 ) 
& = & \frac{ g \nu}{3} \Phi_0 ( 0)
  =  \frac{ g \nu^2}{3} \left[ \ln ( A / \tau ) + \lambda_0 \right],
 \end{eqnarray}
in agreement with Eq.~(\ref{eq:phi1res}).

\section*{APPENDIX B: FRG flow of induced interactions}
\setcounter{equation}{0}
\renewcommand{\theequation}{B\arabic{equation}}
\label{sec:momentum}
\setcounter{subsection}{0}
The  vertex expansion (\ref{eq:Gammatrunc})   of the generating functional $\Gamma_{\Lambda} [ \bar{c} , c , \bar{\psi} , \psi ]$
of the irreducible vertices contains four  different types of four-point vertices, which are defined graphically
in Fig.~\ref{fig:induced4}. In our interaction-momentum cutoff scheme, 
only the effective two-body interaction
$\Gamma_{\Lambda}^{ \bar{\psi} \bar{\psi} \psi \psi } ( P_1^{\prime} , P_2^{\prime} ; P_2 , P_1 )$
between superfluid fluctuations shown in Fig.~\ref{fig:induced4} (d) is finite at the initial scale.
The exact FRG equation for this vertex  is given in 
Eq.~(\ref{eq:flowgamma4}) and is shown graphically in
Fig.~\ref{fig:flowgamma}.
In this appendix, we give the exact FRG flow equations 
for the other three induced interaction vertices shown in Fig.~\ref{fig:induced4}.

\begin{widetext}
In Fig.~\ref{fig:flow4mixed} we show a graphical representation
of the exact FRG flow equation of the induced fermion-boson interaction
vertex 
$ \Gamma^{\bar{c}_{\sigma} {c}_{\sigma}  \bar{\psi} \psi }_{\Lambda} 
( K^{\prime} , K  ; P^{\prime} , P ) $ in our interaction-momentum cutoff scheme.
\begin{figure}[tb]
	\centering
	\vspace{7mm}
	\includegraphics[width=0.99\textwidth]{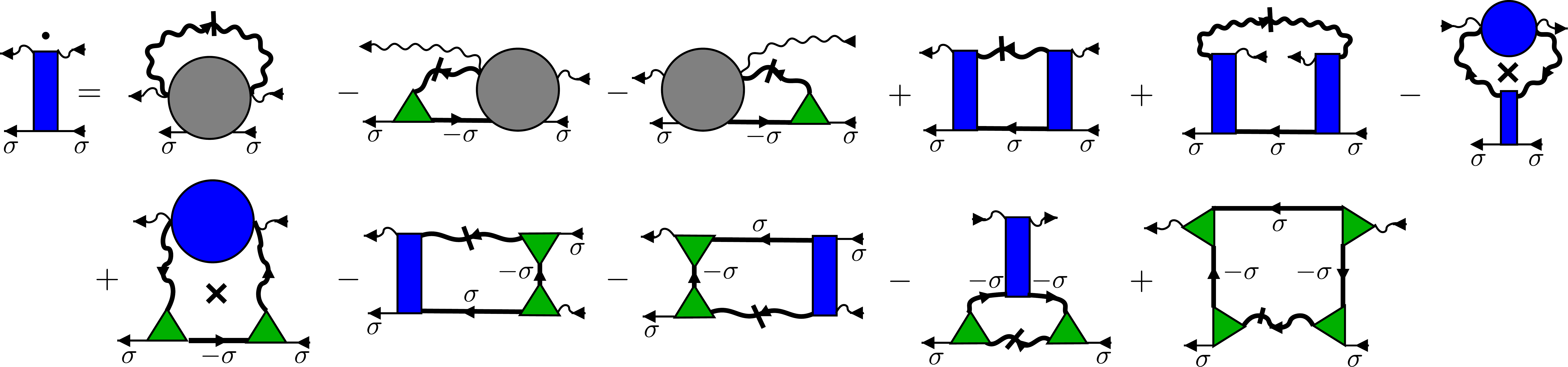}
	\caption{%
			Graphical representation of the exact FRG flow equation for the
			induced fermion-boson interaction
			vertex 
			$ \Gamma^{\bar{c}_{\sigma} {c}_{\sigma}  \bar{\psi} \psi }_{\Lambda} 
			( K^{\prime} , K  ; P^{\prime} , P ) $.
			The cross in the last diagram of the first line and the first diagram in the second line 
			corresponds to the product rule notation Eq.~\eqref{eq:product_rule}.}
	\label{fig:flow4mixed}
\end{figure}
For our purpose, we need only a truncated version of this flow equation where
all vertices which  vanish at the initial scale are neglected
on the right-hand side of the flow equations.
In this limit we obtain the FRG flow equation shown graphically in
Fig.~\ref{fig:flowinduced} (a), which is explicitly given by
	\begin{eqnarray}
		\partial_{\Lambda} \Gamma^{\bar{c}_{\sigma} {c}_{\sigma}  \bar{\psi} \psi }_{\Lambda} 
		( K^{\prime}_1 , K_1  ; P^{\prime}_1 , P_1 ) & = & \int_P  \dot{F}_{\Lambda} ( P ) 
		G_{\Lambda} ( P - K^{\prime}_1 ) G_{\Lambda} ( P - K_1 ) 
		G_{\Lambda} ( P_1 + K_1 - P ) 
		\nonumber
		\\
		& & \times  
		\Gamma_{\Lambda}^{ \bar{c}_{\sigma} \bar{c}_{- \sigma} \psi } ( K_1^{\prime} , P - K_1^{\prime} ; P )
		\Gamma_{\Lambda}^{ {c}_{- \sigma} {c}_{ \sigma} \bar{\psi} } ( P- K_1^{\prime} , P_1 + K_1 - P ; P_1^{\prime} )
		\nonumber
		\\
		& & \times  
		\Gamma_{\Lambda}^{ \bar{c}_{\sigma} \bar{c}_{- \sigma} \psi } ( P_1 + K_1 -P  , P - K_1 ; P_1 )
		\Gamma_{\Lambda}^{ {c}_{- \sigma} {c}_{ \sigma} \bar{\psi} } ( P- K_1 ,  K_1  ; P )
		\nonumber
		\\
		& + & \int_P \left[ F_{\Lambda} ( P ) F_{\Lambda} ( P + K_1 - K_1^{\prime} ) \right]^{\bullet}
		\Gamma_{\Lambda}^{\bar{\psi} \bar{\psi} \psi \psi } ( P_1^{\prime} , P ; P + K_1 - K_1^{\prime} , P_1 )
		\nonumber
		\\
		& & \times 
		\Gamma_{\Lambda}^{ \bar{c}_{\sigma} \bar{c}_{- \sigma} \psi } ( K_1^{\prime} , P - K_1^{\prime} ; P )
		\Gamma_{\Lambda}^{ {c}_{- \sigma} {c}_{ \sigma} \bar{\psi} } 
		( P- K_1^{\prime} , K_1  ; P + K_1 - K_1^{\prime} ) .
		\label{eq:flowgammamixed}
	\end{eqnarray}
Here we have used the product rule notation introduced in Eq.~(\ref{eq:product_rule}),
 \begin{equation}
 [ F_{\Lambda} (P ) F_{\Lambda} ( P^{\prime} ) ]^{\bullet} =
  \dot{F}_{\Lambda} (P ) F_{\Lambda} ( P^{\prime} ) + F_{\Lambda} (P ) \dot{F}_{\Lambda} ( P^{\prime} ).
  \label{eq:product_rule2}
 \end{equation}

Next, consider for completeness the FRG flow equations for the 
two types of purely fermionic induced interaction vertices 
defined in Fig.~\ref{fig:induced4} (a) and (b).
Since these flow equations are rather lengthy, we do not explicitly write them down here
but represent them  graphically 
in Fig.~\ref{fig:flow4fermion_a} and Fig.~\ref{fig:flow4fermion_b}.  
 \begin{figure}[tb]
 \centering
 \includegraphics[width=0.99\textwidth]{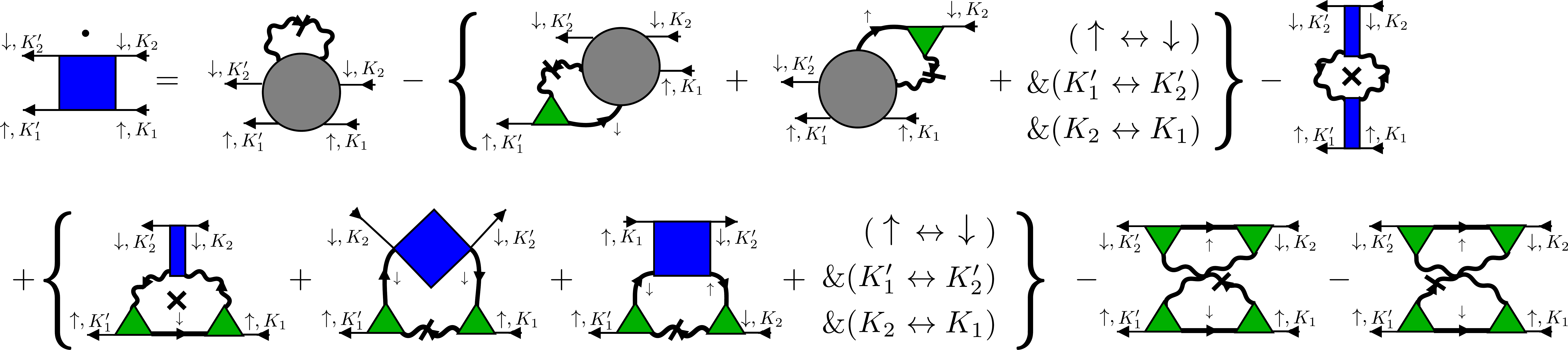}
 \vspace{5mm}
 \vspace{-4mm}
 \caption{%
 Graphical representation of the exact FRG flow equation for the
 induced fermion interaction 
 $ \Gamma^{\bar{c}_{\uparrow} \bar{c}_{\downarrow}  {c}_{\downarrow} {c}_{\uparrow} }_{\Lambda}  
 ( K_1^{\prime} , K_2^{\prime}  ; K_2, K_1 ) $ between electrons with opposite spin.
 The cross in the last diagram of the first line and the first diagram of the second line
 corresponds to our product rule notation. 
 The permutations of the external labels have to be applied on all diagrams in the curly braces.
 Vertices are antisymmetric under permutation of two external fermionic legs corresponding to fields of the same kind.
		}
 \label{fig:flow4fermion_a}
 \end{figure}
  \begin{figure}[tb]
 \centering
 \includegraphics[width=0.99\textwidth]{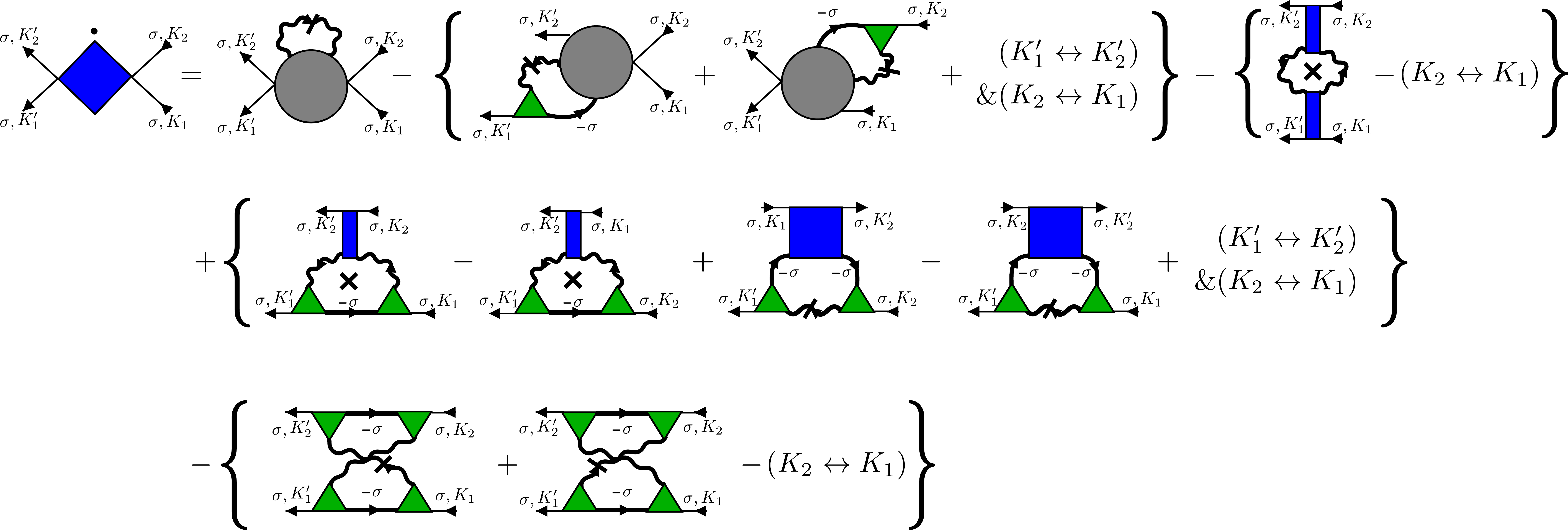}
 \vspace{5mm}
 \vspace{-4mm}
 \caption{%
The flow equation 
for the
induced fermion interaction 
$ \Gamma^{\bar{c}_{\sigma} \bar{c}_{\sigma}  {c}_{\sigma} {c}_{\sigma} }_{\Lambda}  
( K_1^{\prime} , K_2^{\prime}  ; K_2, K_1 ) $ between electrons with parallel spin.
The permutations of the external labels have to be applied on all diagrams in the curly braces.
The cross in the last two diagrams of the first line and the second two diagrams of the second line correspond to our product
rule notation.
Vertices are antisymmetric under permutation of two external fermionic legs corresponding to fields of the same kind
}
 \label{fig:flow4fermion_b}
 \end{figure}
We can close the infinite hierarchy of FRG flow equations by neglecting
all vertices with more than
four external legs on right-hand side.
Then  the FRG flow equation for the induced interaction between two
electrons with opposite spin reduces to
 \begin{eqnarray} 
 & &  \partial_{\Lambda} 
 \Gamma^{ \bar{c}_{\uparrow} \bar{c}_{\downarrow} c_{\downarrow} c_{\uparrow} }_{\Lambda}
 ( K_1^{\prime} , K_2^{\prime} ; K_2 , K_1 ) 
 \nonumber
 \\
& \approx & -
 \int_{P} \left[F_{\Lambda}(P)F_{\Lambda}(P-K_2+K_2^{\prime})\right]^{\bullet}
 \Gamma_{\Lambda}^{\bar{c}_{\downarrow}c_{\downarrow}\bar{\psi}\psi}
 (K_2^{\prime};K_2;P;P-K_2+K_2^{\prime})
 \Gamma_{\Lambda}^{\bar{c}_{\uparrow}c_{\uparrow}\bar{\psi}\psi} (K_1^{\prime};K_1;P-K_2+K_2^{\prime};P)
 \nonumber
 \\
 & + & \bigg\{ \int_{P} \left[ F_{\Lambda}(P) F_{\Lambda}(P-K_1^{\prime}+K_1)  \right]^{\bullet} 
 G_{\Lambda}(P-K_1^{\prime})
 \Gamma_{\Lambda}^{\bar{c}_{\downarrow}c_{\downarrow}\bar{\psi}\psi}(K_2^{\prime};K_2;P;P-K_1^{\prime}+K_1)
 \nonumber
 \\
 &  &\hspace{7mm} \times
 \Gamma_{\Lambda}^{c_{\downarrow}c_{\uparrow}\bar{\psi}}(P-K_1^{\prime},K_1;P-K_1^{\prime}+K_1)
 \Gamma_{\Lambda}^{\bar{c}_{\uparrow}\bar{c}_{\downarrow}\psi}(K_1^{\prime},P-K_1^{\prime};P)
 \nonumber
 \\
 & & + \int_{P} \dot{F}_{\Lambda}(P)  G_{\Lambda}(P-K_1) G_{\Lambda}(P-K_1^{\prime})
 \Gamma_{\Lambda}^{\bar{c}_{\downarrow}\bar{c}_{\downarrow}c_{\downarrow}c_{\downarrow}}
 (P-K_1,K_2^{\prime};P-K_1^{\prime},K_2)
 \nonumber
 \\
 & & \hspace{7mm} \times
 \Gamma_{\Lambda}^{c_{\downarrow}c_{\uparrow}\bar{\psi}}(P-K_1,K_1;P)
 \Gamma_{\Lambda}^{\bar{c}_{\uparrow}\bar{c}_{\downarrow}\psi}(K_1^{\prime},P-K_1^{\prime};P)
 \nonumber
 \\
 & & + \int_{P} \dot{F}_{\Lambda}(P)G_{\Lambda}(P-K_1^{\prime}) G_{\Lambda}(P-K_2)
 \Gamma_{\Lambda}^{\bar{c}_{\uparrow}\bar{c}_{\downarrow}c_{\downarrow}c_{\uparrow}}
 (P-K_2,K_2^{\prime};P-K_1^{\prime},K_1)
 \nonumber
 \\
 & & \hspace{7mm}  \times
 \Gamma_{\Lambda}^{c_{\downarrow}c_{\uparrow}\bar{\psi}}(K_2,P-K_2;P)
 \Gamma_{\Lambda}^{\bar{c}_{\uparrow}\bar{c}_{\downarrow}\psi}(K_1^{\prime},P-K_1^{\prime};P)
 \nonumber
 \\
 & &   
 +[(\uparrow \leftrightarrow \downarrow)\& (K_1^{\prime} \leftrightarrow K_2^{\prime})\& (K_2 \leftrightarrow K_1)]
 \bigg\}
 \nonumber
 \\
 & - & \int_{P} \left[ F_{\Lambda}(P) F_{\Lambda}(P-K_1^{\prime}+K_1) \right] ^{\bullet}
 G_{\Lambda}(P-K_1^{\prime}) G_{\Lambda}(P-K_2)
	\Gamma_{\Lambda}^{\bar{c}_{\uparrow}\bar{c}_{\downarrow} {\psi}}(K_1^{\prime},P-K_1^{\prime};P)
 \nonumber
 \\
 & &   \hspace{4mm} \times
 \Gamma_{\Lambda}^{c_{\downarrow}c_{\uparrow} \bar{\psi}}(P-K_1^{\prime},K_1;P-K_1^{\prime}+K_1)
 \Gamma_{\Lambda}^{\bar{c}_{\uparrow}\bar{c}_{\downarrow} {\psi}}(P-K_2,K_2^{\prime};P+K_1-K_1^{\prime})
 \Gamma_{\Lambda}^{c_{\downarrow}c_{\uparrow} \bar{\psi}}(K_2,P-K_2;P),
 \label{eq:Uindflow_1}
 \end{eqnarray}
while the flow of effective interaction between electrons with parallel spin is given by
 \begin{eqnarray} 
 & & 
 \partial_{\Lambda} 
 \Gamma^{ \bar{c}_{\sigma} \bar{c}_{\sigma} c_{\sigma} c_{\sigma} }_{\Lambda}
 ( K_1^{\prime} , K_2^{\prime} ; K_2 , K_1 ) 
 \nonumber
 \\
 & \approx & 
 \int_{P} \left[F_{\Lambda}(P)F_{\Lambda}(P-K_2^{\prime}+K_1)\right]^{\bullet}
 \Gamma_{\Lambda}^{\bar{c}^{\sigma}c^{\sigma}\bar{\psi}\psi}(K_2^{\prime};K_1;P-K_2^{\prime}+K_1;P)
 \Gamma_{\Lambda}^{\bar{c}^{\sigma}c^{\sigma}\bar{\psi}\psi} (K_1^{\prime};K_2;P;P-K_2^{\prime}+K_1 )
 \nonumber
 \\
 &   -&
 \int_{P} \left[F_{\Lambda}(P)F_{\Lambda}(P-K_2+K_2^{\prime})\right]^{\bullet} 
 \Gamma_{\Lambda}^{\bar{c}_{\sigma}c_{\sigma}\bar{\psi}\psi}(K_2^{\prime};K_2;P;P-K_2+K_2^{\prime})
 \Gamma_{\Lambda}^{\bar{c}_{\sigma}c_{\sigma}\bar{\psi}\psi} (K_1^{\prime};K_1;P-K_2+K_2^{\prime};P)
 \nonumber
 \\
 & +& \bigg\{ \int_{P} \left[ F_{\Lambda}(P) F_{\Lambda}(P-K_1^{\prime}+K_1)  \right]^{\bullet} 
  G_{\Lambda}(P-K_1^{\prime})
 \Gamma_{\Lambda}^{\bar{c}_{\sigma}c_{\sigma}\bar{\psi}\psi}(K_2^{\prime};K_2;P;P-K_1^{\prime}+K_1)
 \nonumber
 \\
 & & \hspace{7mm} \times
 \Gamma_{\Lambda}^{c_{-\sigma}c_{\sigma}\bar{\psi}}(P-K_1^{\prime},K_1;P-K_1^{\prime}+K_1)
 \Gamma_{\Lambda}^{\bar{c}_{\sigma}\bar{c}_{-\sigma}\psi}(K_1^{\prime},P-K_1^{\prime};P)
 \nonumber
 \\
 & &
 -\int_{P} \left[ F_{\Lambda}(P) F_{\Lambda}(P-K_1^{\prime}+K_2) \right]^{\bullet}
 G_{\Lambda}(P-K_1^{\prime})
 \Gamma_{\Lambda}^{\bar{c}^{\sigma}c^{\sigma}\bar{\psi}\psi}(K_2^{\prime},K_1;P,P-K_1^{\prime}+K_2)
 \nonumber
 \\
 & & \hspace{7mm}  \times
 \Gamma_{\Lambda}^{c^{-\sigma}c^{\sigma}\bar{\psi}}(P-K_1^{\prime},K_2;P-K_1^{\prime}+K_2)
 \Gamma_{\Lambda}^{\bar{c}^{\sigma}\bar{c}^{-\sigma}\psi}(K_1^{\prime},P-K_1^{\prime};P)
 \nonumber
 \\
 & & - \int_{P} \dot{F}_{\Lambda}(P)  G_{\Lambda}(P-K_1) G_{\Lambda}(P-K_1^{\prime})
 \Gamma_{\Lambda}^{\bar{c}_{-\sigma}\bar{c}_{\sigma}c_{\sigma}c_{-\sigma}}
  (P-K_1,K_2^{\prime};K_2,P-K_1^{\prime})
 \nonumber
 \\
 & & \hspace{7mm} \times
 \Gamma_{\Lambda}^{c_{-\sigma}c_{\sigma}\bar{\psi}}(P-K_1,K_1;P)
 \Gamma_{\Lambda}^{\bar{c}_{\sigma}\bar{c}_{-\sigma}\psi}(K_1^{\prime},P-K_1^{\prime};P)
 \nonumber
 \\
 & & + \int_{P} \dot{F}_{\Lambda}(P)G_{\Lambda}(P-K_1^{\prime}) G_{\Lambda}(P-K_2)
 \Gamma_{\Lambda}^{\bar{c}_{-\sigma}\bar{c}_{\sigma}c_{\sigma}c_{-\sigma}}
 (P-K_2,K_2^{\prime};K_1,P-K_1^{\prime})
 \nonumber
 \\
 & & \hspace{7mm}  \times
 \Gamma_{\Lambda}^{c_{-\sigma}c_{\sigma}\bar{\psi}}(P-K_2,K_2;P)
 \Gamma_{\Lambda}^{\bar{c}_{\sigma}\bar{c}_{-\sigma}\psi}(K_1^{\prime},P-K_1^{\prime};P)
 \nonumber
 \\
 & &   +[(\uparrow \leftrightarrow \downarrow)\& (K_1^{\prime} \leftrightarrow K_2^{\prime})
 \& (K_2 \leftrightarrow K_1)]
 \bigg\}
 \nonumber
 \\
 & -&  \int_{P} \left[ F_{\Lambda}(P) F_{\Lambda}(P-K_1^{\prime}+K_1) \right] ^{\bullet}
 G_{\Lambda}(P-K_1^{\prime}) G_{\Lambda}(P-K_2)
 \Gamma_{\Lambda}^{\bar{c}_{\sigma}\bar{c}_{-\sigma} {\psi}}(K_1^{\prime},P-K_1^{\prime};P)
 \nonumber
 \\
 & &  \hspace{4mm} \times  \Gamma_{\Lambda}^{c_{-\sigma}c_{\sigma} \bar{\psi}}
 (P-K_1^{\prime},K_1;P-K_1^{\prime}+K_1)
 \Gamma_{\Lambda}^{\bar{c}_{-\sigma}\bar{c}_{\sigma} {\psi}}(P-K_2,K_2^{\prime};P+K_1-K_1^{\prime})
 \Gamma_{\Lambda}^{c_{\sigma}c_{-\sigma} \bar{\psi}}(K_2,P-K_2;P)
 \nonumber
 \\
 &  + & \int_{P}
 \left[ F_{\Lambda}(P) F_{\Lambda}(P-K_1^{\prime}+K_2) \right]^{\bullet} G_{\Lambda}(P-K_1^{\prime}) G_{\Lambda}(P-K_1)  \Gamma_{\Lambda}^{\bar{c}^{\sigma}\bar{c}^{-\sigma} {\psi}}(K_1^{\prime},P-K_1^{\prime};P)
 \nonumber
 \\
 & &  \hspace{4mm} \times 
 \Gamma_{\Lambda}^{c^{-\sigma}c^{\sigma} \bar{\psi}}(P-K_1,K_1;P)	
  \Gamma_{\Lambda}^{\bar{c}^{\sigma}\bar{c}^{-\sigma} {\psi}}(K_2^{\prime},P-K_1;P-K_1^{\prime}+K_2)
 \Gamma_{\Lambda}^{c^{-\sigma}c^{\sigma} \bar{\psi}}
 (P-K_1^{\prime},K_2;P-K_1^{\prime}+K_2).
 \label{eq:Uindflow_2}
 \end{eqnarray}
As a first step in an iterative solution of these flow equations,
we may set all vertices which vanish at the initial scale equal to zero.
Then the FRG flow equation (\ref{eq:Uindflow_1})
for the effective interaction between electrons with opposite
spin reduces to
	\begin{eqnarray} 
		\partial_{\Lambda} 
		\Gamma^{ \bar{c}_{\uparrow} \bar{c}_{\downarrow} c_{\downarrow} c_{\uparrow} }_{\Lambda}
		( K_1^{\prime} , K_2^{\prime} ; K_2 , K_1 ) & \approx & 
		- \int_K 
		\bigr[ 
		{F}_{\Lambda} ( K + K_2 ) F_{\Lambda} ( K + K_2^{\prime} )  \bigl]^{\bullet}
		{G}_{\Lambda} ( K ) G_{\Lambda} ( K + K_2 - K_1^{\prime} )  
		\nonumber
		\\
		& & \hspace{7mm} \times \Gamma_{\Lambda}^{\bar{c}_{\uparrow} \bar{c}_{\downarrow} \psi } ( K_1^{\prime} , K + K_2 - K_1^{\prime} ; K + K_2 )
		\Gamma_{\Lambda}^{{c}_{\downarrow} {c}_{\uparrow} \bar{\psi} } (  K + K_2 - K_1^{\prime} ,  K_1; K + K_2^{\prime} )
		\nonumber
		\\
		& & \hspace{7mm} \times 
		\Gamma_{\Lambda}^{{c}_{\downarrow} {c}_{\uparrow} \bar{\psi} } (  K_2,  K; K+K_2  )
		\Gamma_{\Lambda}^{\bar{c}_{\uparrow} \bar{c}_{\downarrow} \psi } ( K , K_2^{\prime} ; K + K_2^{\prime} ),
		\label{eq:Uindflow}
	\end{eqnarray}
\end{widetext}
which is shown graphically in Fig.~\ref{fig:flowinduced} (b).
Another approximation strategy is
to replace  the
three-point and bosonic four-point vertices on the right-hand sides of the flow equations
in Fig.~\ref{fig:flowinduced} by their initial values 
$\Gamma_{\Lambda}^{ \bar{c}_{\uparrow} \bar{c}_{\downarrow} \psi } = 
\Gamma_{ \Lambda}^{ c_{\downarrow} c_{\uparrow} \bar{\psi}}
\approx 1 $ and
$\Gamma_{\Lambda}^{\bar{\psi} \bar{\psi} \psi \psi } \approx 
\Gamma_{\Lambda_0}^{\bar{\psi} \bar{\psi} \psi \psi }$.
Then the FRG flow of the mixed boson-fermion interaction reduces to
\begin{eqnarray}
	&&\partial_{\Lambda} \Gamma^{\bar{c}_{\sigma} {c}_{\sigma}  \bar{\psi} \psi }_{\Lambda} 
	( K^{\prime}_1 ; K_1  ; P^{\prime}_1 ; P_1 ) 
	\nonumber
	\\
	\nonumber
	& = & \int_P  \dot{F}_{\Lambda} ( P ) 
	G_{\Lambda} ( P - K^{\prime}_1 ) G_{\Lambda} ( P - K_1 ) 
	G_{\Lambda} ( P_1 + K_1 - P ) 
	\\
	& + & \int_P \left[ F_{\Lambda} ( P ) F_{\Lambda} ( P + K_1 - K_1^{\prime} ) \right]^{\bullet}
	\nonumber
	\\
	&  & \hspace{4mm} \times \Gamma_{\Lambda_0}^{\bar{\psi} \bar{\psi} \psi \psi } ( P_1^{\prime} , P ; P + K_1 - K_1^{\prime} , P_1 ),
	\label{eq:flowgammamix}
\end{eqnarray}
while  the induced interaction between fermions with opposite spin determined by the
truncated flow equation
\begin{eqnarray} 
 & & \partial_{\Lambda}  \Gamma^{ \bar{c}_{\uparrow} \bar{c}_{\downarrow} 
 c_{\downarrow} c_{\uparrow} }_{\Lambda}( K_1^{\prime} , K_2^{\prime} ; K_2 , K_1 )  
 \nonumber
 \\
 & = & - \int_P \bigr[  {F}_{\Lambda} ( P ) F_{\Lambda} ( P + K_1 - K_1^{\prime} )
 \bigl]^{\bullet}
 \nonumber
 \\
 & & \hspace{4mm} \times 
 {G}_{\Lambda} ( P - K_1^{\prime} ) G_{\Lambda} ( P + K_1 - K_1^{\prime} - K_2^{\prime}) .
 \hspace{7mm}
 \label{eq:Uindflow_red}
\end{eqnarray}

\section*{APPENDIX C: 
Particle-particle bubble at finite momentum and frequency}
\setcounter{equation}{0}
\renewcommand{\theequation}{C\arabic{equation}}
 \label{sec:ppbubble}

To calculate the expansion of the
particle-particle bubble for small momenta and frequencies, it is
convenient to expand in powers of external momenta and frequencies before
carrying out the Matsubara sums.
Therefore we write Eq.~(\ref{eq:phi0int}) as
 \begin{eqnarray}
 \Phi_0 ( \bd{p} , i \bar{\omega} ) & = &     \int_K G_0 ( K ) G_0 ( P-K)
 \nonumber
 \\
& = &     T \sum_{\omega} \int_{\bd{k}}
 \frac{1}{ i {\omega} - \xi_{\bd{k}} }
 \frac{1}{ i \bar{\omega} - i {\omega}  - \xi_{\bd{p} - \bd{k} }  }.
 \label{eq:phimatsubara}
 \end{eqnarray}
In the book by
Larkin and Varlamov \cite{Larkin05} one can find
an approximate evaluation of Eq.~(\ref{eq:phimatsubara}) 
in the regime $v_F p \ll T \ll E_F$ where the momentum integral is dominated by
 states with energies close to the Fermi energy. In this regime the energy dependence
of the density of states $\nu ( \epsilon )$ can be neglected so that we may approximate
$\nu ( \epsilon ) \approx \nu ( E_F ) \equiv \nu $  under the integral. 
Using the $T$-matrix regularization defined via Eq.~(\ref{eq:phi0reg})
we then obtain for the regularized particle-particle bubble
 \begin{eqnarray}
 \Phi^{\rm reg}_0 ( \bd{p} , i \bar{\omega} ) & \approx &  \nu 
 \Biggl[ \ln \left( \frac{A E_F}{ T } \right) 
 + \psi \left( \frac{1}{2} \right)  -
\psi \left( \frac{1}{2} + \frac{ | \bar{\omega} |}{ 4 \pi T } \right) 
 \nonumber
 \\
 & & \hspace{7mm} + \frac{ \langle ( \bd{v}_{F} \cdot {\bd{p}} )^2 \rangle }{ 2 ( 4 \pi T )^2 }
  \psi^{\prime \prime} \left( \frac{1}{2} + \frac{ | \bar{\omega} |}{ 4 \pi T } \right) 
 \Biggr],
 \label{eq:PhiLV}
 \end{eqnarray}
where for a spherical
Fermi surface in $D$  dimensions the Fermi surface average in Eq.~(\ref{eq:PhiLV}) is
 \begin{equation}
\langle ( \bd{v}_{F} \cdot {\bd{p}} )^2 \rangle = \frac{ v_F^2 p^2}{D}.
 \end{equation}
The Digamma function $\psi ( z )$ has the representation
 \begin{equation}
 \psi ( z ) = \frac{ d \ln \Gamma ( z ) }{dz} = - \gamma_E + \sum_{n=0}^{\infty}
 \left[ \frac{1}{n+1} - \frac{1}{n+z } \right],
 \end{equation}
where $\gamma_E \approx 0.577$ is the Euler-Mascheroni constant.
The numerical constant
 $A = 8 / ( \pi e^{ 2 - \gamma_E } )$ in the argument of the logarithm in Eq.~(\ref{eq:PhiLV})
has already been introduced in Eq.~(\ref{eq:Adef}).
Note that
 \begin{subequations}
 \begin{eqnarray}
 \psi ( 1/2 ) & = & - \gamma_E - 2 \ln 2,
 \\
  \psi^{\prime} ( 1/2) & = & 3\zeta ( 2 ) = \frac{ \pi^2}{2},
 \\
 \psi^{\prime \prime} ( 1/2) & = & - 14 \zeta ( 3 ) \approx - 16.8,
  \end{eqnarray}
 \end{subequations}
and that for large $| z |$ the Digamma function has the asymptotic expansion
 \begin{equation}
 \psi ( z+1 ) \sim \ln z + \frac{1}{2 z} + {\cal{O}} ( z^{-2} ).
 \end{equation}
The mean-field critical temperature $T_{c0}$ is determined by
 \begin{equation}
 g^{-1} - \left. \Phi^{\rm reg}_0 ( 0 , 0 ) \right|_{ T_{c0}} =0,
 \label{eq:tcmf}
 \end{equation}
which yields the well-known weak-coupling result quoted in Eq.~(\ref{eq:tcweak}).
Setting
  \begin{eqnarray}
 r_0 & \equiv &  g^{-1} - \Phi^{\rm reg}_0 ( 0 , 0 )  = \left. \Phi_0^{\rm reg} ( 0 , 0 ) \right|_{ T_{c0}}
 - \Phi_0^{\rm reg} ( 0 , 0 ) 
 \nonumber
 \\
 & \approx & \nu \ln \left( \frac{T}{  T_{c0}} \right) \approx \nu \frac{ T - T_{c0}}{T_{c0}}
 \equiv \nu t_0,
 \label{eq:r0t0}
 \end{eqnarray}
we see that in the weak coupling regime and for small momenta
the inverse bosonic propagator can be written as
 \begin{eqnarray}
 F_0^{-1} ( \bd{p} , i \bar{\omega} )
 & = & \nu \biggl[ t_0 + \psi \left( \frac{1}{2} + \frac{ | \bar{\omega} |}{ 4 \pi T } \right)  - \psi \left( \frac{1}{2} \right) 
 \nonumber
 \\
 & & \hspace{4mm}
   -   \frac{ v_F^2 p^2  }{ 2 D ( 4 \pi T )^2 }
  \psi^{\prime \prime} \left( \frac{1}{2} + \frac{ | \bar{\omega} |}{ 4 \pi T } \right) 
 \biggr]. \hspace{7mm}
 \end{eqnarray}
The corresponding retarded propagator can be obtained via analytic continuation,
 $ | \bar{\omega } | = - i ( i \bar{\omega} ) {\rm sgn} \bar{\omega} \rightarrow - i \omega $.
Assuming $ | \omega | \ll 4 \pi T$ we may expand the Digamma functions in powers of
frequencies and obtain for the inverse retarded propagator
 \begin{equation}
  F_0^{-1} ( \bd{p} , \omega + i 0^+ ) \approx \nu \left[ t_0 - i \omega / \omega_0  + p^2 / p_0^2 \right],
 \label{eq:F0inv}
 \end{equation}
where
 \begin{equation}
 \omega_0 = 8 T / \pi ,
 \label{eq:omega0def}
 \end{equation}
and
 \begin{equation}
 p_0^2 =  \frac{ 16 D }{ 7 \zeta ( 3 ) }  \left( \frac{  \pi T}{v_F } \right)^2
 \label{eq:p0def}
 \end{equation}
can be identified with the square of the inverse coherence length.
In particular, in the static limit the Gaussian propagator of the
pairing field can be written as
 \begin{equation}
 F_0 ( \bd{p} , 0 ) \approx \frac{1}{\nu [ t_0 + p^2 / p^2_0 ]}.
 \label{eq:Fstat}
 \end{equation}

\begin{figure}[tb]
  \centering
\vspace{7mm}
 \includegraphics[width=0.45\textwidth]{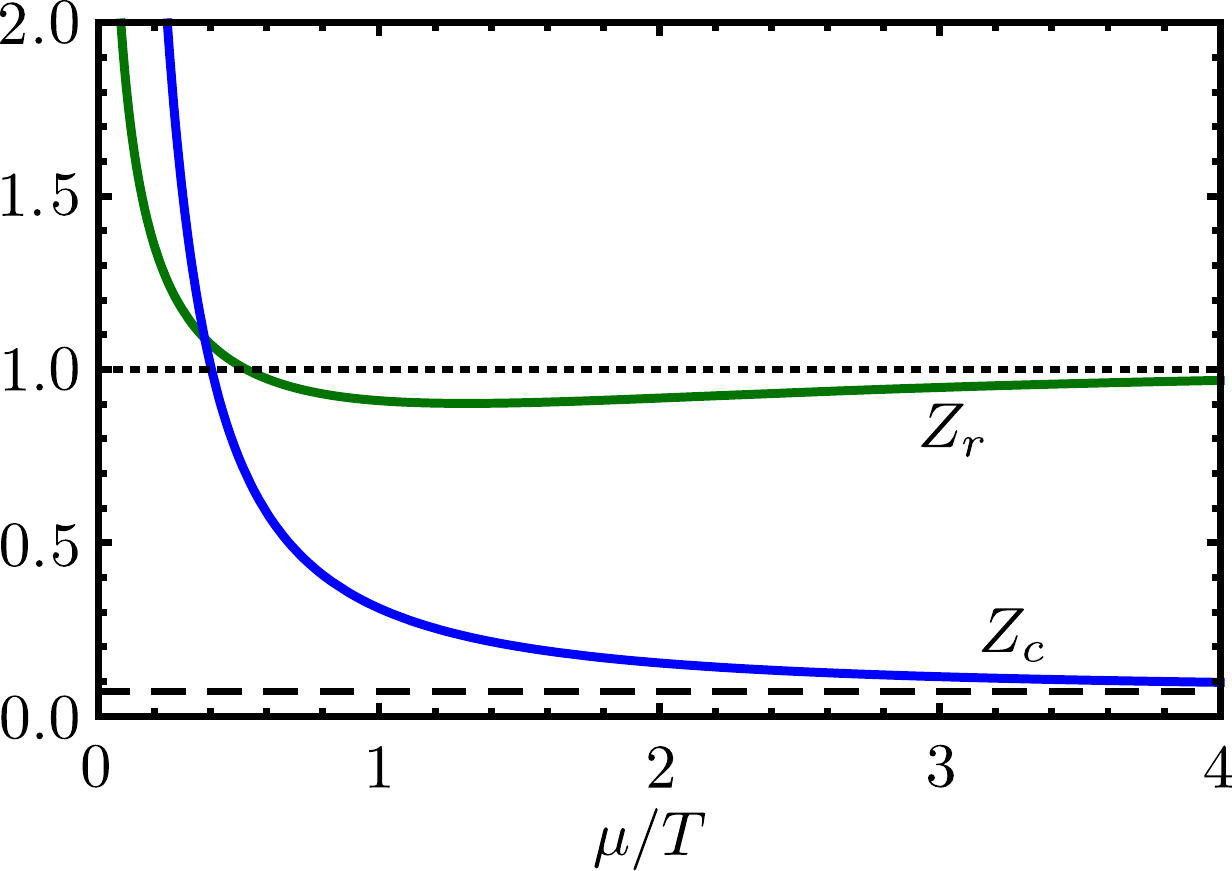}
  \caption{%
Graph of the functions $Z_r ( \alpha )$  and $Z_c ( \alpha )$
defined in Eqs.~(\ref{eq:Zrx}) and (\ref{eq:Zcx}).
While for large $\alpha = \mu /T$ these functions approach finite constants,
$Z_r ( \infty ) =1$ and $Z_c ( \infty ) \approx 0.071$, the asymptotic behavior for
 $\alpha \rightarrow 0$ is $Z_r ( \alpha ) \propto \alpha^{-1/2}$ and
$Z_c ( \alpha ) \propto \alpha^{-3/2}$.
}
\label{fig:ZZgraph}
\end{figure}

The above expressions are only valid in the weak coupling BCS limit at low temperatures,
where 
$\tilde{g} = \nu g  \ll 1$,
 $T  \ll E_F$, and
$ p \lesssim p_0  \ll k_F$.
On the other hand, when $\tilde{g} $ is of the order of unity 
the energy dependence of the
density of states cannot be neglected, so that Eq.~(\ref{eq:PhiLV})
is not valid. 
Setting for simplicity $ \bar{\omega} =0 $ 
(which is sufficient for our purpose because we are 
only interested in 
classical long-wavelength  fluctuations) we write the static propagator
of the order parameter field as
 \begin{equation}
 F_0 ( \bd{p} , 0 ) = \frac{1}{ r_0 + c_0 {p}^2 }.
 \label{eq:F0rc}
 \end{equation}
In the BCS limit we obtain from Eq.~(\ref{eq:F0inv})
 \begin{eqnarray}
 r_0 & \approx & \nu t_0,
 \\
 c_0 & \approx & \nu /p_0^2.
 \end{eqnarray}
More generally, for arbitrary values of $g$ the coefficients $r_0$ and $c_0$ can be obtained directly from
Eq.~(\ref{eq:phimatsubara}). For the parameter $r_0$ which measures the distance to the
critical point we obtain
   \begin{eqnarray}
 r_0 & = &   \left. \Phi_0^{\rm reg} ( 0 , 0 ) \right|_{ T_c}
 - \Phi_0^{\rm reg} ( 0 , 0 )
 \nonumber
 \\
 & = & \int_0^{\infty} d \epsilon \nu ( \epsilon )
 \frac{ \tanh \bigl( \frac{ \epsilon - \mu }{2 T_c } \bigr) -
 \tanh ( \frac{ \epsilon - \mu }{ 2 T } ) }{ 2 ( \epsilon - \mu ) } 
 \nonumber
 \\
 & = & 
\int_0^{\infty} d \epsilon \nu ( \epsilon )
 \frac{ \sinh  \bigl( \frac{ (\epsilon - \mu) }{ 2 T}  t_0 \bigr) }{ 2 ( \epsilon - \mu ) 
 \cosh \bigl( \frac{ \epsilon - \mu }{2 T_{c0} } \bigr) 
 \cosh \bigl( \frac{ \epsilon - \mu }{2 T } \bigr)    } 
, \hspace{7mm}
 \end{eqnarray}
where the energy-dependent density of states (per spin projection)
is in three dimensions given by
 \begin{equation}
 \nu ( \epsilon ) = \int_{\bd{k}} \delta ( \epsilon - \epsilon_{\bd{k}} ) = \frac{m \sqrt{2 m \epsilon}}{2 \pi^2 } = K_3  m \sqrt{2m \epsilon}.
 \end{equation}
Assuming $ \mu > 0$ and introducing the dimensionless integration variable
$x = \epsilon / \mu $ we obtain to leading order in the reduced temperature
$t_0 = ( T - T_{c0})/T_{c0}$,
 \begin{equation}
 r_0 = Z_r (  \mu / T  )  \nu ( \mu )  t_0   ,
 \end{equation}
where the dimensionless function $Z_r ( \alpha )$ is given by
 \begin{equation}
 Z_r ( \alpha ) = \frac{\alpha}{4} \int_0^{\infty} d x  \frac{\sqrt{x}}{\cosh^2 
 \bigl( \alpha \frac{  x-1 }{ 2 }  \bigr)  }.
 \label{eq:Zrx}
 \end{equation}
In the BCS limit where $\mu \approx E_F \gg T$ we may approximate
$Z_r ( \mu / T  ) \approx Z_r ( \infty ) =1$ and obtain $r_0 = \nu t_0$,
in agreement with Eq.~(\ref{eq:r0t0}).
A graph of the function $Z_r ( \mu / T  )$ is shown in Fig.~\ref{fig:ZZgraph}.
Finally, consider the coefficient $c_0$ in Eq.~(\ref{eq:F0rc}), which can be written as
 \begin{eqnarray}
 c_0 & = & - \left. \frac{ \partial \Phi^{\rm reg}_0 ( \bd{p} , 0 ) }{\partial p^2 } \right|_{ p=0}
  \nonumber
 \\
 & = &
\frac{ T}{2m} 
  \sum_{\omega} \int_{\bd{k}} \biggl[ 
 \frac{ \frac{2}{D} \frac{ k^2}{m}}{ ( i \omega - \xi_{\bd{k}} ) ( i \omega + \xi_{\bd{k}} )^3}
 \nonumber
 \\
 & &  \hspace{20mm}
  -\frac{1}{( i \omega - \xi_{\bd{k}} ) ( i \omega + \xi_{\bd{k}} )^2 }
 \biggr] .
 \end{eqnarray}
After carrying out the Matsubara sums we obtain
 \begin{equation}
 c_0 =  Z_c (  \mu / T  )    \frac{\nu ( \mu ) \mu }{2 m T^2} ,
 \end{equation}
with 
 \begin{eqnarray}
 Z_c ( \alpha ) & = & \alpha \int_0^{\infty} d x \sqrt{x} \biggl\{ (x-1 ) S_2 ( \alpha (x-1) ) 
 \nonumber
 \\
&  & \hspace{-15mm} + \frac{4}{3} x \Bigl[ S_2 ( \alpha (x-1)) - 2 \alpha^2 (x-1)^2 S_3 ( \alpha (x-1)) \Bigr] \biggr\}.
 \nonumber
 \\
 & &
 \label{eq:Zcx}
 \end{eqnarray}
Here $S_2 (a )$ and $S_3 ( a )$ are defined by the following fermionic Matsubara sums,
 \begin{eqnarray}
 S_2 ( a ) & = & \sum_{ n = - \infty}^{\infty} \frac{1}{ \left[
 ( \pi ( 2n +1 ))^2 + a^2 \right]^2 } 
 \nonumber
 \\
 & = &  
 \frac{   \sinh a - a   }{ 8 a^3 \cosh^2 ( a/2 )    },
 \\
 S_3 ( a )  & =  & \sum_{ n = - \infty}^{\infty} \frac{1}{ \left[
 ( \pi ( 2n +1 ))^2 + a^2 \right]^3 } 
 \nonumber
 \\
 & = & \frac{1}{32 a^5} \biggl[  6 \tanh (a/2 )    -
 \frac{ 3 a }{  \cosh^2 ( a/2 ) }
 \nonumber
 \\
 & & \hspace{10mm}
 - \frac{ 8 a^2 \sinh^4 ( a/2 )}{ \sinh^3 a } 
 \biggr] .
 \end{eqnarray}
A graph of the function $Z_c ( \alpha )$ is shown in Fig.~\ref{fig:ZZgraph}.
In the BCS limit where $ \mu / T \gg 1$ we obtain to leading order in $D$ dimensions
 \begin{equation}
 Z_c ( \mu /T  ) \sim Z_c ( \infty ) = \frac{ 7 \zeta ( 3 ) }{4 \pi^2 D}.
  \end{equation}

\section*{APPENDIX D: Justification of the classical approximation
close to $T_c$}

\setcounter{equation}{0}
\renewcommand{\theequation}{D\arabic{equation}}
 \label{sec:justification}

To justify the static approximation for pairing fluctuations in the evaluation of 
the electronic self-energy given
in Eq.~(\ref{eq:sigmasing})
we go back to Eq.~(\ref{eq:sigmak}), introduce  the spectral representation  of the
Gaussian pairing propagator, and explicitly carry our the Matsubara sum.
Therefore it is useful to write the Gaussian pairing propagator as
 \begin{equation}
 F_0 ( P ) = \frac{g_0}{ 1 -  g_0 \Phi_0 ( P ) } = g_0 +  g_0^2 \frac{\Phi_0 ( P )}{ 1 -  g_0 \Phi_0 ( P ) }.
 \label{eq:F0split}
 \end{equation}
 Given the fact that particle-particle bubble 
$\Phi_0 ( \bd{p} , i \bar{\omega} )$ vanishes for large $ | \bar{\omega } |$ as $ 1/ | \bar{\omega } |$
we see that the corresponding resummed bubble 
 $\Phi_0 ( P ) / [ 1 - g_0 \Phi_0 ( P ) ]$ vanishes also for large $ | \bar{\omega } |$, so that
it has a spectral representation
 \begin{equation}
 \frac{\Phi_0 ( \bd{p} , i \bar{\omega}  )}{ 1 - g_0 \Phi_0 ( \bd{p} , i \bar{\omega} ) }
 = \int_{- \infty}^{\infty} d \omega^{\prime} \frac{ S ( \bd{p} , \omega^{\prime} )}{ 
    \omega^{\prime} -  i \bar{\omega}  }.
 \label{eq:phispec}
 \end{equation}
The inverse relation is
 \begin{equation}
S ( \bd{p} , \omega ) =  \frac{1}{\pi} 
 {\rm Im} \left[ \frac{\Phi_0 ( \bd{p} ,  \omega + i0^+  )}{ 1 - g_0 
 \Phi_0 ( \bd{p} , \omega + i 0^+ ) }
 \right].
 \end{equation}
From the expansion  (\ref{eq:F0inv}) we see that 
for small frequencies $ | \omega | \ll T $,
 \begin{equation}
\frac{\Phi_0 ( \bd{p} ,  \omega + i0^+  )}{ 1 - g_0 
 \Phi_0 ( \bd{p} , \omega + i 0^+ ) }
 \approx \frac{ 1}{\nu g_0^2} \frac{1}{t_0 - i  \omega / \omega_0 + c_0 p^2  },
 \end{equation}
where  in the BCS regime 
$\omega_0 = 8 T / \pi$ and $p_0 = \sqrt{ 3 / (7 \zeta (3 )) } 4 \pi T / v_F$, see
Eqs.~(\ref{eq:omega0def}) and (\ref{eq:p0def}).
The spectral function is therefore
 \begin{equation}
 S ( \bd{p} , \omega ) = \frac{1}{\pi  \nu g_0^2} 
 \frac{ \omega / \omega_0 }{  ( \omega/ \omega_0 )^2 +  ( t_0 + p^2/ p_0^2 )^2 }.
 \end{equation}
Substituting Eqs.~(\ref{eq:F0split}) and (\ref{eq:phispec}) into
Eq.~(\ref{eq:sigmak}) we may carry out the Matsubara sum and obtain
 \begin{eqnarray}
 \Sigma ( \bd{k} , i \omega ) & = & - g_0 \rho_0 + g_0^2 \int_{\bd{p}} 
\int_{ - \infty}^{\infty} d \omega^{\prime} 
 S ( \bd{p} , \omega^{\prime} )
 \nonumber
 \\
 & & \times  \frac{ b ( \omega^{\prime} ) + f ( \xi_{ \bd{p} - \bd{k}} ) -1 }{
 i \omega - \omega^{\prime} + \xi_{\bd{p} - \bd{k} } }.
 \end{eqnarray}
Here $b ( \omega^{\prime} ) = 1 / [ e^{ \omega^{\prime} / T } -1 ]$ is the
Bose function and $f ( \xi ) = 1 / [ e^{ \xi / T } +1 ]$ is the Fermi function.
From this expression we can now justify the static approximation for temperatures close to
$T_c$. In this regime, the dynamics of the boson is much slower that the dynamics of the fermions because the 
typical value of the boson frequency is $\omega^{\prime} \approx
 \omega_0 ( t_0 + p^2 / p_0^2 ) $, whereas the typical value of the fermion energy
is of order $ v_F p  \ll T$ . In this regime we may approximate the Bose function by its classical 
limit $b ( \omega^{\prime} ) \approx T / \omega^{\prime} $ and neglect the
term  $f ( \xi_{ \bd{p} - \bd{k}} ) -1$ and the constant $- g_0 n$.
In this approximation
 \begin{eqnarray}
 \Sigma ( \bd{k} , i \omega ) & = &  g_0^2  \int_{\bd{p}} 
\int_{ - \infty}^{\infty} d \omega^{\prime} 
 \frac{S ( \bd{p} , \omega^{\prime} )}{\omega^{\prime}}
\frac{ T }{
 i \omega + \xi_{\bd{p} - \bd{k} } }.
 \nonumber
 \\
 & &
 \end{eqnarray}
By definition the  frequency integral gives 
 \begin{equation}
\int_{ - \infty}^{\infty} d \omega^{\prime} 
 \frac{S ( \bd{p} , \omega^{\prime} )}{\omega^{\prime}} =
  \frac{\Phi_0 ( \bd{p} , 0 )}{ 1 - g_0 \Phi_0 ( \bd{p} , 0 ) }.
 \end{equation}
For $p \ll T / v_F$ we may approximate in the numerator
$\Phi_0 ( \bd{p} , 0 ) \approx  1/ g_0$ so that we finally arrive at Eq.~(\ref{eq:sigmasing}).

\section*{APPENDIX E: FRG calculation with
vertex corrections}
\setcounter{equation}{0}
\renewcommand{\theequation}{E\arabic{equation}}
\label{subsec:FRGdamp2}

In the FRG calculation of the quasi-particle damping presented in Sec.~\ref{subsec:FRGdamp}
we have set the mixed fermion-boson interaction 
$\Gamma^{\bar{c}_{\sigma} {c}_{\sigma} \bar{\psi} \psi}_{\Lambda} ( K^{\prime} , K; P^{\prime} , P ) $ equal to zero and
ignored the FRG flow  of the three-point vertices
$\Gamma_{\Lambda}^{ \bar{c}_{\uparrow} \bar{c}_{\downarrow}  \psi } ( K_1^{\prime} , K_2^{\prime} ; P )   $ and
$\Gamma_{\Lambda}^{c_{\downarrow} c_{\uparrow} \bar{\psi} } ( K_1 , K_2 ; P )$.
However, if we assume the usual Fermi liquid scaling in the fermionic sector
and Gaussian critical scaling in the bosonic sector,
it is easy to see that the coupling
\begin{eqnarray}
  & &  \Gamma_{\Lambda}^{ \bar{c}_{\uparrow} \bar{c}_{\downarrow}  \psi } 
( \bd{k}_F , i0^+ , -\bd{k}_F , - i 0^+  ; 0 ,0 ) 
 \nonumber
 \\
 & =  & \Gamma_{\Lambda}^{c_{\downarrow} c_{\uparrow} \bar{\psi} } 
(- \bd{k}_F , - i 0^+ , \bd{k}_F, i 0^+   ; 0 , 0 ) \equiv v_{\Lambda}
 \end{eqnarray}
is relevant at the critical point with scaling dimension $  2- D/2  = 1/2 $ in three dimensions.
In the above definition of $v_{\Lambda}$  
it is understood that the bosonic momenta and frequencies are set equal to zero,
while the fermionic frequency is analytically continued to the real axis and then set
equal to zero with an infinitesimal imaginary part as indicated.
Similarly, the coupling $w_{\Lambda}$ defined by
 \begin{equation}
\Gamma^{\bar{c}_{\sigma} {c}_{\sigma} \bar{\psi} \psi}_{\Lambda} 
( \bd{k}_F , \pm i 0^+ ,  \bd{k}_F  , \pm i 0^+; 0,0 ) \equiv \pm i w_{\Lambda}
 \label{eq:wdef}
 \end{equation}
has scaling dimensions $3-D$ and is therefore marginal in three dimensions.
Hence, for the calculation of  the feedback of critical order parameter fluctuations on the
electronic properties, the RG flow of these two couplings has to be taken into account.
If we neglect all other (irrelevant) vertices, the RG flow of $w_{\Lambda}$ is given by the
truncated flow equation shown graphically in Fig.~\ref{fig:flowinduced}~(a), which is
explicitly written down in Eq.~(\ref{eq:flowgammamixed}).
Introducing the dimensionless rescaled couplings
 \begin{eqnarray}
 \tilde{v}_l & = & \sqrt{ \frac{ K_3 T}{v_F^2 c_{\Lambda} \Lambda}} v_{\Lambda}
 \\
\tilde{w}_l & = & \frac{K_3 T}{v_F c_{\Lambda}} w_{\Lambda} ,
\end{eqnarray}
and  the rescaled  damping
 \begin{equation}
 \tilde{\gamma}_l = \frac{\gamma_{\Lambda}}{v_F \Lambda},
 \end{equation}
we find that Eq.~(\ref{eq:flowgammamixed}) reduces to
 \begin{eqnarray}
 \partial_l \tilde{w}_l & = & - \eta_l \tilde{w}_l +  \frac{ \tilde{v}_l^4}{2 (1 +
 \tilde{r}_l ) \tilde{\gamma}_l^2 } 
 \left[  \arctan (1/\tilde{\gamma}_l) + \frac{ \tilde{\gamma}_l}{ 1 + \tilde{\gamma}_l^2 } \right]
 \nonumber
 \\
 & & + \frac{ \tilde{u}_l  \tilde{v}_l^2 }{ ( 1 + \tilde{r}_l )^2 } \arctan (1/ \tilde{\gamma_l}) .
 \label{eq:vertex4flow_rescaled}
 \end{eqnarray}
Similarly, we obtain 
from the FRG flow equations for the three-point vertices  given in
Eqs.~(\ref{eq:vertex3flow1}, \ref{eq:vertex3flow2}) 
which are 
shown graphically in Fig.~\ref{fig:flowinduced}~(b),
 \begin{equation}
 \partial_l \tilde{v}_l = \frac{1 - \eta_l}{2} \tilde{v}_l +  \frac{ 2 \tilde{v}_l \tilde{w}_l }{ 1 + \tilde{r}_l }
 \arctan (1 / \tilde{\gamma}_l ) .
 \label{eq:vertex3flow_rescaled}
 \end{equation}
Taking into account the couplings $v_{\Lambda}$ and $w_{\Lambda}$,
we obtain from Eq.~(\ref{eq:flowself}) for the flow of the quasiparticle damping,
 \begin{equation}
 \partial_{\Lambda} \gamma_{\Lambda} = K_3 T \left[ \frac{w_{\Lambda}  \Lambda^2}{
 r_{\Lambda} + c_{\Lambda} \Lambda^2 }
 - \frac{  \Lambda v_{\Lambda}^2  \arctan ( v_F \Lambda / \gamma_{\Lambda} )   }{v_F ( r_{\Lambda} + c_{\Lambda} \Lambda^2 ) } \right].
 \end{equation}
In terms of the  rescaled couplings introduced above this can be written as
 \begin{equation}
 \partial_l \tilde{\gamma}_l = \tilde{\gamma}_l - \frac{ \tilde{w}_l }{ 1 + \tilde{r}_l }
 + \frac{\tilde{v}_l^2 \arctan ( 1 / \tilde{\gamma}_l ) }{ 1 + \tilde{r}_l }.
 \label{eq:gammaflow_rescaled}
 \end{equation}
The above system of flow equations should be integrated with the following initial
conditions,
 \begin{eqnarray} 
 \Lambda_0 & = & p_0 \propto \frac{  T }{v_F } ,
 \\
 \tilde{u}_0 & = &    \frac{K_3 T u_0}{c_0^2 \Lambda_0} \propto  
 \left( \frac{T }{ E_F} \right)^2
,
 \\
 \tilde{v}_0 & = & \sqrt{ \frac{K_3 T }{c_0 v_F^2 \Lambda_0}} 
\propto \frac{T }{ E_F},
 \\
 \tilde{w}_0 & = & 0,
 \\
\tilde{\gamma}_0 & = &  \frac{\gamma_{\rm FL}}{ v_F p_0 } \propto \frac{T }{ E_F}.
 \end{eqnarray}
The finite initial value of $\tilde{\gamma}_l$ takes into account
the usual Fermi liquid contribution to the quasi-particle damping which is not related to
critical pairing fluctuations, see the discussion in Sec.\ref{subsec:gauss}.
For our model defined in Eq.~(\ref{eq:Sbare}) 
second order perturbation theory in the regularized interaction $g$ yields
for the quasiparticle damping at low temperatures\cite{Abrikosov63} 
\begin{align}
 \gamma_{\rm FL}=\frac{m^3}{8\pi} g^2 T^2 = \frac{\pi^2}{4} \tilde{g}^2 T^2 / E_F .
\end{align}
More generally, we may use the generic form of the quasiparticle damping of a Fermi liquid
$\gamma_{\rm FL} = C_{\rm FL} T^2 / E_F$ [see Eq.~(\ref{eq:dampFL})], where the
value of the numerical constant $C_{\rm FL}$ is determined by
all types of interaction processes into account, 
including those  which are not included in our effective low-energy
model (\ref{eq:Sbare}).
For simplicity we choose $C_{\rm FL}$ such that the initial condition
for the rescaled damping is given by $\tilde{\gamma}_0=T/E_F$.

\begin{figure}[tb]
  \centering
 \includegraphics[width=0.48\textwidth]{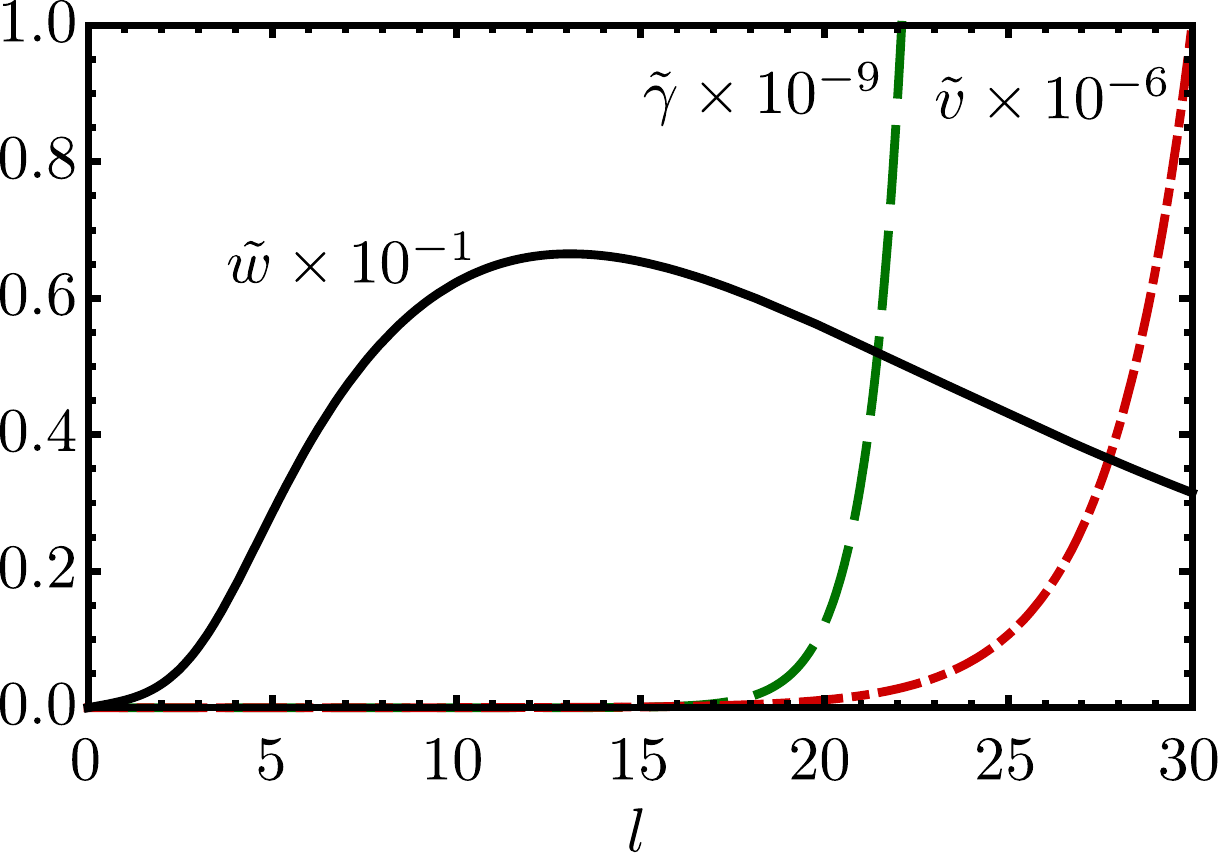}%
  \caption{%
RG flow
of the rescaled damping $\tilde{\gamma}_l$ (green dashed line),  three-legged vertex
$\tilde{v}_l$ (red dashed-dotted line), and mixed four-legged vertex 
$\tilde{w}_l$ (black solid line) for $T=T_c=0.13E_F$.
For large $l$ the rescaled damping $\tilde{\gamma}_l$ is proportional to
$e^{l}$, so that $e^{-l} \tilde{\gamma}_l$ approaches a finite limit.
}
\label{fig:tilde_fermionic_flow}
\end{figure}

\begin{figure}[tb]
  \centering
 \includegraphics[width=0.48\textwidth]{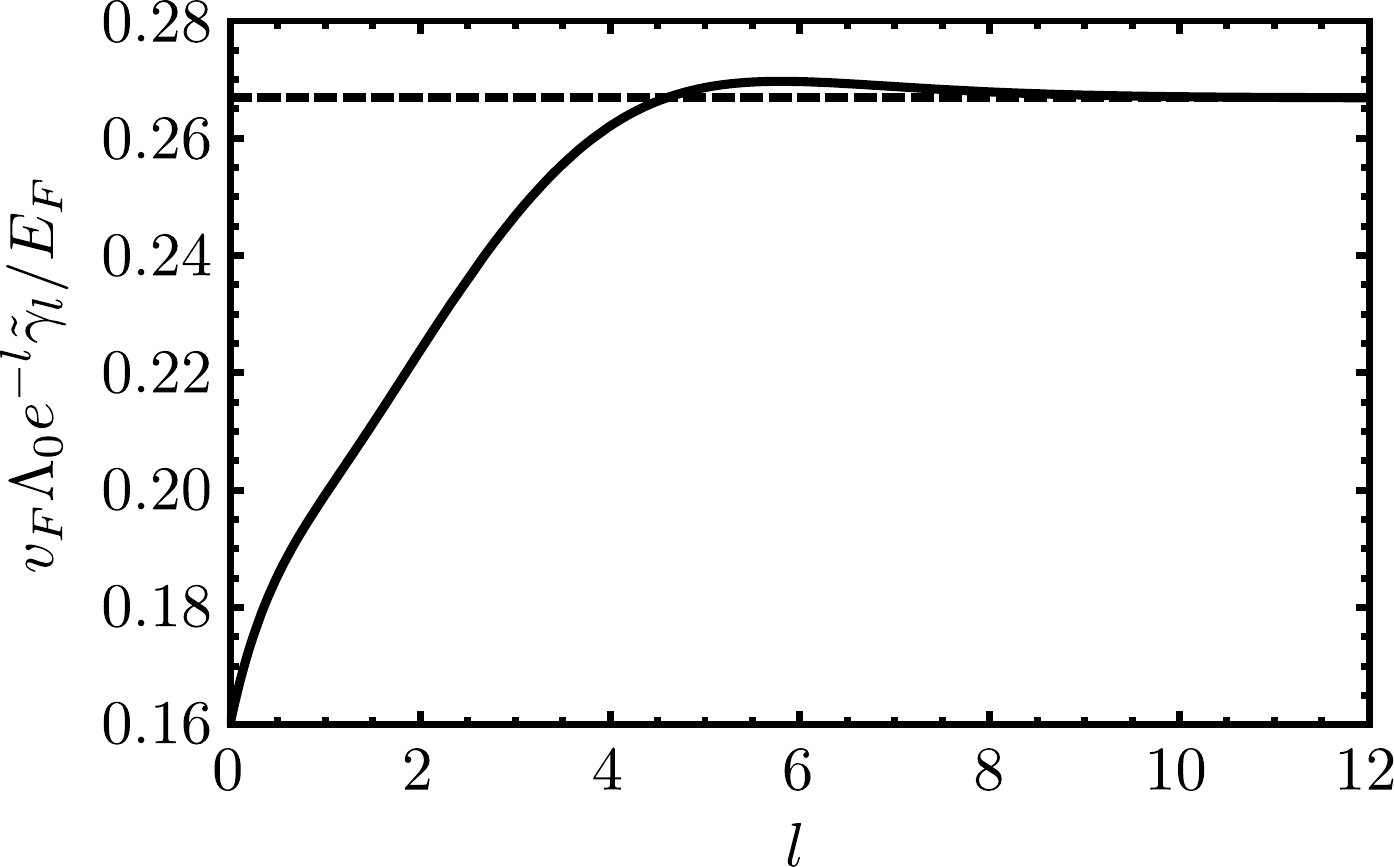}
  \caption{%
RG flow  of the physical damping 
$\gamma_{\Lambda} = v_F \Lambda \tilde{\gamma}_l$
as a function of $l = \ln ( \Lambda_0 / \Lambda )$ for $T=T_c=0.13E_F$.
The damping converges against a constant value for large $l$ (dashed black line).
}
\label{fig:gamma_flow}
\end{figure}

\begin{figure}[tb]
  \centering
 \includegraphics[width=0.48\textwidth]{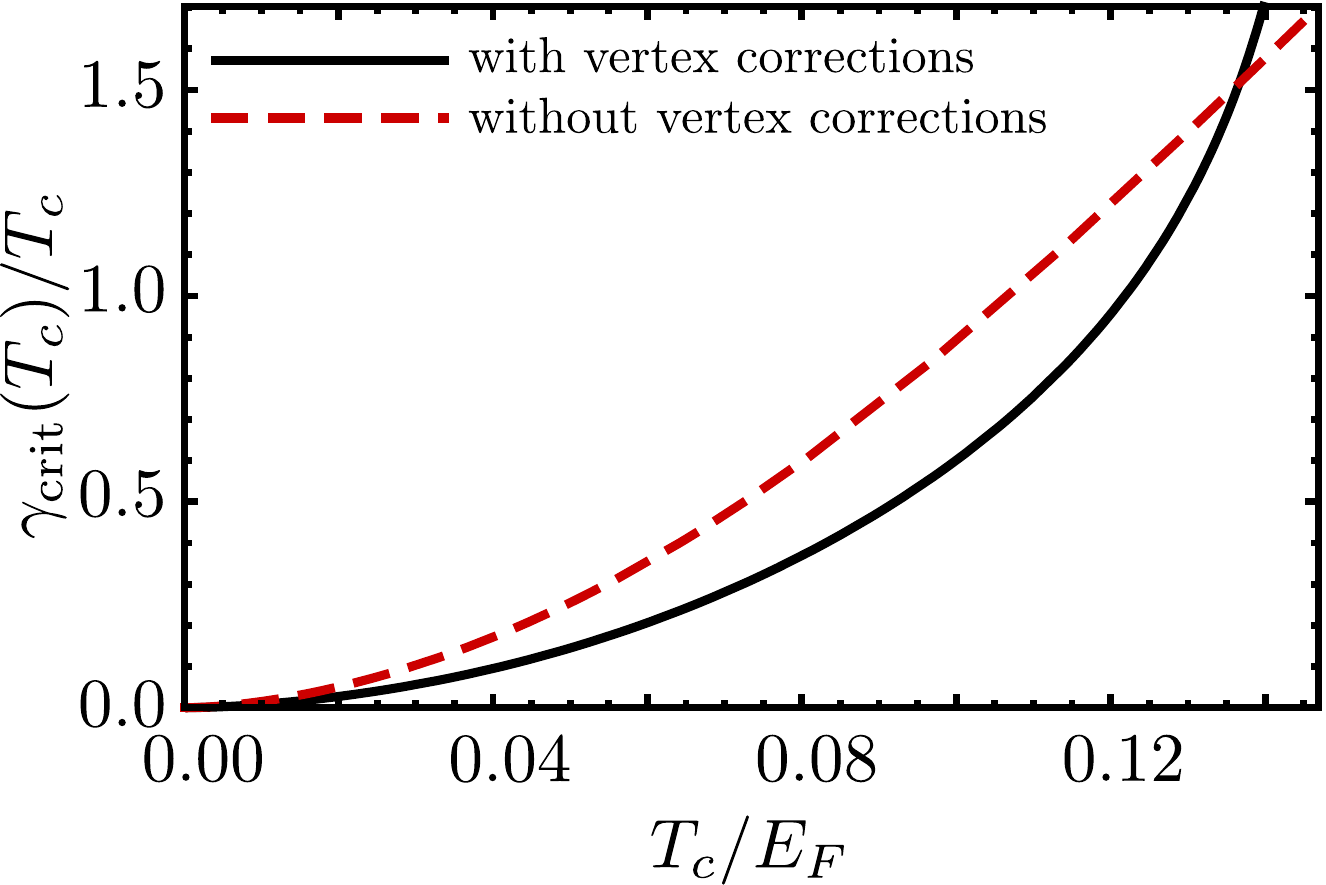}%
  \caption{%
The black solid line represents our numerical 
result for the damping $\gamma_{\rm crit} ( T_c) $ due to classical critical pairing fluctuations
including vertex corrections,  while the
red dashed line is the corresponding result without vertex corrections
discussed in Sec.~\ref{subsec:FRGdamp}.
}
\label{fig:damping_w}
\end{figure}

After solving the
flow equations (\ref{eq:rlflow}--\ref{eq:etaflow}) in the bosonic sector
for various temperatures, we may substitute 
the result into the flow equations  (\ref{eq:vertex4flow_rescaled}) 
and (\ref{eq:vertex3flow_rescaled}) for the four-legged and three-legged vertices, and into
the  flow equation (\ref{eq:gammaflow_rescaled}) for the quasiparticle damping.
In Fig.~\ref{fig:tilde_fermionic_flow} we plot the flow of 
the rescaled couplings  $\tilde{v}_l$ and
$\tilde{w}_l$ as well as the rescaled quasiparticle damping $\tilde{\gamma}_l$ 
for $T=T_c=0.13E_F$. 
From the solution for the dimensionless rescaled damping
$\tilde{\gamma}_l$
we can reconstruct the physical quasiparticle damping due classical  pairing
parameter fluctuations as follows,
 \begin{equation}
 \gamma_{\rm crit} ( T )  =  v_F \Lambda_0   \lim_{l \rightarrow \infty} e^{-l} \tilde{\gamma}_l 
 - \gamma_{\rm FL} ,
 \end{equation}
where the   subtraction of the initial condition $\gamma_{\rm FL} = v_F \Lambda_0 \tilde{\gamma}_0 $ is necessary to isolate the contribution from classical 
pairing fluctuations.
The rescaled damping  $\tilde{\gamma}_l$ is proportional to $ \propto e^l$ for large $l$,
so that  $e^{-l}\tilde{\gamma}_l$ converges
against a constant value, as shown in  Fig.~\ref{fig:gamma_flow}.
Our final result for the contribution from classical critical fluctuations to the
quasi-particle damping $\gamma_{\rm crit} ( T_c )$ at the  $T_c$ 
is shown in Fig.~\ref{fig:damping_w}, where we also show the corresponding expression
without vertex corrections derived in Sec.~\ref{subsec:FRGdamp}.
Comparing the two curves we can see, that
in the weak coupling limit  $T_c \ll E_F$ the qualitative behavior 
is not modified by vertex corrections, while for larger values of the interaction  
(corresponding to $T_c \approx 0.1 E_F$)
vertex corrections do have a significant effect.
One should keep in mind, however, that in the derivation of the flow equations
(\ref{eq:vertex4flow_rescaled})
and
(\ref{eq:vertex3flow_rescaled}) we have made several simplifications 
(for example, we have projected all external momenta of the vertices on the Fermi surface)
which
can only be expected to be quantitatively accurate  in the weak coupling BCS regime.
Hence, quantitative accuracy of our FRG calculation including vertex corrections 
can only be expected  for $T_c / E_F \ll 1$.
In this regime our FRG result for the quasiparticle damping
shown in Fig.~\ref{fig:damping_w} can be fitted by
\begin{equation}
 \gamma_{\rm crit} (T_c) \approx C \frac{T_c^3}{E_F^2} \ln(E_F / T_c),
\end{equation}
with $C \approx 18$.
The above weak-coupling result including vertex corrections
confirms our result of Sec.~\ref{subsec:FRGdamp} given in Eq.~(\ref{eq:gammaint}).
Note, however, that vertex corrections reduce the numerical value of the prefactor $C$ from
$34$ to $18$, which is still large compared with unity.
A controlled calculation of the precise numerical value of $C$ is 
beyond the scope of this work; in the calculation including vertex corrections
of the numerical value of $C$ is also sensitive to the choice
of the numerical coefficient $C_{\rm FL}$ in the expression for the
Fermi liquid damping $\gamma_{\rm FL} = C_{\rm FL} T^2 / E_F$
defining the initial condition for the FRG flow equations.

\begin{figure}[t!]
\vspace{7mm}
  \centering
 \includegraphics[width=0.44\textwidth]{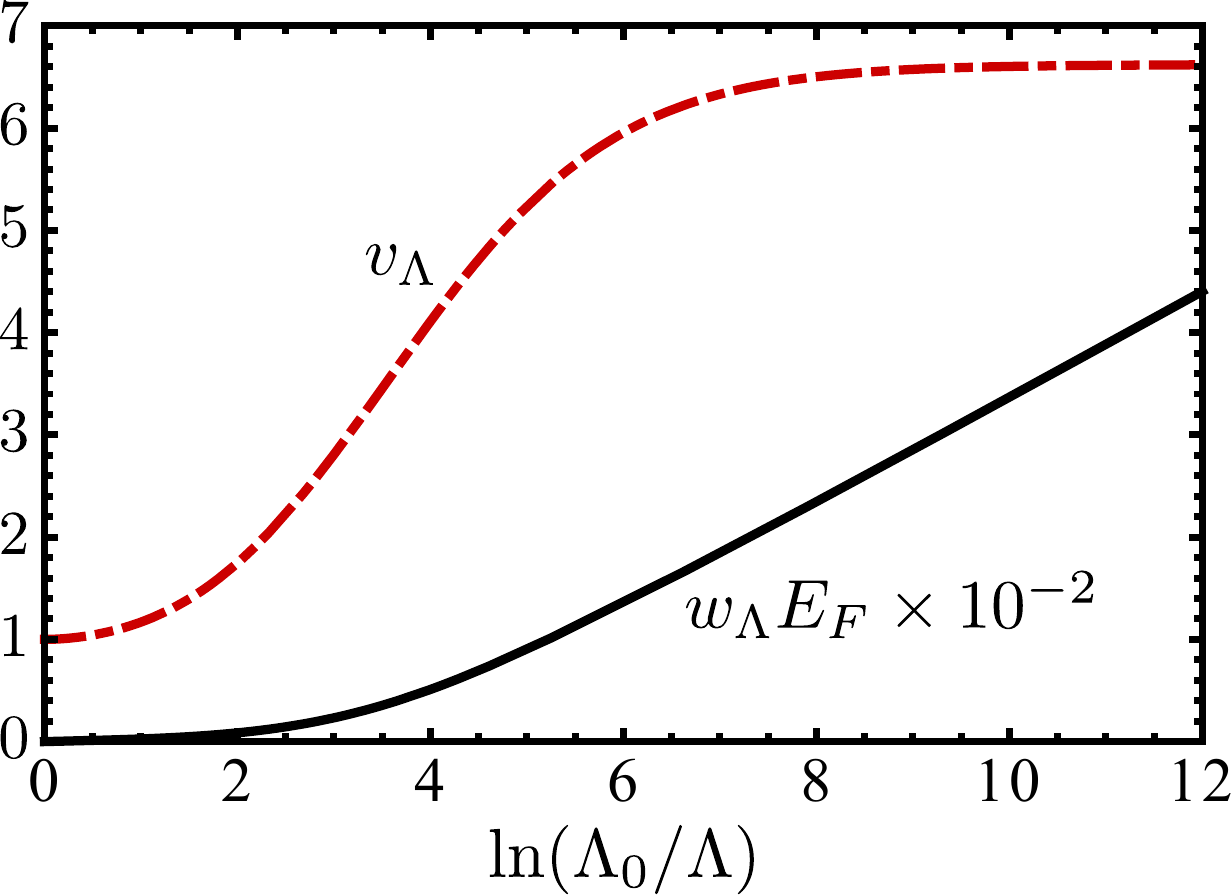}
  \caption{%
RG flow of the physical couplings $v_\Lambda$ and $w_\Lambda$ 
as a function of $l = \ln ( \Lambda_0 / \Lambda )$ for $T=T_c=0.13E_F$.
}
\label{fig:v_w_flow}
\end{figure}
Finally, let us point out that our FRG calculation predicts
that at the critical point the marginal part $w_{\Lambda}$
of the mixed four-point vertex defined in Eq.~(\ref{eq:wdef})
diverges logarithmically for
vanishing cutoff $\Lambda \rightarrow 0$, while the
relevant part $v_{\Lambda}$ of the three-legged vertex approaches a finite value 
in this limit.  To see this, we plot these un-rescaled couplings in Fig.~\ref{fig:v_w_flow}
for $T=T_c=0.13 E_F$ as function of the logarithmic 
flow parameter $l = \ln ( \Lambda_0 / \Lambda )$.
Our  numerical result for $w_{\Lambda}$ is for small $\Lambda$ (corresponding to large $l$ of the form  $w_\lambda \propto \log (\Lambda_0/\Lambda)$.
The logarithmic growth of the vertex correction at $T_c$
can be understood analytically
from the flow equation \eqref{eq:vertex4flow_rescaled}, which implies that
$\tilde{w}_l \propto e^{-\eta_{\star} l}$ for large $l$, where
$\eta_{\star}$ is the fixed point value of the anomalous dimension.
In the limit of large $l$ the flow equation \eqref{eq:vertex4flow_rescaled} therefore  
reduces to
\begin{eqnarray}
\partial_l \tilde{w}_l & \approx & - \eta_\star \tilde{w}_l  + \textrm{const}\
e^{-\eta_{\star} l} .
\label{eq:vertex4flow_rescaled_large_l}
\end{eqnarray}
The analytic solution to this inhomogeneous differential equation is
given by $\tilde{w}_l \propto l e^{-\eta_\star l}$. If we
scale back to the physical coupling we find $w_l\propto l$ for large $l$.
The logarithmic divergence of a vertex correction associated with a marginal
coupling at the critical point should not be surprising.
Our FRG approach automatically takes care of this divergence and its feedback
to the other scale-dependent couplings in the problem.
The non-analytic form of the quasiparticle damping
is not modified by this divergence.

\end{appendix}

\end{document}